\documentclass[12pt,reqno]{amsart}
\allowdisplaybreaks[4]
\usepackage{graphicx} 
\usepackage{amssymb}
\usepackage{color}
\usepackage{amsmath}
\usepackage{cite}
\usepackage{subfigure}
\usepackage{graphicx}
\usepackage{epstopdf}
\usepackage{bibentry}

\begin{document}
\title{Rational and semi-rational solutions of the nonlocal Davey-Stewartson equations}
\author{Jiguang Rao$\dag$, Yi Cheng$\dag$, Jingsong He$\ddag$}
\thanks{$^*$ Corresponding author: hejingsong@nbu.edu.cn, jshe@ustc.edu.cn}
\dedicatory {$\dag$ School of Mathematical Sciences, USTC, Hefei, Anhui 230026, P. R. China\\
$\ddag$ Department of Mathematics, Ningbo University,
Ningbo, Zhejiang 315211, P.\ R.\ China\\
}
\begin{abstract}
In this paper, the  partially party-time ($PT$) symmetric  nonlocal Davey-Stewartson (DS) equations with respect to $x$  is called $x$-nonlocal DS equations, while  a fully $PT$ symmetric  nonlocal  DSII equation is called nonlocal DSII equation.
Three kinds of solutions, namely breather,  rational and semi-rational solutions for these nonlocal DS equations are derived by employing the bilinear method. For the $x$-nonlocal DS equations, the usual ($2+1$)-dimensional breathers are periodic in $x$ direction and localized in $y$ direction. Nonsingular rational solutions are lumps, and semi-rational solutions are composed of lumps, breathers and periodic line waves. For the nonlocal DSII equation, line breathers are periodic in both $x$ and $y$ directions with parallels in profile, but localized in time. Nonsingular rational solutions are ($2+1$)-dimensional line rogue waves, which arise from a constant background and disappear into the same  constant background, and this process only lasts for a short period of time. Semi-rational solutions describe interactions of line rogue waves and periodic line waves.
\end{abstract}
 \maketitle \vspace{-0.9cm}
 \noindent{{\bf Keywords}: Party-time-symmetry,  Nonlocal Davey-Stewartson equations, Rogue waves, \\ \indent  Rational solution, Semi-rational solution.}

\section{Introduction}

Physical systems exhibiting $PT$-symmetry is an important  subject of intense investigations in the past few years\cite{prl-1,prl-2,prl-3,prl-4,prl-5}. In general, the name of $PT$-symmetry was derived based on the works of Bender and Boettcher\cite{prl-1,prl-5}, who pointed out that in quantum mechanics a wide class of non-Hermitain but $PT$-symmetry Hamiltonians can possess entirely real spectra as long as the $PT$ symmetry is not spontaneously broken. This concept has since been applied
to optics \cite{2,3}, Bose-Einstein condensation \cite{4}, electric circuits
\cite{5}, mechanical systems \cite{6}, magnetics \cite{magnetics} and other settings.
Basically, the non-Hermitian Hamiltonian $H=\widehat{P}^{2}/2+V(x)$ is $PT$ symmetric when it satisfies the condition $V(x)=V^{*}(-x)$, where $\widehat{P}$ denotes the momentum operator and $V(x)$ is the complex potential \cite{prl-1,prl-5}.
Recently,  the concept of $PT$ symmetry in multidimensions has been generalized to include partial-parity-time symmetry, and it is shown that partial-parity-time-symmetric systems share most of the properties of $PT$ systems\cite{yang}. Besides, the partially $PT$ symmetry has been shown possible interesting applications in optics \cite{yang, par1,par2,par3,par4,par5,par6}. Thus more researches about partially $PT$ symmetry are also inevitable and worthwhile,  such as the integrable models and nonlinear waves in this system.

In \cite{fokas-pt} Fokas, and in \cite{ab-pt} Ablowitz and Musslimani,   introduced the following nonlocal DS equation
\begin{equation} \label{DSIIeq}
\begin{aligned}
&iA_{t}=A_{xx}+\gamma^2 A_{yy}+(\epsilon V-2Q)A,\\
&Q_{xx}-\gamma^2 Q_{yy}=(\epsilon V)_{xx}, \epsilon=\pm1\,, \\
\end{aligned}
\end{equation}
where $\gamma^2=\pm1$, $A$ and $Q$ are two functions of $x,y,t$. With different definition of the potential $V$, this equation can be classified into two types of nonlocal nonlinear equation:
\begin{itemize}
\item (i) When
\begin{equation} \label{v1}
\begin{aligned}
V=A(x,y,t)\,[A(-x,y,t)]^{*}\,,
\end{aligned}
\end{equation}
namely the potential $V$ satisfies the partial $PT$ symmetry requirement, then the equation given by \eqref{DSIIeq} and \eqref{v1} is  a partially PT symmetric DS equation \cite{fokas-pt,ab-pt} with respect to x, which is called $x$-nonlocal DS eqation.  The case $\gamma^2=1$ is called the $x$-nonlocal DSI eqation, while $\gamma^2=-1$ is the $x$-nonlocal DSII equation.
\item(ii) When
\begin{equation} \label{v2}
\begin{aligned}
V=A(x,y,t)\,[A(-x,-y,t)]^{*}\,,
\end{aligned}
\end{equation}
thus the potential $V$ satisfies the fully PT symmetry requirement. In this case, equation given by \eqref{DSIIeq} and \eqref{v2} is a fully PT symmetric DS equation, which is  denoted by the nonlocal DS equation \cite{fokas-pt,ab-pt}.   In the
case $\gamma^2=1$,  this system defined in \eqref{DSIIeq} and \eqref{v2}  is called the nonlocal DSI equation, whereas in the case $\gamma^2=-1$, the nonlocal DSII equation.
\end{itemize}
These nonlocal DS equations are  natural two-dimensional extensions of
the nonlocal nonlinear Schr\"odinger (NLS) equation proposed by Ablowitz and Musslimani \cite{ablowitz}.
Note that two types of rogue wave solutions for the nonlocal DSI equation were obtained formally in \cite{jigu}, and Darboux transformations and global explicit solutions for the $x$-nonlocal DSI equation were given in \cite{zzx2}.  Besides, a lot of work were done after that for these equations and other nonlocal equations\cite{zhang,chow,xutao,zuo2,zuo1,ab-pt,fokas-pt,ab-pt1,ab-pt2,zzx1,zzx2,lou,zhiwei,wei,taojpsj}.

Motivated by the physical importance of partially and fully $PT$-symmetry systems in the  multi-dimension space \cite{yang, par1,par2,par3,par4,par5,par6}, we investigate above two types of the  nonlocal DS equations.  In this work, we focus  on following aspects:
\begin{itemize}
\item (i) What type of solutions exist for the partially and fully $PT$ symmetric nonlocal DS equations respectively.
\item (ii) The difference between the solutions of the partially and fully $PT$ symmetric nonlocal DS equations.
\end{itemize}

The outline of the paper is organized as follows: In section \ref{2}, several kinds of  solutions for the partially  $PT$ symmetric nonlocal DS equations, namely breathers, lumps and mixed solutions consisting of lumps, breathers and periodic line waves are derived by the bilinear method and a long wave limit, their typical dynamics are analyzed and illustrated.
In section \ref{3}, rogue wave solutions and semi-rational solutions composed of rogue waves and periodic line waves of the fully $PT$ symmetric nonlocal DSII equation are derived and demonstrated.   In section \ref{4}, the difference between the solutions of partially $PT$ symmetric nonlocal DS equations and fully $PT$ symmetric nonlocal DSII equation are listed,  and further discussion are given.

\section{Solutions of the $x$-nonlocal DS equations}\label{2}$\\$
In this section, we focus on solutions of the $x$-nonlocal DS equations defined in \eqref{DSIIeq} and \eqref{v1} by employing the bilinear method.
The $x$-nonlocal DS equations is translated into the following bilinear form :
\begin{equation}\label{DSbi}
\begin{aligned}
&(D_{x}^{2}+\gamma^2 D_{y}^{2}-iD_{t})g \cdot f =0,\\
&(D_{x}^{2}-\gamma^2 D_{y}^{2})f \cdot f =2\epsilon f^{2}-2\epsilon gg^*(-x,y,t),\\
\end{aligned}
\end{equation}
through the variable transformation
\begin{equation} \label{BTtr}
\begin{aligned}
A=\sqrt{2}\frac{g}{f}\,,Q=\epsilon-2( {\rm log} f)_{xx}.
\end{aligned}
\end{equation}
Here $f\,,g$ defined in (\ref{DSbi}) are functions with respect to three variables $x\,,y$ and $t$, and function $f$ satisfies the condition
\begin{equation} \label{gh}
\begin{aligned}
&[f(-x,y,t)]^{*}=f(x,y,t)\,,
\end{aligned}
\end{equation}
and the operator $D$ is the Hirota's bilinear differential operator\cite{hirota} defined by
\begin{equation}
\begin{aligned}
&P(D_{x},D_{y},D_{t}, )F(x,y,t\cdot\cdot\cdot)\cdot G(x,y,t,\cdot\cdot\cdot)\\
=&P(\partial_{x}-\partial_{x^{'}},\partial_{y}-\partial_{y^{'}},
\partial_{t}-\partial_{t^{'}},\cdot\cdot\cdot)F(x,y,t,\cdot\cdot\cdot)G(x^{'},y^{'},t^{'},\cdot\cdot\cdot)|_{x^{'}=x,y^{'}=y,t^{'}=t},\nonumber
\end{aligned}
\end{equation}
where P is a polynomial of $D_{x}$,$D_{y}$,$D_{t},\cdot\cdot\cdot$.

\subsection{Solitons, breathers of the $x$-nonlocal DS equations}$\\$

By the Hirota's bilinear method with the perturbation expansion \cite{hirota}, and set
 $f$ and $g$ be the forms of
 \begin{equation}\label{rfg}
\begin{aligned}
f=&\sum_{\mu=0,1}\exp(\sum_{k<j}^{(N)}\mu_{k}\mu_{j}A_{kj}+\sum_{k=1}^{N}\mu_{k}\eta_{k}),\\
g=&\sum_{\mu=0,1}\exp(\sum_{i<j}^{(N)}\mu_{k}\mu_{j}A_{kj}+\sum_{k=1}^{N}\mu_{k}(\eta_{k}+i\phi_{k})),\\
\end{aligned}
\end{equation}
then \eqref{BTtr} produces the $N$-soliton solutions of the $x$-nonlocal DS equations. Here
\begin{equation}\label{cs1}
\begin{aligned}
\eta_{j}&=iP_{j}\,x+Q_{j}\,y+\Omega_{j}t+\eta_{j}^{0}\,,\\
\Omega_{j}&=\sqrt{-1-\frac{4\epsilon}{P_{j}^{2}+\gamma^2Q_{j}^{2}}}(P_{j}^{2}-\gamma^2Q_{j}^{2})\,,\\
\exp(A_{jk})&=\frac{(P_{j}-P_{k})^{2}+\gamma^2(Q_{j}-Q_{k})^{2}+2\epsilon-2\epsilon\cos(\phi_{j}-\phi_{k})}{(P_{j}+P_{k})^{2}+\gamma^2(Q_{j}+Q_{k})^{2}+2\epsilon-2\epsilon\cos(\phi_{j}+\phi_{k})}\,,\\
\cos(\phi_{j})&=1+\frac{P_{j}^{2}+\gamma^2Q_{j}^{2}}{2\epsilon}\,,\sin(\phi_{j})=-\frac{(P_{j}^{2}+\gamma^2Q_{j}^{2})\sqrt{-1-\frac{4\epsilon}{P_{j}^{2}+\gamma^2Q_{i}^{2}}}}{2\epsilon},
\end{aligned}
\end{equation}
and $P_{j}\,,Q_{j}\,$ are freely real parameters, and $\eta_{k}^{0}$ is an arbitrary complex constant. The natation $\sum_{\mu=0}$ indicates summation over all possible combinations of $\mu_{1}=0,1\,,\mu_{2}=0,1\,,...,\mu_{N}=0,1$; the $\sum_{j<k}^{N}$ summation is over all possible combinations of the $N$
elements with the specific condition $j<k$.

\begin{figure}[!htbp]
\centering
\subfigure[]{\includegraphics[height=3cm,width=5.2cm]{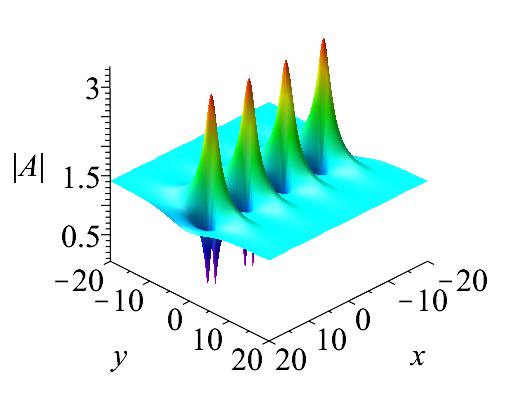}}\quad
\subfigure[]{\includegraphics[height=3cm,width=5.2cm]{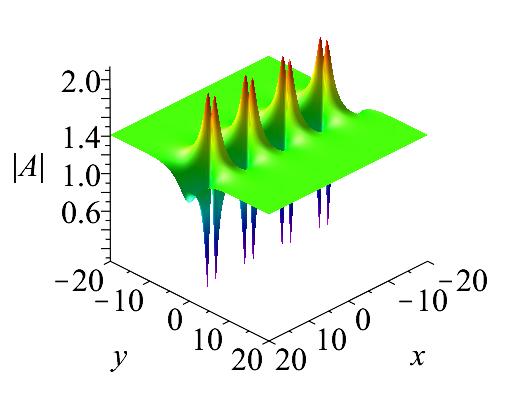}}\quad
\subfigure[]{\includegraphics[height=3cm,width=5.2cm]{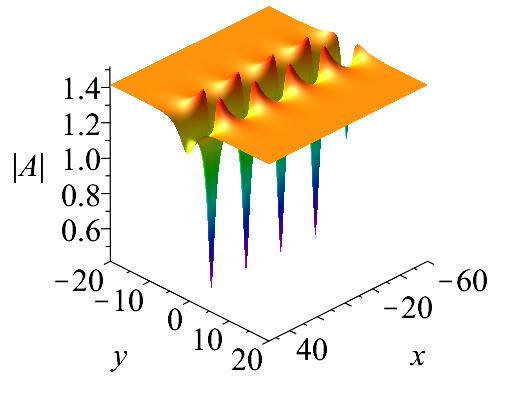}}
\caption{Three types of one-breather solution $|A|$  of the $x$-nonlocal DSI equation given by (\ref{br}) at time $t=0$  in the $(x,y)$-plane with parameters $\gamma^2=1,\epsilon=-1$: (a) $P=\frac{2}{3}\,,Q_{0}=\frac{1}{4}\,,\eta^{0}=0$\,, (b) $P=\frac{2}{3}\,,Q_{0}=\frac{1}{2}\,,\eta^{0}=0$\,, (c)$P=\frac{1}{4}\,,Q_{0}=\frac{1}{2}\,,\eta^{0}=0$.~}\label{fig1}
\end{figure}

Here, it should be emphasized that the constraint $-1-\frac{4\epsilon}{P_{k}^{2}+\gamma^2Q_{k}^{2}}>0$ must be hold for $\Omega_{k}$ to be real and $|\sin(\phi_{k})|<1$, hereafter we set $\epsilon=-1$ when $\gamma^2=1$ in this paper. In particular, when $Q_{j}=0$, the corresponding solutions are periodic line waves, which are periodic in $x$ direction and localized in $y$ direction on the $(x,y)$-plane. Although these soliton solutions also have singularities as soliton solutions of the nonlocal NLS equation \cite{ablowitz},  but general $n$-breather solutions for the $x$-nonlocal DS equations can be generated by
the following parameter constraints
\begin{equation}\label{pa-01}
\begin{aligned}
N=2n\,,P_{n+j}=-P_{j}\,,Q_{n+j}=Q_{j}\,,\eta_{n+j}^{0}=\eta_{j}^{0^{*}}.
\end{aligned}
\end{equation}
Next, we consider the one-breather solutions. For simplicity, we set $n=1\,,P_{1}=-P_{2}=P\,,Q_{1}=Q_{2}=Q_{0}\,,\eta_{1}^{0}=\eta_{2}^{0}=\eta^{0}$, where $P\,,Q_{0}\,,\eta^{0}$ are real.
The one breather can also be expressed in terms of hyperbolic and trigonometric functions as
\begin{equation} \label{br}
\begin{aligned}
A=\sqrt{2}\frac{g_{1}}{f_{1}}\,,Q=\epsilon-2( {\rm log} f_{1})_{xx},
\end{aligned}
\end{equation}
where
\begin{equation} \label{bfg}
\begin{aligned}
f_{1}=&\sqrt{M}\cosh {\Theta}+\cos(P\,x)\,,\\
g_{1}=&\sqrt{M}[\cos^{2}\phi \cosh {\Theta}+\sin^{2}{\phi}\sinh{\Theta}+i\cos\phi\sin\phi(\cosh{\Theta}-\sinh{\Theta})]\\
&+\cos(P\,x)(\cos\phi+i\,\sin\phi)\,,
\end{aligned}
\end{equation}
 $M\,,\Theta\,,\phi$ and $\eta_{0}$ are defined by
 \begin{equation}
\begin{aligned}
&M=-\frac{4\epsilon P^2}{(\gamma^2Q_{0}^2+P^2)^2+4\epsilon P^2}\,,\exp(\eta_{0})=\sqrt{M}\exp(\eta^{0}),\\ &\exp(i\phi)=\frac{1}{2\epsilon}(P^{2}+\gamma^2Q_{0}^{2}+2\epsilon)+\frac{1}{2\epsilon}i\sqrt{-1-\frac{4\epsilon}{P^{2}+\gamma^2Q_{0}^{2}}}(P^{2}+\gamma^2Q_{0}^{2})\,,\\
&\Theta=-(Q_{0}y+\sqrt{-1-\frac{4\epsilon}{P^{2}+Q_{0}^{2}}}(P^{2}-\gamma^2Q_{0}^{2})t+\eta_{0}).\nonumber
\end{aligned}
\end{equation}
The period of $|A|$ is $\frac{2\pi}{P}$ along the $x$ direction on the $(x,y)$-plane. The dynamical profiles of three types of breathers are illustrated in Fig.\ref{fig1}. The high-order breathers can also be generated from \eqref{rfg} under parameter constrains \eqref{pa-01}, which still keep periodic in x direction and localized in y direction.

\begin{figure}[!htbp]
\centering
\subfigure[]{\includegraphics[height=3cm,width=5.2cm]{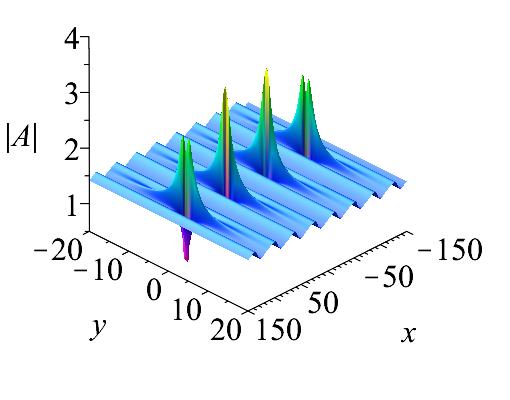}}
\subfigure[]{\includegraphics[height=3cm,width=5.2cm]{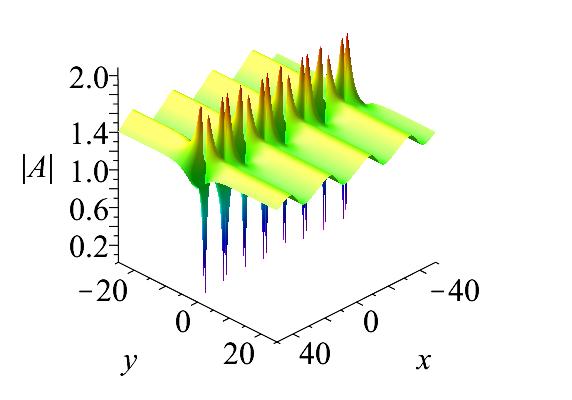}}
\subfigure[]{\includegraphics[height=3cm,width=5.2cm]{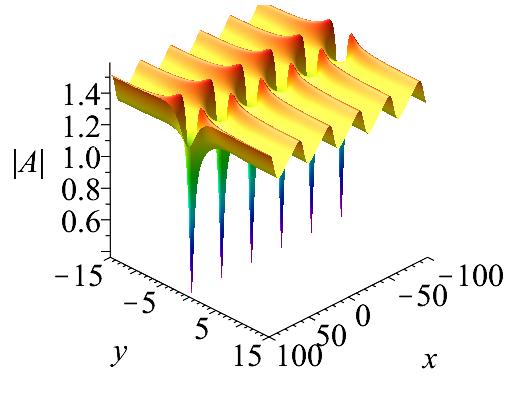}}
\caption{Hybrid solutions $|A|$ of the $x$-nonlocal DSI equation composed of one breather and periodic line waves at time $t=0$ with parameters $\gamma^2=1,\epsilon=-1$: (a) $P_{0}=\frac{1}{12}\,,Q^{'}_{0}=\frac{1}{20}\,,P_{3}=\frac{1}{5}$ \,; (b) $P_{0}=\frac{1}{2}\,,Q^{'}_{0}=\frac{1}{2}\,,P_{3}=\frac{1}{4}$ \,; (c) $P_{0}=\frac{1}{6}\,,Q^{'}_{0}=\frac{1}{6}\,,P_{3}=\frac{1}{4}$.~}\label{fig02}
\end{figure}

In addition to breather solutions, a family of analytical solutions consisting of breathers and periodic line waves for the $x$-nonlocal DS equations can also be generated by taking
\begin{equation}\label{pa-1}
\begin{aligned}
N=2n+1\,,P_{n+j}=-P_{j}\,,Q_{n+j}=Q_{j}\,,\eta_{n+j}^{0}=\eta_{j}^{0^{*}}\,,Q_{2n+1}=0
\end{aligned}
\end{equation}
in \eqref{rfg}. This subclass of hybrid solutions describes $n$-breathers on a background of periodic line waves,  the period of those line waves is $\frac{2\pi}{P_{2n+1}}$ along $x$ direction. For example, with parameter choices
\begin{equation}\label{pa-11}
\begin{aligned}
N=3\,,P_{1}=-P_{2}=P_{0}\,,Q_{1}=Q_{2}=Q^{'}_{0}\,,\eta_{1}^{0}=\eta_{2}^{0^{*}}=0\,,\eta_{03}=\frac{\pi}{3}\,,Q_{3}=0,
\end{aligned}
\end{equation}
the corresponding solutions consisting of a breather and periodic line waves is shown in Fig. \ref{fig02}. It is seen that both of the breather and periodic line waves are periodic in $x$ direction and localized in $y$ direction, and the periodic of these line waves is $\frac{2\pi}{P_{3}}$.

\subsection{Rational solutions of the $x$-nonlocal DS equations}$\\$

Now, we are in a position to construct rational solutions of the $x$-nonlocal DS equations.
Following earlier works in the literatures \cite{long1,long2}, a family of rational solutions can typically be generated by taking a long wave limits of the obtained N-soliton solution \eqref{rfg}. Indeed, with parameter choices
\begin{equation}
\begin{aligned}
Q_{j}=\lambda_{j}P_{j}\,,\eta_{j}^{0}=i\,\pi\,(1\leq j\leq N)
\end{aligned}
\end{equation}
in \eqref{rfg}, and taking the limit as $P_{j} \rightarrow 0$, then exponential functions $f$ and $g$  are translated into pure rational functions.
 Thus rational solutions of the $x$-nonlocal DS  equations can be presented in the following Theorem.

$\\$
\noindent\textbf{Theorem 1.} {\sl The $x$-nonlocal DS equations given by \eqref{DSIIeq} and \eqref{v1} have $Nth-$order rational solutions
\begin{equation} \label{ration}
\begin{aligned}
A=\sqrt{2}\frac{g_{N}}{f_{N}}\,,Q=\epsilon-2( {\rm log} f_{N} )_{xx},
\end{aligned}
\end{equation}
where
\begin{equation}\label{fg}
\begin{aligned}
f_{N}=&\prod_{j=1}^{N}\theta_{j}+\frac{1}{2}\sum_{j,k}^{N}\alpha_{jk}\prod_{l\neq j,k}^{N}\theta_{l}+\cdots\\&+\frac{1}{M!2^{M}}\sum_{j,k,...,m,n}^{N}\overbrace{\alpha_{jk}\alpha_{sl}\cdots\alpha_{mn}}^{M}\prod_{p\neq j,k,...m,n}^{N}\theta_{p}+\cdots,\\
g_{N}=&\prod_{j=1}^{N}(\theta_{j}+b_{j})+\frac{1}{2}\sum_{j,k}^{N}\alpha_{jk}\prod_{l\neq j,k}^{N}(\theta_{l}+b_{l})
+\cdots\\&+\frac{1}{M!2^{M}}\sum_{j,k,...,m,n}^{N}\overbrace{\alpha_{jk}\alpha_{sl}\cdots\alpha_{mn}}^{M}\prod_{p\neq j,k,...m,n}^{N}(\theta_{p}+b_{p})+\cdots,
\end{aligned}
\end{equation}
with
\begin{equation}\label{rt}
\begin{aligned}
\theta_{j}=&i x+\lambda_{j} y+2\delta_{j}\sqrt{-\frac{\epsilon}{\gamma^2\lambda_{j}^{2}+1}}(-\gamma^2\lambda_{j}^{2}+1)t,\\
b_{j}=&-\frac{i\delta_{j}(\gamma^2\lambda_{j}^{2}+1)}{\epsilon}\sqrt{-\frac{\epsilon}{\gamma^2\lambda_{j}^{2}+1}}\,,\\
a_{jk}=&\frac{(\gamma^2\lambda_{j}^{2}+1)(\gamma^2\lambda_{k}^{2}+1)}{2\delta_{j}\delta_{k}\sqrt{-\frac{\epsilon}{\gamma^2\lambda_{j}^{2}+1}}\sqrt{-\frac{\epsilon}{\gamma^2\lambda_{k}^{2}+1}}(\gamma^2\lambda_{j}^{2}+1)(\gamma^2\lambda_{k}^{2}+
1)+2\epsilon(1+\gamma^2\lambda_{j}\lambda_{k})},
\end{aligned}
\end{equation}
 the two positive integers $j$ and $k$ are not large than $N$, $\lambda_{j}\,,\lambda_{k}$ are arbitrary real constants, and $\delta_{j}\,,\delta_{k}=\pm1$.
 The first four of \eqref{fg} are written as
 \begin{equation} \label{fgn}
\begin{aligned}
f_{1}=&\theta_{1}\,,\\
f_{2}=&\theta_{1}\,\theta_{2}+a_{12}\,,\\
f_{3}=&\theta_{1}\theta_{2}\theta_{3}+a_{12}\theta_{3}+a_{13}\theta_{2}+a_{23}\theta_{1}\,,\\
f_{4}=&\theta_{1}\theta_{2}\theta_{3}\theta_{4}+a_{12}\theta_{3}\theta_{4}+a_{13}\theta_{2}\theta_{4}+a_{14}\theta_{2}\theta_{3}+a_{23}\theta_{1}\theta_{4}+a_{24}\theta_{1}\theta_{3}\\
&+a_{34}\theta_{1}\theta_{2}+a_{12}a_{34}+a_{13}a_{24}+a_{14}a_{23}\,,\\
g_{1}=&\theta_{1}+b_{1}\,,\\
g_{2}=&(\theta_{1}+b_{1})\,(\theta_{2}+b_{2})+a_{12}\,,\\
g_{3}=&(\theta_{1}+b_{1})(\theta_{2}+b_{2})(\theta_{3}+b_{3})+a_{12}(\theta_{3}+b_{3})+a_{13}(\theta_{2}+b_{2})+a_{23}(\theta_{1}+b_{1})\,,\\
g_{4}=&(\theta_{1}+b_{1})(\theta_{2}+b_{2})(\theta_{3}+b_{3})(\theta_{4}+b_{4})+a_{12}(\theta_{3}+b_{3})(\theta_{4}+b_{4})+a_{13}(\theta_{2}\\&
+b_{2})(\theta_{4}+b_{4})
+a_{14}(\theta_{2}+b_{2})(\theta_{3}+b_{3})+a_{23}(\theta_{1}+b_{1})(\theta_{4}+b_{4})+a_{24}(\theta_{1}\\&
+b_{1})(\theta_{3}+b_{3})+a_{34}(\theta_{1}+b_{1})(\theta_{2}+b_{2})+a_{12}a_{34}+a_{13}a_{24}+a_{14}a_{23}\,.\\
\end{aligned}
\end{equation}
In what follows the above formulae for $f$ and $g$ will be used in order to construct explicit form of rational solutions.
}\\

\noindent\textbf{Remark 1.} The constraint $-\frac{\epsilon}{\gamma^2\lambda_{j}^{2}+1}>0$ must hold for $\sqrt{-\frac{\epsilon}{\gamma^2\lambda_{j}^{2}+1}}$ to be real, hereafter we set $\epsilon=-1$ when $\gamma^2=1$ in this paper. When $N = 2n , \lambda_{n+j} = -\lambda_{j} \neq 0 \,, \delta_{n+j}\delta_{j} = -1$,  these rational solutions are nonsingular, which can be proved by a similar way in Ref. \cite{long2}, thus we omit its proof in this paper.

To generate smooth rational solutions for the $x$-nonlocal DS equations defined in \eqref{DSIIeq} and \eqref{v1}, hereafter we take parameter constraints in \eqref{fg}
\begin{equation} \label{constrain}
\begin{aligned}
N=2n\,,\lambda_{j}=-\lambda_{n+j}\,,\delta_{j}\delta_{n+j}=-1\,(1\leq j\leq n).
\end{aligned}
\end{equation}
To demonstrate these nonsingular rational solutions, we first consider the case of $N=2\,,\lambda_{1}=-\lambda_{2}\,,\delta_{1}\delta_{2}=-1$. For simplicity, taking $\lambda_{1}=-\lambda_{2}=\lambda$ in (\ref{fg}), then corresponding solutions can be rewritten as
\begin{equation}\label{rw-1}
\begin{aligned}
A(x,y,t)&=\sqrt{2}\frac{(\theta_{1}+b_{1})(\theta_{2}+b_{2})+a_{12}}{\theta_{1}\theta_{2}+a_{12}}\\
&=\sqrt{2}[1+\frac{2i\sqrt{-\frac{\gamma^2\lambda^{2}+1}{\epsilon}}(\lambda\,y+2\sqrt{-\frac{\epsilon}{\gamma^2\lambda^{2}+1}}(-\gamma^2\lambda^{2}+1)t)+\frac{\gamma^2\lambda^{2}+1}{\epsilon}}
{x^{2}+(\lambda\,y+2\sqrt{-\frac{\epsilon}{\gamma^2\lambda^{2}+1}}(-\gamma^2\lambda^{2})t)^2-\frac{(\gamma^2\lambda^{2}+1)^{2}}{4\epsilon}}]\,,\\
Q(x,y,t)&=\epsilon-2( {\rm log} (\theta_{1}\theta_{2}+a_{12}))_{xx}\\
&=\epsilon+\frac{4(\lambda\,y+2\sqrt{-\frac{\epsilon}{\gamma^2\lambda^{2}+1}}(1-\gamma^2\lambda^{2})t)^{2}-4x^{2}-\frac{(\gamma^2\lambda^{2}+1)^{2}}{\epsilon}}{[x^{2}+(\lambda\,y+2\sqrt{-\frac{\epsilon}{\gamma^2\lambda^{2}+1}}(1-\gamma^2\lambda^{2})t)^2-\frac{(\gamma^2\lambda^{2}+1)^{2}}{4\epsilon}]^{2}}\,.
\end{aligned}
\end{equation}
It is easy to find that solutions $A$ and $Q$ given by (\ref{rw-1}) are nonsingular when $\epsilon=-1,\gamma^2\lambda^2+1>0$. In this situation, $A$ and $Q$ are constant along the trajectory where
\begin{equation}
\begin{aligned}
x=0\,,\qquad \lambda\,y+2\sqrt{-\frac{\epsilon}{\gamma^2\lambda^{2}+1}}(-\gamma^2\lambda^{2}+1)t=0.\nonumber
\end{aligned}
\end{equation}
Besides, at any fixed time, $(A,Q)\rightarrow (\sqrt{2},-1)$ when $(x,y)$ goes to infinity. Thus this solution is nothing else but a localized lump moving on the constant background in the $(x,y)$-plane.

\begin{figure}[!htbp]
\centering
\subfigure[]{\includegraphics[height=3cm,width=5.2cm]{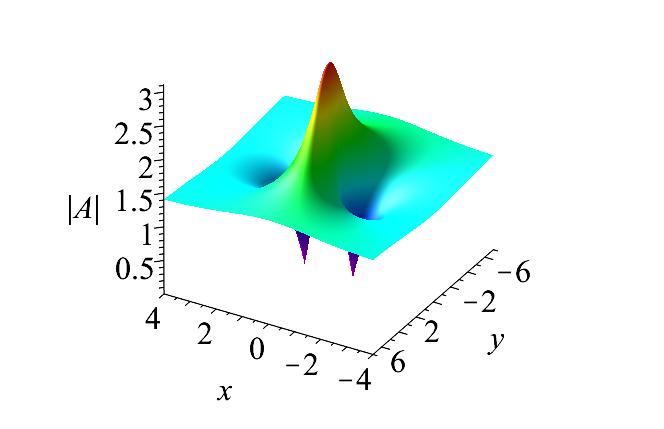}}
\subfigure[]{\includegraphics[height=3cm,width=5.2cm]{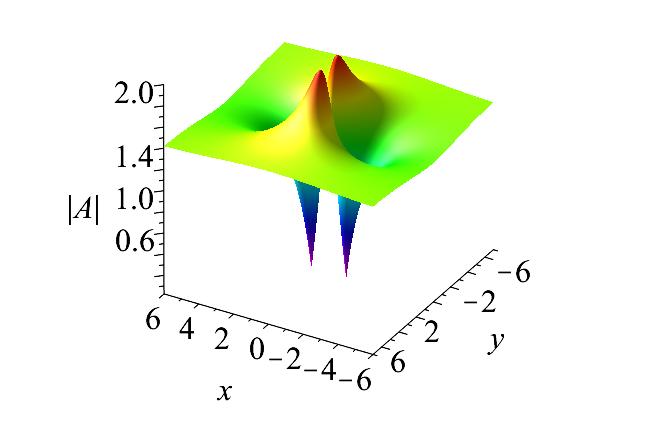}}
\subfigure[]{\includegraphics[height=3cm,width=5.2cm]{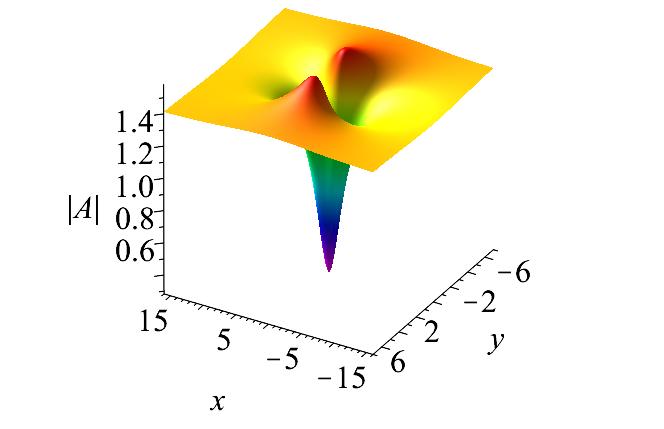}}
\caption{Three kinds of first-order lump $|A|$ of the $x$-nonlocal DSI equation defined in \eqref{rw-1} at time $t=0$ with parameters $\epsilon=-1,\gamma^2=1$: (a) A bright lump with parameters $\lambda=\frac{1}{2}$\,, (b) A four-petaled lump with parameters $\lambda=1$\,, (c) A dark lump with parameters $\lambda=2$\,.~}\label{1-lump}
\end{figure}

After a shift of $t$,  patterns of lump solutions do not change, thus we can discuss the different patterns of lump solutions at time $t=0$ without loss of generality.  In the case of $\gamma^2=1$, namely in the $x$-nonlocal DSI equation, the solution $|A|$ has critical points\\
$A_{1}=(x_{1},y_{1})=(0,0)$,\\
$A_{2}=(x_{2},y_{2})=(\frac{1}{2}\sqrt{-\lambda^{4}+2\lambda^{2}+3},0)$\,,\\
$A_{3}=(x_{3},y_{3})=(-\frac{1}{2}\sqrt{-\lambda^{4}+2\lambda^{2}+3},0)$\,,\\
$A_{4}=(x_{4},y_{4})=(0,\frac{\sqrt{3\lambda^{4}+2\lambda^{2}-1}}{2\lambda})$\,,\\
$A_{5}=(x_{5},y_{5})=(0,-\frac{\sqrt{3\lambda^{4}+2\lambda^{2}-1}}{2\lambda})$\,.\\
Based on the analysis of these critical points for rational solutions \eqref{rw-1}, lump solutions for the $x$-nonlocal DSI equation can be classified into three patterns:

(a) Bright lump: when $0 <\lambda^{2}\leq\frac{1}{3}$, $|A|$ has one global maximum point (point $A_{1}$) and two global minimum points (point $A_{2}$ and point $A_{3}$), see Fig. \ref{1-lump}(a) .

(b) Four-petaled lump: when $ \frac{1}{3}< \lambda^{2}< 3$, $|A|$ has two global maximum points point ($A_{4}$ and point $A_{5}$), and two global minimum points point ($A_{2}$ and point $A_{3})$, see Fig. \ref{1-lump}(b).

(c) Dark lump: when $\lambda^{2}\geq 3$, $|A|$ has two global maximum points (point $A_{4}$ and point $A_{5}$), and one global minimum point (point $A_{1}$), see Fig. \ref{1-lump}(c).\\
However, in the case of $\gamma^2=-1, 0<\lambda^2<1$, i.e., in the $x$-nonlocal DSII equation, the solution $|A|$ has one global maximum point (point $ (0,0) $) and two global minimum points (point $(0, \frac{3\lambda^4-2\lambda^2-1}{2\lambda}) $ and point $(0, -\frac{3\lambda^4-2\lambda^2-1}{2\lambda})$, thus there are only bright lumps in the $x$-nonlocal DSII equation. That is a difference between fundamental lumps of the $x$-nonlocal DSI equation and $x$-nonlocal DSII equation.

\begin{figure}[!htbp]
\centering
\subfigure[]{\includegraphics[height=3cm,width=5cm]{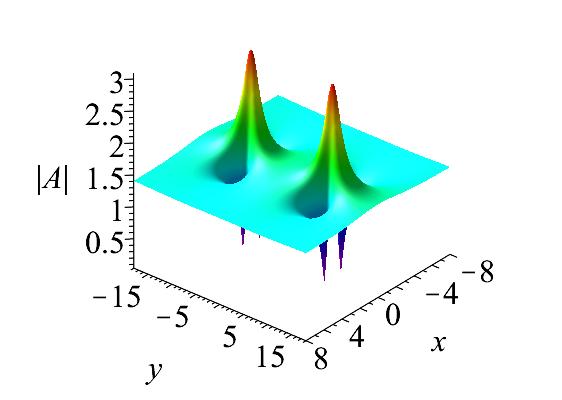}}
\subfigure[]{\includegraphics[height=3cm,width=5cm]{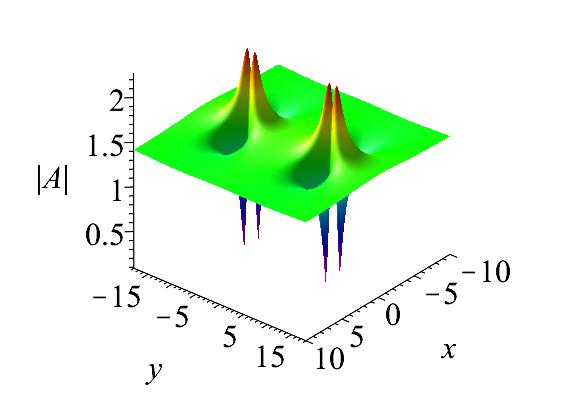}}
\subfigure[]{\includegraphics[height=3cm,width=5cm]{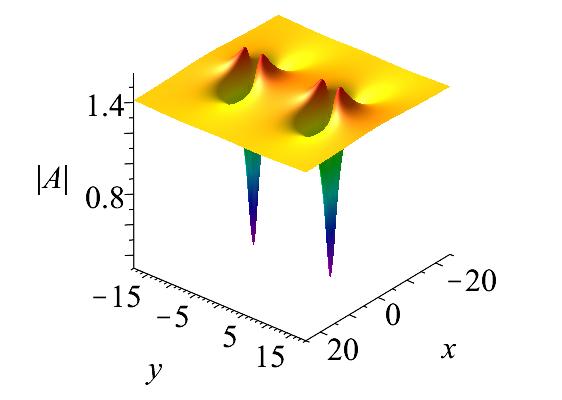}}

\subfigure[]{\includegraphics[height=3cm,width=4.2cm]{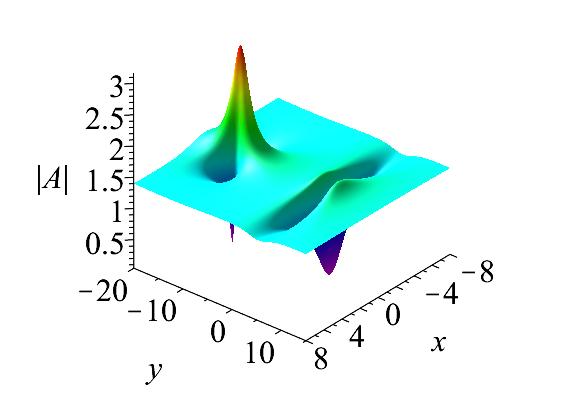}}\qquad\quad
\subfigure[]{\includegraphics[height=3cm,width=4.2cm]{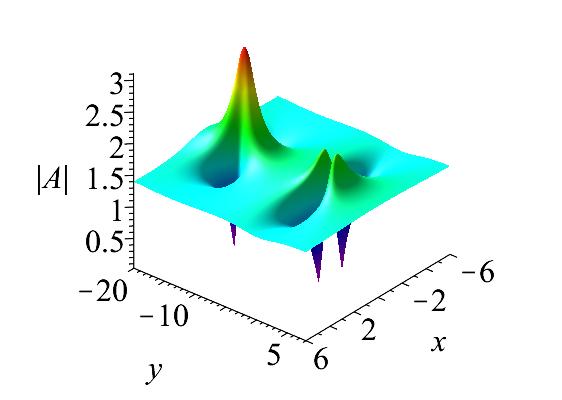}}\qquad\quad
\subfigure[]{\includegraphics[height=3cm,width=4.2cm]{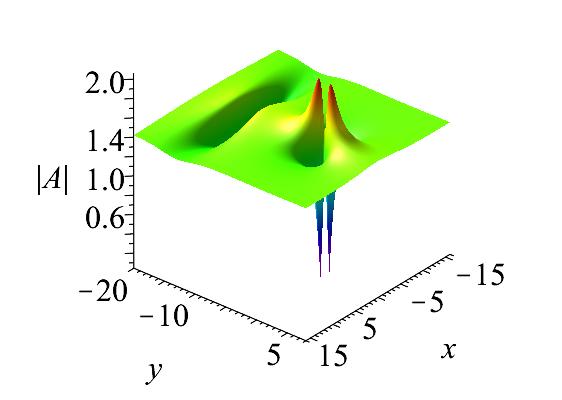}}\qquad
\caption{(Color online) The second-order lumps $|A|$  of the $x$-nonlocal DSI equation given by \eqref{2-ration} at time $t=5$ and parameters $\gamma^2=1,\epsilon=-1$. (a) The coexistence of two bright lumps with parameters $\lambda_{1}=\frac{1}{2}\,,\lambda_{2}=-\frac{1}{2}\,,\lambda_{3}=-\frac{1}{2}\,,\lambda_{4}=\frac{1}{2}$\,. (b) The coexistence of two four-petaled lumps with parameters $\lambda_{1}=\frac{4}{3}\,,\lambda_{2}=-\frac{4}{3}\,,\lambda_{3}=-\frac{4}{3}\,,\lambda_{4}=\frac{4}{3}$\,. (c) The coexistence of two dark lumps with parameters $\lambda_{1}=2\,,\lambda_{2}=-2\,,
\lambda_{3}=-2\,,\lambda_{4}=2$\,. (d) The coexistence of a bright lump and a dark lump with parameters $\lambda_{1}=\frac{1}{2}\,,\lambda_{2}=-\frac{1}{2}\,,\lambda_{3}=-2\,,\lambda_{4}=2$\,. (e) The coexistence of a bright lump and a four-petaled lump with parameters $\lambda_{1}=\frac{1}{2}\,,\lambda_{2}=-\frac{1}{2}\,,\lambda_{3}=-\frac{2}{3}\,,\lambda_{4}=\frac{2}{3}$\,. (f) The coexistence of a dark lump and a four-petaled lump with parameters $\lambda_{1}=2\,,\lambda_{2}=-2\,,\lambda_{3}=-\frac{2}{3}\,,\lambda_{4}=\frac{2}{3}$\,. }\label{fig3}
\end{figure}

More generally, this classification for first-order lumps is also suitable for high-order lump solutions in the $x$-nonlocal DS equations.
For larger $N=2n\,(n\geq2)$ and other parameters meet the parameter constraints defined in \eqref{constrain}, higher-order lumps can be derived, which describe the interaction of $n$ individual fundamental lumps.
For instance, taking
\begin{equation} \label{pa-3}
\begin{aligned}
N=4\,,\lambda_{1}=-\lambda_{3}\,,\lambda_{2}=-\lambda_{4}\,,\delta_{1}\delta_{3}=-1\,,\delta_{2}\delta_{4}=-1,
\end{aligned}
\end{equation}
 the second-order lump solutions  can be obtained from \eqref{ration}, which is
\begin{equation} \label{2-ration}
\begin{aligned}
A=\sqrt{2}\frac{g_{4}}{f_{4}}\,,Q=\epsilon-2( {\rm log} f_{4} )_{xx}.
\end{aligned}
\end{equation}
 Here, any patterns of first-order lumps could coexist with any patterns of first-order lumps, thus there are six patterns of second-order lumps, see Fig. \ref{fig3}. As interactions between rogue waves in $(1+1)$ dimension, interactions between lumps can also generated interesting localized waves patterns. These patterns of localized waves have also been shown in three-component NLS equation \cite{3-NLS}. To the best of authors' knowledge, these localized waves have not been shown in any nonlocal systems  with the self-induced $PT$-symmetric potential before.
For larger $N$, higher-order lumps can be derived. They have qualitatively similar behaviours, but more lumps interact with each other,  and more complicated localized wave patterns could be generated.

Note that fundamental lumps in the $x$-nonlocal DS equations and fundamental rogue waves in some $(2+1)$ dimensional systems may share the same wave structure \cite{KP1,KP2,3D,kkp}, i.e., one upper peak and two down peaks. However, they are different in high-order cases: the $n$th-order lumps in the $x$-nonlocal DS equations are composed of $n$ fundamental lumps, while the $n$th-order rogue waves in some $(2+1)$ dimensional systems are composed of $\frac{n(n+1)}{2}$ fundamental rogue waves. That is a visual difference between rogue waves and lumps in $(2+1)$ dimensional systems.

\subsection{Semi-rational solutions  of the $x$-nonlocal DS equations}$\\$

To understand the solution behaviors of these newly-proposed $x$-nonlocal DS equations defined by (\ref{DSIIeq}) and (\ref{v1}), we consider semi-rational solutions for the $x$-nonlocal DS equations. By taking  a long wave limit of a part of exponential functions in $f$ and $g$ given by (\ref{rfg}), a family of semi-rational solutions consisting of breathers, lumps and periodic line waves, can typically be generated.  Indeed, taking parameters in (\ref{rfg})
\begin{equation} \label{pa-6}
\begin{aligned}
Q_{k}=\lambda_{k}P_{k}\,\,,\eta_{k}^{0}=i\pi\,,0<2j<N\,,1\leq k\leq 2j
\end{aligned}
\end{equation}
and a suitable limit as $P_{k}\rightarrow 0$ for all $k$,  then the functions  $f$ and $g$  defined in  (\ref{rfg}) become
 combinations of polynomial  and exponential functions, which generate semi-rational solutions $A$ and $Q$ for
 the $x$-nonlocal DS equations.
Furthermore, taking parameter constraints
\begin{equation} \label{pa-5}
\begin{aligned}
\lambda_{k}=-\lambda_{j+k}\,,\delta_{k}\delta_{j+k}=-1\,,Q_{s}=0\,(2j+1 \leq s\leq N)\,,
\end{aligned}
\end{equation}
the corresponding semi-rational solutions are a hybrid of lumps, breathers and periodic line waves. Next, we mainly consider the following three types of semi-rational solutions for the $x$-nonlocal DS equations:
$\\$

\noindent\textbf{Type 1: A hybrid of lumps and periodic line waves }$\\$

\begin{figure}[!htbp]
\centering
\subfigure[]{\includegraphics[height=3cm,width=5.2cm]{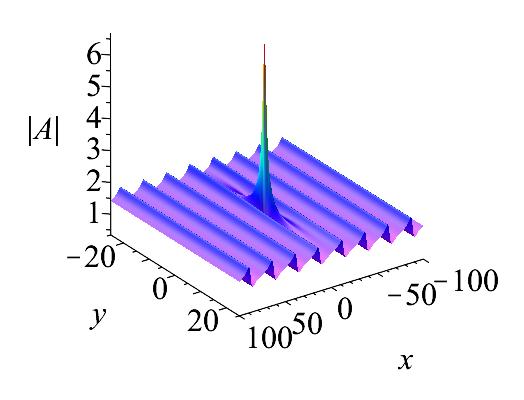}}
\subfigure[]{\includegraphics[height=3cm,width=5.2cm]{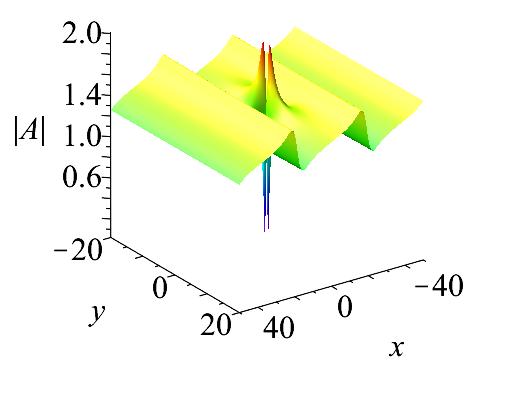}}
\subfigure[]{\includegraphics[height=3cm,width=5.2cm]{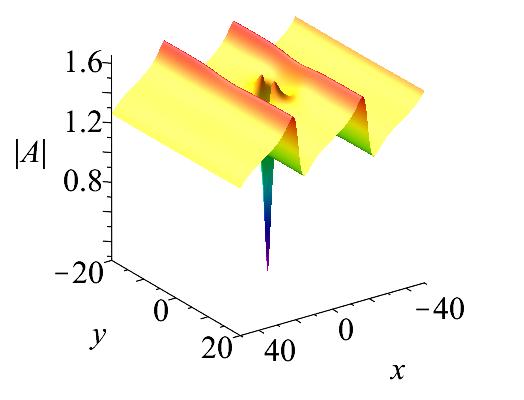}}
\caption{Semi-rational solutions constituting of a lump and periodic line waves for the $x$-nonlocal DSI  equation with parameters $t=0,\gamma^2=1,\epsilon=-1$: (a) A bright lump and periodic line waves with parameters $\lambda_{0}=\frac{1}{2}\,,P_{3}=\frac{1}{4}\,,\eta_{3}^{0}=\frac{\pi}{6}$\,. (b) A four-petaled lump and periodic line waves with parameters $\lambda_{0}=1\,,P_{3}=\frac{1}{6}\,,\eta_{3}^{0}=\frac{\pi}{6}$\,. (c) A dark lump and periodic line waves with parameters $\lambda_{0}=2\,,P_{3}=2\,,\eta_{3}^{0}=\frac{\pi}{6}$.~}\label{fig5}
\end{figure}

To demonstrate semi-rational solutions composed of lumps and periodic line waves, we first consider the simplest semi-rational solutions generated from 3-soliton solutions. Taking parameters choices in (\ref{rfg})
\begin{equation} \label{pa-6}
\begin{aligned}
N=3\,,Q_{1}=\lambda_{1}P_{1}\,,Q_{2}=\lambda_{2}P_{2}\,,\eta_{1}^{0}=\eta_{2}^{0}=i\,\pi\,,Q_{3}=0,
\end{aligned}
\end{equation}
and $P_{1}\,,P_{2}\rightarrow 0$,
then functions $f$ and $g$ of solutions can be rewritten as
\begin{equation} \label{hy-1}
\begin{aligned}
f=&(\theta_{1}\theta_{2}+a_{12})+(\theta_{1}\theta_{2}+a_{12}+a_{13}\theta_{2}+a_{23}\theta_{1}+a_{12}a_{23})e^{\eta_{3}}\,,\\
g=&[(\theta_{1}+b_{1})(\theta_{2}+b_{2})+a_{12}]+[(\theta_{1}+b_{1})(\theta_{2}+b_{2})+a_{12}+a_{13}(\theta_{2}+b_{2})\\
&+a_{23}(\theta_{1}+b_{1})+a_{12}a_{23}]e^{\eta_{3}+i\phi_{3}}\,,
\end{aligned}
\end{equation}
where $$a_{s3}=\frac{P_{3}(\gamma^2\lambda_{s}^{2}+1)}{P_{3}\delta_{s}\sqrt{-\frac{\epsilon}{\gamma^2\lambda_{s}^{2}+1}}\sqrt{-\frac{P_{3}^{2}+4\epsilon}{P_{3}^{2}}}(\gamma^2\lambda_{s}^{2}+1)+2\epsilon}(\,s=1,2\,)\,,$$  and $a_{12}\,,b_{s}\,,\phi_{s},\eta_{3}$ are given by \eqref{rt} and \eqref{cs1}. Here the constrain $-\frac{P_{3}+4\epsilon}{P_{3}^{2}}>0$ has to hold for  $\sqrt{-\frac{P_{3}+4\epsilon}{P_{3}^{2}}}$ to be real. Further,
taking
\begin{equation} \label{py-1}
\begin{aligned}
\lambda_{1}=-\lambda_{2}=\lambda_{0}\,, \delta_{1}\delta_{2}=-1\,,
\end{aligned}
\end{equation}
then the semi-rational solutions composed of a lump and periodic line waves are derived, see Fig.\ref{fig5}.  As can be seen, these periodic line waves can coexist with any
patterns of lumps, and they keep periodic in $x$ direction and localized in $y$ direction and the period is $\frac{2\pi}{P_{3}}$.  An interesting phenomenon is that, the interaction between the lump and periodic line waves can generated much higher peaks.  Comparing to these first-order lumps shown in Fig. \ref{1-lump} (a) and Fig.\ref{fig5} (a), which are constructed by same parameters, but the maximum amplitude of the lump in Fig \ref{fig5} (a) can reach $5\sqrt{2}$, while the maximum amplitude of the lump in Fig \ref{1-lump}(a) does not exceed $3\sqrt{2}$ (see Fig. \ref{1-lump}(a) and  Fig. \ref{fig5} (a)). Thus energy transfer between the first-order lump and periodic line waves has occurred.  Similar researches about localized waves on a background of periodic line waves for the NLS equation have been reported in Ref. \cite{eur}, which demonstrate rogue waves on a cnoidal background.

\begin{figure}[!htbp]
\centering
\subfigure[]{\includegraphics[height=3cm,width=6cm]{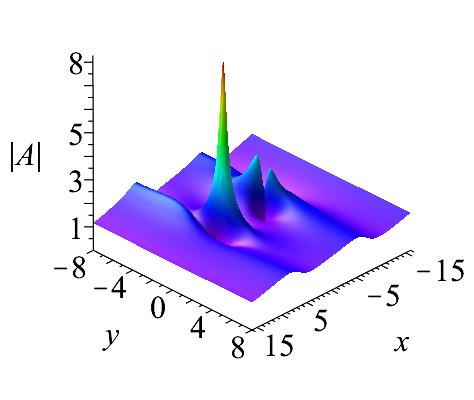}}
\subfigure[]{\includegraphics[height=3cm,width=6cm]{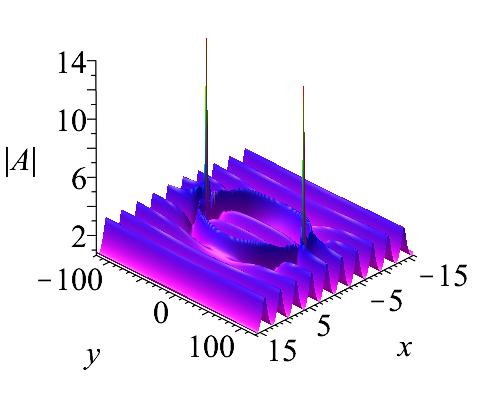}}

\subfigure[]{\includegraphics[height=3cm,width=6cm]{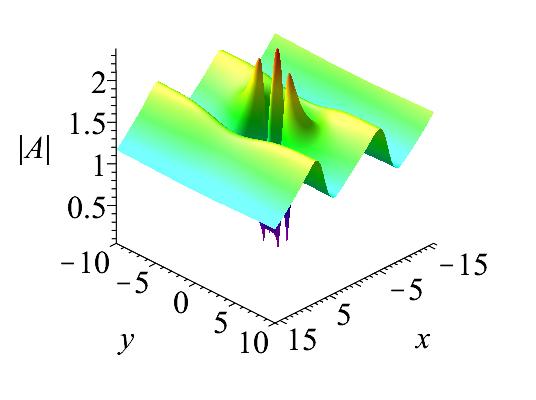}}
\subfigure[]{\includegraphics[height=3cm,width=6cm]{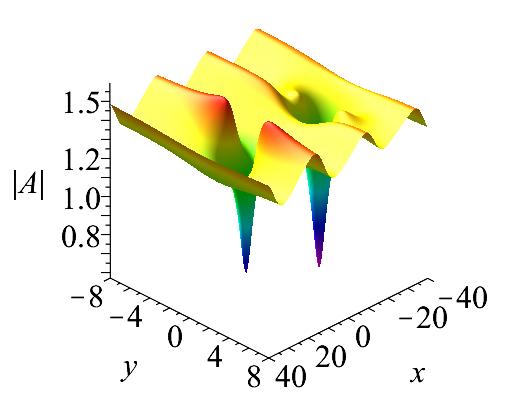}}
\caption{Semi-rational solutions constituting of two lumps and periodic line waves for the $x$-nonlocal DSI equation with parameters $\gamma^2=1,\epsilon=-1,t=0$: (a) A four-petaled lump, a bright lump and periodic line waves with parameters $\lambda_{1}^{0}=\frac{4}{5}\,,\lambda_{2}^{0}=-\frac{3}{2}\,,P_{5}=\frac{1}{2}\,,\eta_{3}^{0}=\frac{\pi}{6}$. (b) Two bright lump and periodic line waves with parameters $\lambda_{1}^{0}=\frac{1}{12}\,,\lambda_{2}^{0}=-\frac{1}{12}\,,P_{5}=2\,,\eta_{3}^{0}=\frac{\pi}{6}$. (c) Two four-petaled lump and periodic line waves with parameters $\lambda_{1}^{0}=\frac{5}{4}\,,\lambda_{2}^{0}=-\frac{5}{4}\,,P_{5}=\frac{1}{2}\,,\eta_{3}^{0}=\frac{\pi}{6}$. (d) Two dark lump and periodic line waves with parameters $\lambda_{1}^{0}=-2\,,\lambda_{2}^{0}=3\,,P_{5}=\frac{1}{4}\,,\eta_{3}^{0}=\frac{\pi}{6}$.~}\label{fig6}
\end{figure}

Higher-order semi-rational rational solutions composed of more lumps and periodic line waves can also be generated by a similar way for larger $N$ in (\ref{rfg}). For example, a hybrid of two lumps and periodic line waves can be derived from 5-soliton. By setting
\begin{equation} \label{pa-6}
\begin{aligned}
N=5\,,Q_{j}=\lambda_{j}P_{j}\,,\eta_{j}^{0}=i\,\pi\,,Q_{5}=0\,(j=1,2,3,4)\,,
\end{aligned}
\end{equation}
and taking $P_{j}\rightarrow 0$ in \eqref{rfg},
then functions $f$ and $g$ of solutions can be rewritten as
\begin{equation} \label{ff3}
\begin{aligned}
f=&(\theta_{1}\theta_{2}\theta_{3}\theta_{4}+a_{12}\theta_{3}\theta_{4}+a_{13}\theta_{2}\theta_{4}+a_{14}\theta_{2}\theta_{3}+a_{23}\theta_{1}\theta_{4}
+a_{24}\theta_{1}\theta_{3}+a_{34}\theta_{1}\theta_{2}+a_{12}a_{34}+\\
&a_{13}a_{24}+a_{14}a_{23})+e^{\eta_{5}}[\theta_{1}\,\theta_{2}\,\theta_{3}\,\theta_{4}+a_{45}\,\theta_{1}\,\theta_{2}\,\theta_{3}+a_{35}\,\theta_{1}\,\theta_{2}\,\theta_{4}
+a_{25}\,\theta_{1}\,\theta_{3}\,\theta_{4}+a_{15}\theta_{2}\,\theta_{3}\,\theta_{4}+\\
&(a_{35}a_{45}+a_{34})\theta_{1}\,\theta_{2}+(a_{25} a_{45}+a_{24})\,\theta_{1}\,\theta_{3}+(a_{25}a_{35}+a_{23})\,\theta_{1}\,\theta_{4}+(a_{15}a_{45}+a_{14})\,\theta_{2}\,\theta_{3}\\
&+(a_{15}a_{35}+a_{13})\,\theta_{2}\,\theta_{4}+(a_{15}a_{25}+a_{12})\theta_{3}\,\theta_{4}+(a_{25}a_{35}\,a_{45}+a_{23}a_{45}+a_{25}a_{34}+a_{24}a_{35})\,\theta_{1}\\&
+(a_{15}a_{35}a_{45}+a_{14}a_{35}+a_{13}a_{45}+a_{15}a_{34})\,\theta_{2}+(a_{15} a_{25}a_{45}+a_{14}a_{25}+a_{15}a_{24}+a_{12}a_{45})\,\theta_{3}\\
&+(a_{15}a_{25}a_{35}+a_{15}a_{23}+a_{13}a_{25}+a_{12}a_{35})\,\theta_{4}+a_{12}a_{34}+a_{13}\,a_{24}+a_{14}a_{23}+a_{12}\,a_{35}\,a_{45}+\\
&a_{13}a_{25}a_{45}+a_{14}a_{25}a_{35}+a_{15}a_{24}a_{35}+a_{15}a_{25}a_{34}\,+a_{15}a_{23}a_{45}+a_{15}a_{25} a_{35} a_{45}]\,,\\
g=&[(\theta_{1}+b_{1})(\theta_{2}+b_{2})(\theta_{3}+b_{3})(\theta_{4}+b_{4})+a_{12}(\theta_{3}+b_{3})(\theta_{4}+b_{4})+a_{13}(\theta_{2}+b_{2})(\theta_{4}+b_{4})+\\
&a_{14}(\theta_{2}+b_{2})(\theta_{3}+b_{3})+a_{23}(\theta_{1}+b_{1})(\theta_{4}+b_{4})
+a_{24}(\theta_{1}+b_{1})(\theta_{3}+b_{3})+a_{34}(\theta_{1}+b_{1})(\theta_{2}\\
&+b_{2})+a_{12}a_{34}+a_{13}a_{24}+a_{14}a_{23}]+e^{\eta_{5}+i\phi_{5}}[(\theta_{1}+b_{1})(\theta_{2}+b_{2})(\theta_{3}+b_{3})(\theta_{4}+b_{4})+\\
&a_{45}\,(\theta_{1}+b_{1})(\theta_{2}+b_{2})(\theta_{3}+b_{3})+a_{35}\,(\theta_{1}+b_{1})(\theta_{2}+b_{2})(\theta_{4}+b_{4})
+a_{25}\,(\theta_{1}+b_{1})(\theta_{3}+\\
&b_{3})(\theta_{4}+b_{4})+a_{15}(\theta_{2}+b_{2})(\theta_{3}+b_{3})(\theta_{4}+b_{4})+
(a_{35}a_{45}+a_{34})(\theta_{1}+b_{1})(\theta_{2}+b_{2})+\\
&(a_{25} a_{45}+a_{24})\,(\theta_{1}+b_{1})(\theta_{3}+b_{3})+(a_{25}a_{35}+a_{23})(\theta_{2}+b_{2})(\theta_{4}+b_{4})+(a_{15}a_{45}+a_{14})\\
&\,(\theta_{2}+b_{2})(\theta_{3}+b_{3})+(a_{15}a_{35}+a_{13})\,(\theta_{2}+b_{2})(\theta_{4}+b_{4})+(a_{15}a_{25}+a_{12})(\theta_{3}+b_{3})(\theta_{4}+b_{4})\\
&+(a_{25}a_{35}\,a_{45}+a_{23}a_{45}+a_{25}a_{34}+a_{24}a_{35})\,(\theta_{1}+b_{1})+(a_{15}a_{35}a_{45}+a_{14}a_{35}+a_{13}a_{45}+a_{15}a_{34})\\
&(\theta_{2}+b_{2})+(a_{15} a_{25}a_{45}+a_{14}a_{25}+a_{15}a_{24}+a_{12}a_{45})\,(\theta_{3}+b_{3})+(a_{15}a_{25}a_{35}+a_{15}a_{23}+a_{13}a_{25}\\
&+a_{12}a_{35})(\theta_{4}+b_{4})+a_{12}(a_{34}+a_{35}\,a_{45})+a_{13}(\,a_{24}+a_{25}a_{45})+a_{14}(a_{23}+a_{25}a_{35})+\\
&a_{15}(a_{24}a_{35}+a_{25}a_{34}\,+a_{23}a_{45}+a_{25} a_{35} a_{45})],
\end{aligned}
\end{equation}
where $$a_{j5}=\frac{P_{5}(\gamma^2\lambda_{j}^{2}+1)}{P_{5}\delta_{j}\sqrt{-\frac{\epsilon}{\gamma^2\lambda_{j}^{2}+1}}\sqrt{-\frac{P_{5}^{2}+4\epsilon}{P_{5}^{2}}}(\gamma^2\lambda_{j}^{2}+1)+2\epsilon},$$ and $\theta_{j}\,,b_{j}\,,a_{ij}\,(\,1\leq i<j\leq4)$ are given by \eqref{rt}, $\eta_{5}\,,\phi_{5}$ are are given by  \eqref{cs1}. Under parameter constraints
\begin{equation} \label{pa-7}
\begin{aligned}
\lambda_{1}=-\lambda_{3}=\lambda_{1}^{0}\,,\lambda_{2}=-\lambda_{4}=\lambda_{2}^{0}\,,\delta_{1}\delta_{3}=-1\,,\delta_{2}\delta_{4}=-1\,,
\end{aligned}
\end{equation}
then a family of semi-rational solutions are generated,  which describe two lumps on a  background of periodic line waves.  As discussed above, the combination of two fundamental lumps exists six patterns (see Fig. \ref{fig3}), and these two lumps could also interact with periodic line waves, thus the wave patterns of these semi-rational solutions could be more various. Four kinds of localized waves for these semi-rational solutions are illustrated in Fig.\ref{fig6}.  In the view of mathematics, the coexistence of one eye-shaped (i.e.,bright state ) wave and
one four-petaled wave were shown in Fig.1 of Ref.\cite{2-nls},  but our results are more general. Because the coexistence demonstrated in this paper does not only exist on a constant background but also on a background with periodic line waves  (see Fig. \ref{fig3} and Fig. \ref{fig6} (a)), while the former one just exist on a constant background.
The superposition of two bright lump, two four-petaled lump or two dark lump on a periodic line waves background are shown in Fig. \ref{fig6} (b), Fig. \ref{fig6} (c), and Fig. \ref{fig6} (d) respectively.  Interestingly, the  superposition of two bright lumps on the periodic line waves background generates a ring shaped wave pattern, see Fig. \ref{fig6} (b).

\noindent\textbf{Type 2: A hybrid of lumps and breathers }$\\$

\begin{figure}[!htbp]
\centering
\subfigure[$\eta_{3}^{0}=-2\pi$]{\includegraphics[height=3.0cm,width=3.0cm]{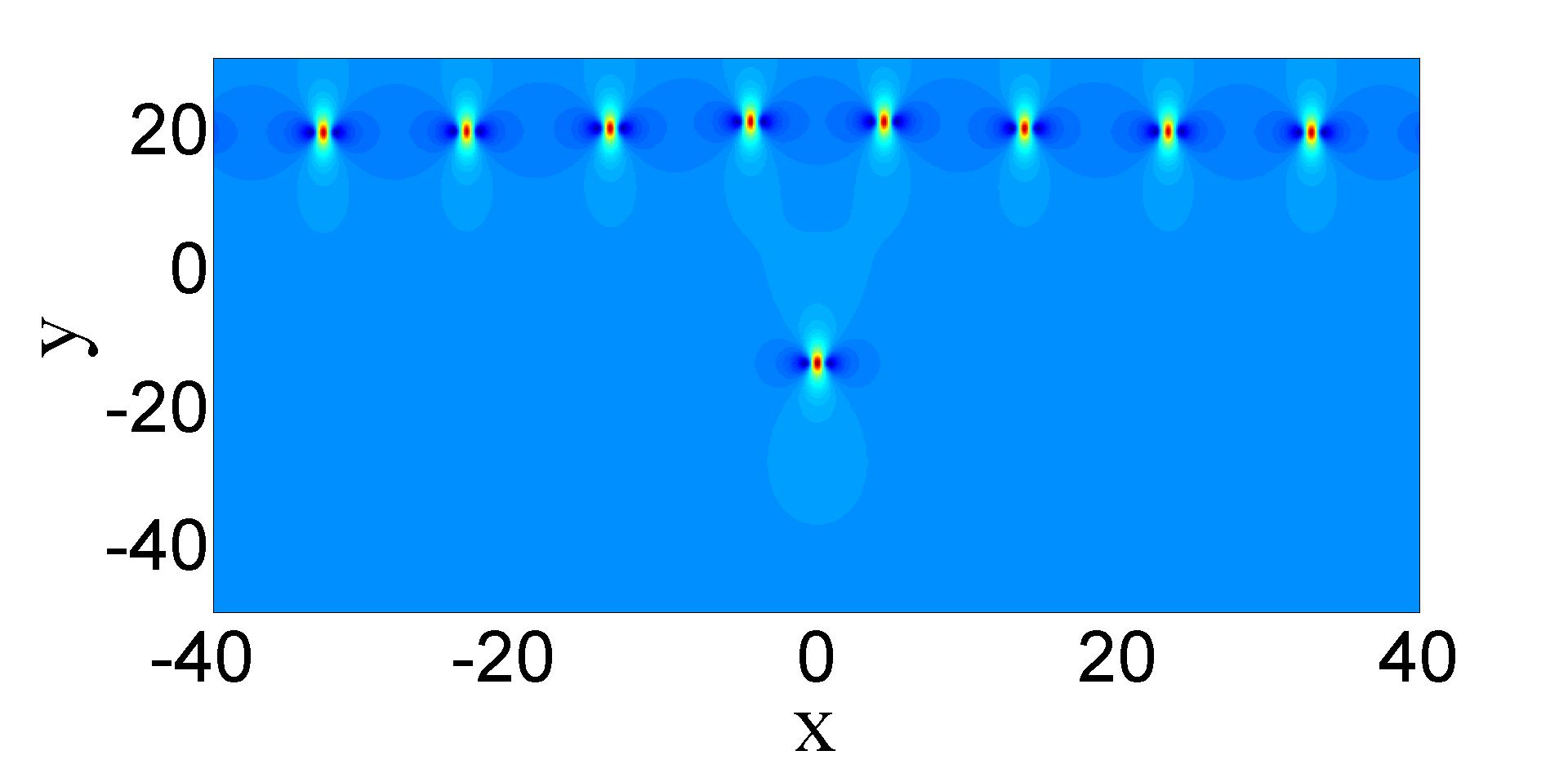}}
\subfigure[$\eta_{3}^{0}=-\pi$]{\includegraphics[height=3.0cm,width=3.0cm]{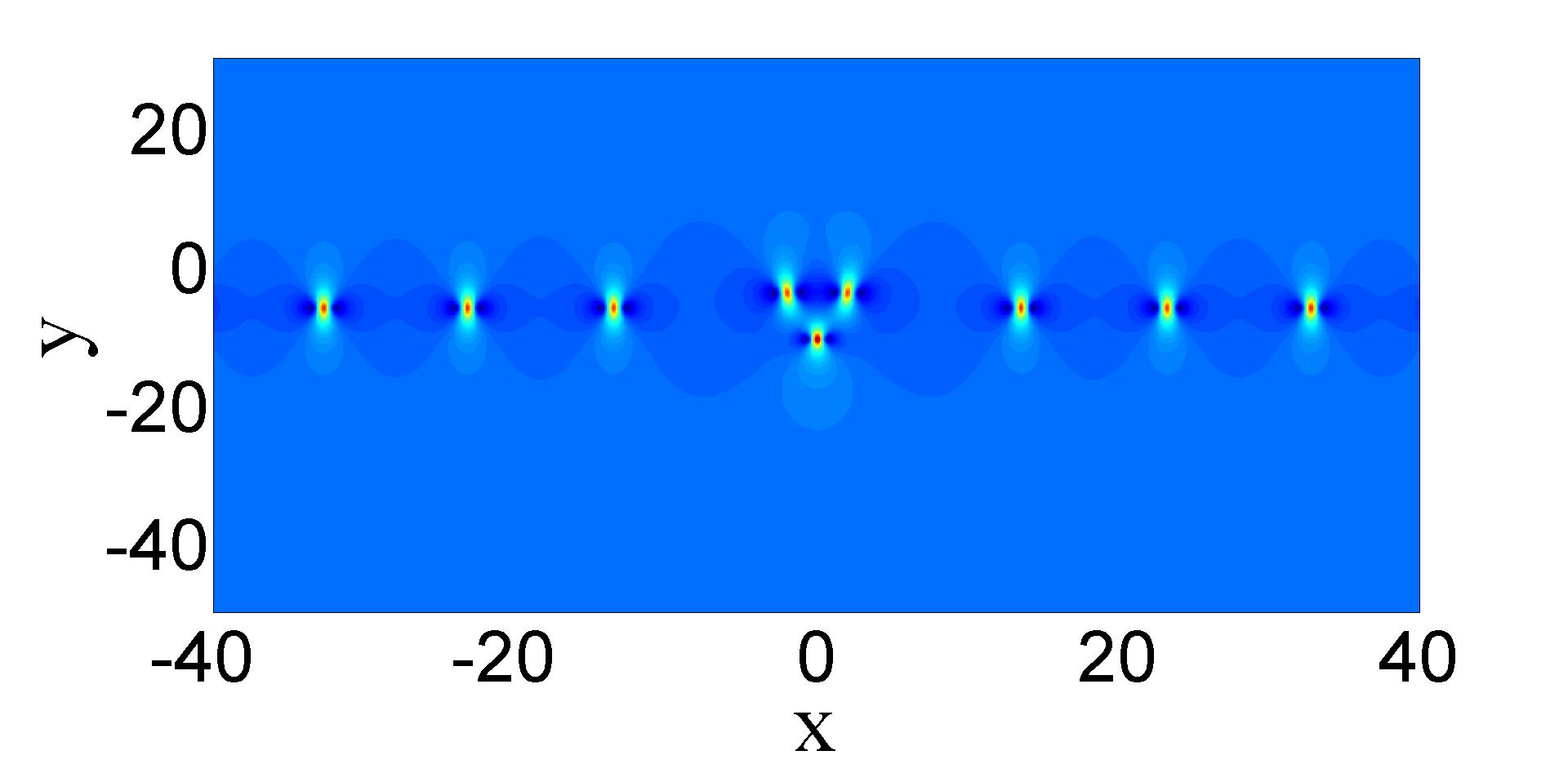}}
\subfigure[$\eta_{3}^{0}=0$]{\includegraphics[height=3.0cm,width=3.0cm]{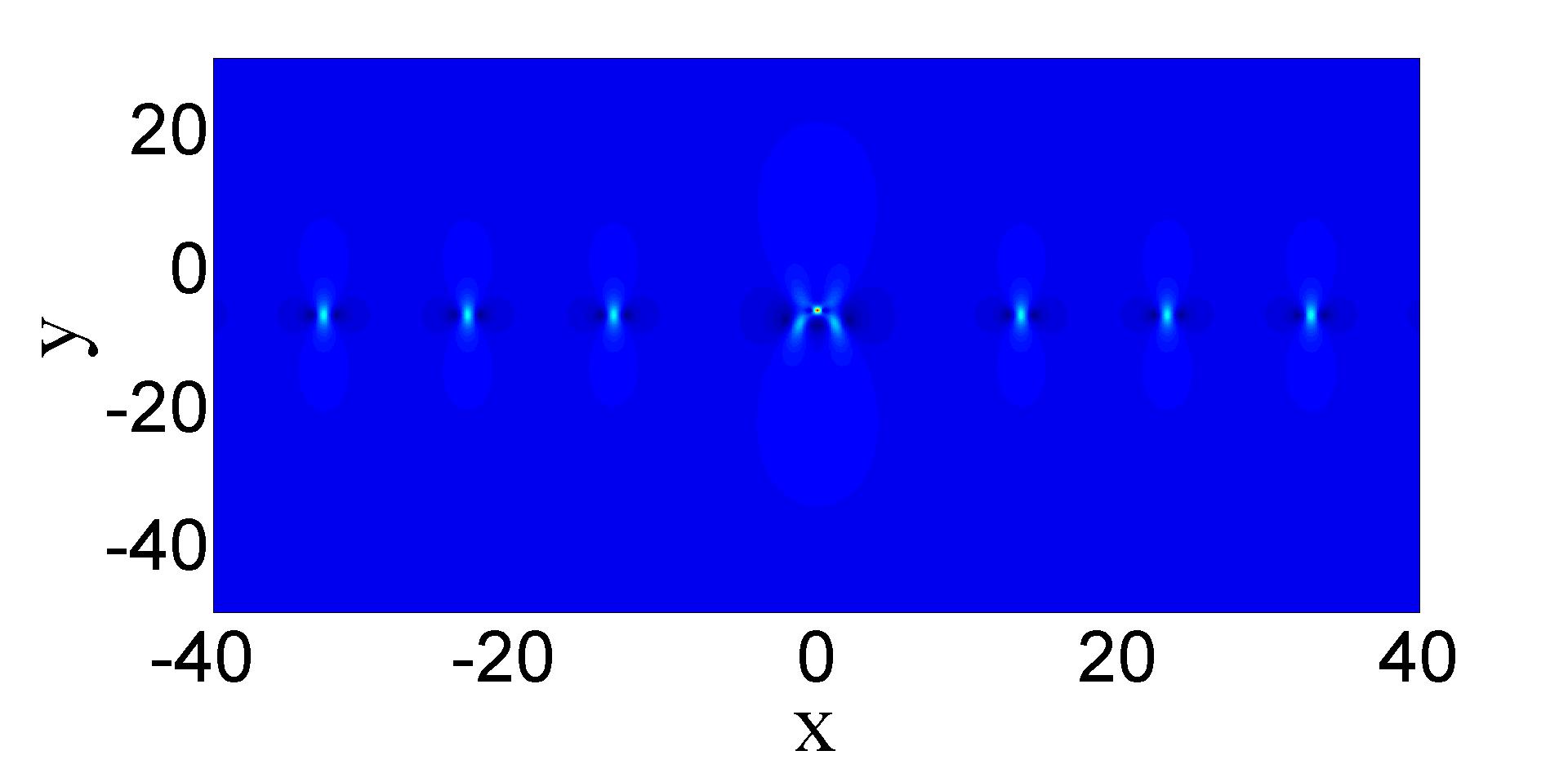}}
\subfigure[$\eta_{3}^{0}=\pi$]{\includegraphics[height=3.0cm,width=3.0cm]{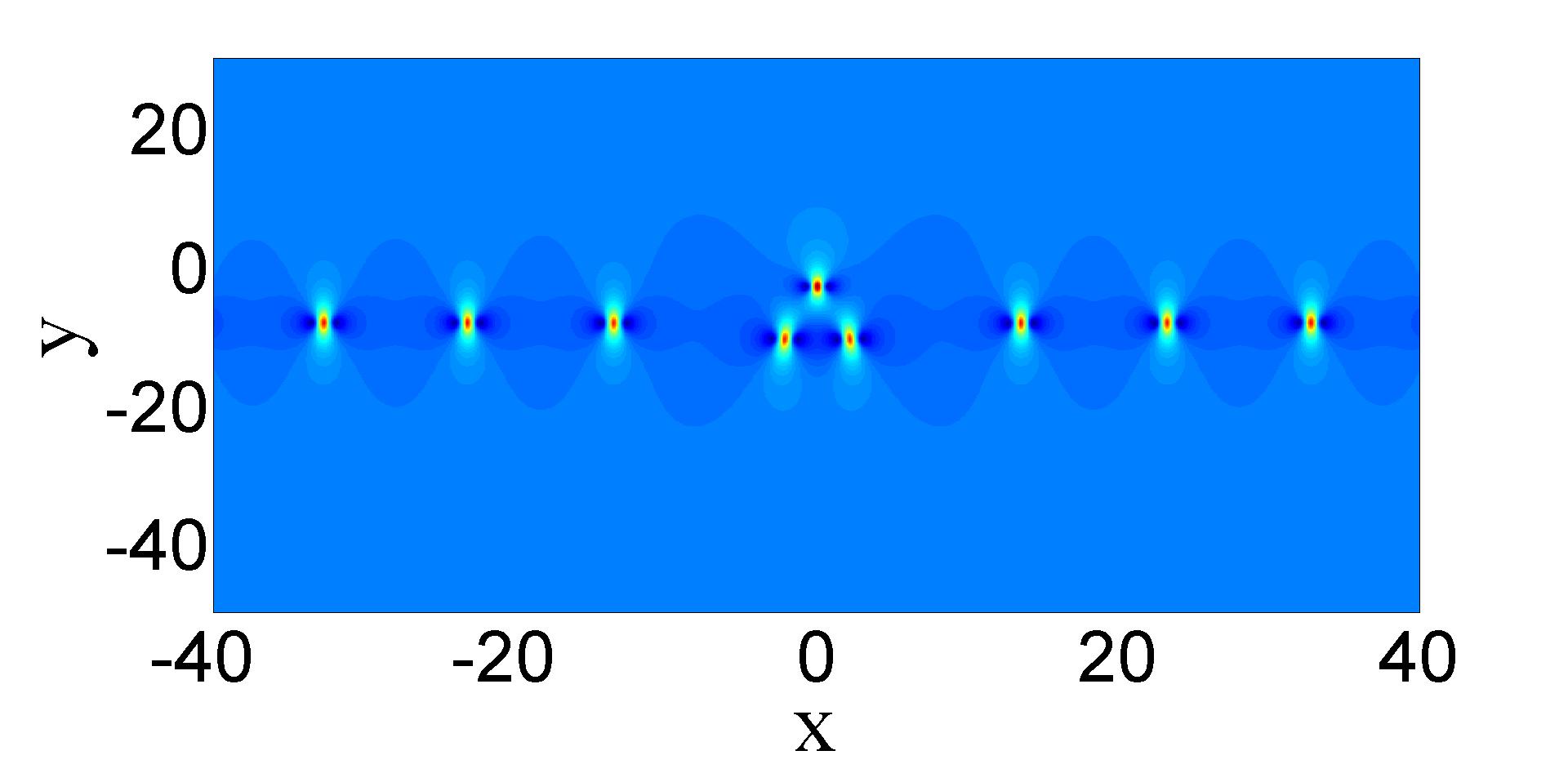}}
\subfigure[$\eta_{3}^{0}=\frac{3}{2}\pi$]{\includegraphics[height=3.0cm,width=3.0cm]{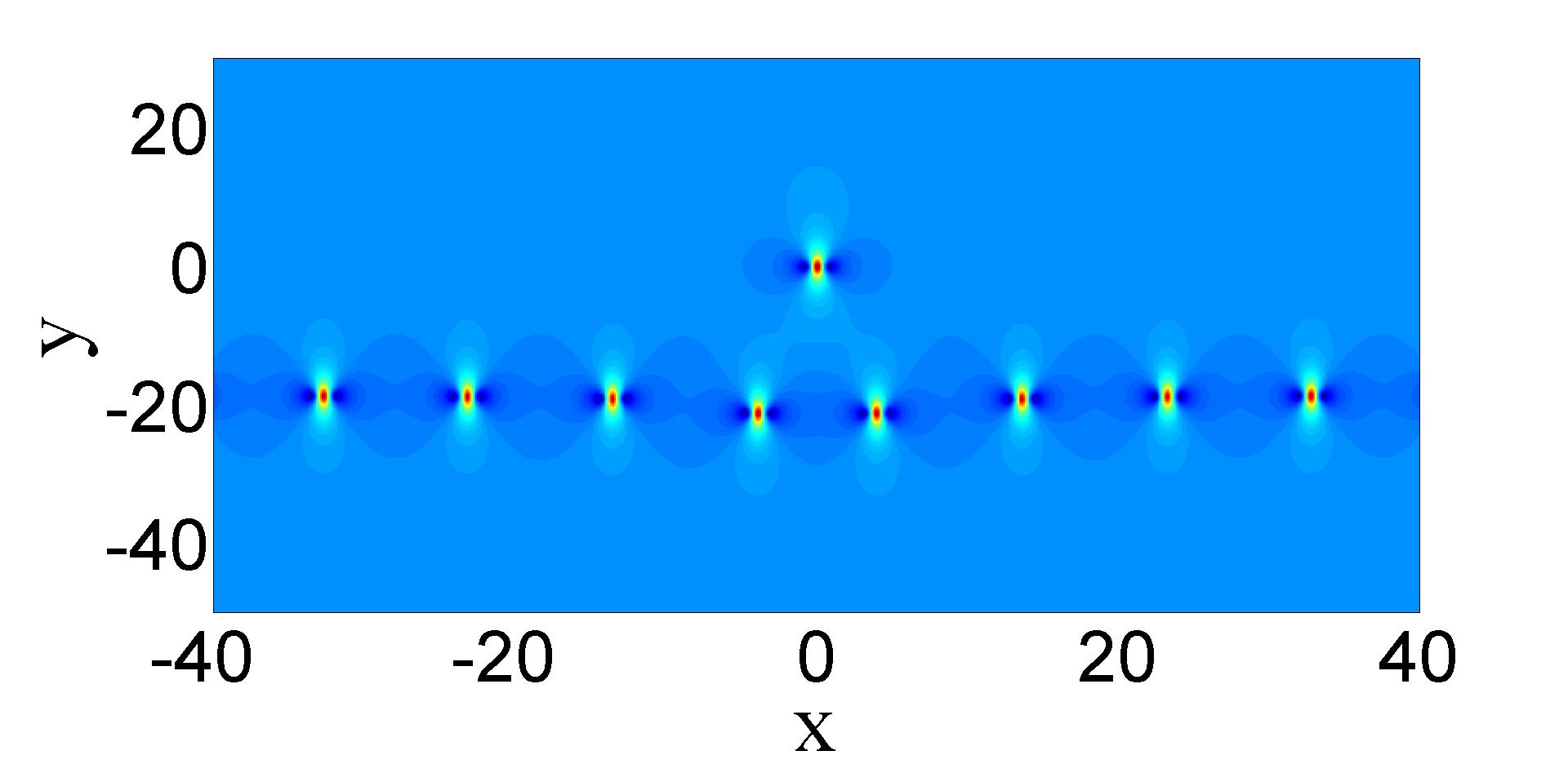}}
\caption{Semi-rational solutions of the $x$-nonlocal DSII equation constituting of one lump and one breather with parameters  $\gamma^2=-1,\epsilon=-1,\lambda_{1}=\frac{1}{2}\,,\lambda_{2}=-\frac{1}{3}\,,P_{3}=1\,,P_{4}=-1\,,Q_{3}=\frac{1}{4}\,,Q_{4}=\frac{1}{4}$. The plotted is $|A|$ field. The constant background value is $\sqrt{2}$.~}
\label{lump+br}
\end{figure}

To construct a type of mixed solution consisting of one lump and one breather, we have to take parameters in \eqref{rfg}
\begin{equation} \label{pa-6}
\begin{aligned}
N=4\,,Q_{1}=\lambda_{1}P_{1}\,,Q_{2}=\lambda_{2}P_{2}\,,\eta_{1}^{0}=\eta_{2}^{0}=i\,\pi\,,
\end{aligned}
\end{equation}
and take a limit as $P_{1}\,,P_{2}\rightarrow 0$,
then functions $f$ and $g$ of solutions can be rewritten as
\begin{equation} \label{hy-11}
\begin{aligned}
f=&e^{A_{34}}(a_{13}a_{23}+a_{13}a_{24}+a_{13}\theta_{2}+a_{14}a_{23}+a_{14}a_{24}+a_{14}\theta_{2}+a_{23}\theta_{1}+a_{24}\theta_{1}+\theta_{1}\theta_{2}\\
&+a_{12})e^{\eta_{3}+\eta_{4}}+(a_{13}a_{23}+a_{13}\theta_{2}+a_{23}\theta_{1}+\theta_{1}\theta_{2}+a_{12})e^{\eta_{3}}+(a_{14}a_{24}+a_{14}\theta_{2}+a_{24}\theta_{1}\\
&+\theta_{1}\theta_{2}+a_{12})e^{\eta_{4}}+\theta_{1}\theta_{2}+a_{12}\,,\\
g=&e^{A_{34}}[a_{13}a_{23}+a_{13}a_{24}+a_{13}(\theta_{2}+b_{2})+a_{14}a_{23}+a_{14}a_{24}+a_{14}(\theta_{2}+b_{2})+a_{23}(\theta_{1}+b_{1})\\
&+a_{24}(\theta_{1}+b_{1})+(\theta_{1}+b_{1})(\theta_{2}+b_{2})+a_{12}]e^{\eta_{3}+i\phi_{3}+\eta_{4}+i\phi_{4}}+[a_{13}a_{23}+a_{13}(\theta_{2}+b_{2})\\
&+a_{23}(\theta_{1}+b_{1})+(\theta_{1}+b_{1})(\theta_{2}+b_{2})+a_{12}]e^{\eta_{3}+i\phi_{3}}+[a_{14}a_{24}+a_{14}(\theta_{2}+b_{2})+a_{24}(\theta_{1}\\
&+b_{1})+(\theta_{1}+b_{1})(\theta_{2}+b_{2})+a_{12}]e^{\eta_{4}+i\phi_{4}}+(\theta_{1}
+b_{1})(\theta_{2}+b_{2})+a_{12},
\end{aligned}
\end{equation}
where $$a_{sl}=\frac{P_{l}(\gamma^2\lambda_{s}^{2}+1)}{P_{l}\delta_{s}\sqrt{-\frac{\epsilon}{\gamma^2\lambda_{s}^{2}+1}}\sqrt{-\frac{P_{l}^{2}+4\epsilon}{P_{l}^{2}}}(\gamma^2\lambda_{s}^{2}+1)+2\epsilon}(\,s=1,2\,,l=3,4\,),$$  and $a_{12}\,,b_{s}\,,\phi_{l},\eta_{l}\,,e^{A_{34}}$ are given by \eqref{rt} and \eqref{cs1}. Further, taking  parameters
\begin{equation} \label{co-hy-2}
\begin{aligned}
\lambda_{1}=-\lambda_{2}\,,\delta_{1}\delta_{2}=-1\,,P_{3}=-P_{4}\,,Q_{3}=Q_{4}\,,
\eta_{3}^{0}=\eta_{4}^{0*},
\end{aligned}
\end{equation}
thus corresponding semi-rational solutions consisting of a fundamental lump and one breather are obtained, see Fig.\ref{lump+br}.  It is emphasized that, in order to make clear the combination of this semi-rational solutions, we fix all the parameters except $\eta_{3}^0$. By altering the value of $\eta_{3}^{0}$, the location of the breather would be changed. When $|\eta_{3}^0|>>0$,
the lump and breather would separate completely, and the period and the amplitude of the breather are stable. When $\eta_3^0\rightarrow 0$, the breather get closer to the lump, and the breather possesses a higher amplitude and a smaller period. After  the lump immerses into the breather, the
superimposition between the lump and the breather would excite a peak whose height is larger than the maximum amplitudes of the breather or the lump.  Semi-rational solutions composed of multi-lumps and multi-breathers can be generated by a similar way with larger $N$ in (\ref{rfg}).

$\\$

\noindent\textbf{Type 3: A hybrid of lumps, breathers and periodic line waves }$\\$

\begin{figure}[!htbp]
\centering
\subfigure[]{\includegraphics[height=3.5cm,width=7.0cm]{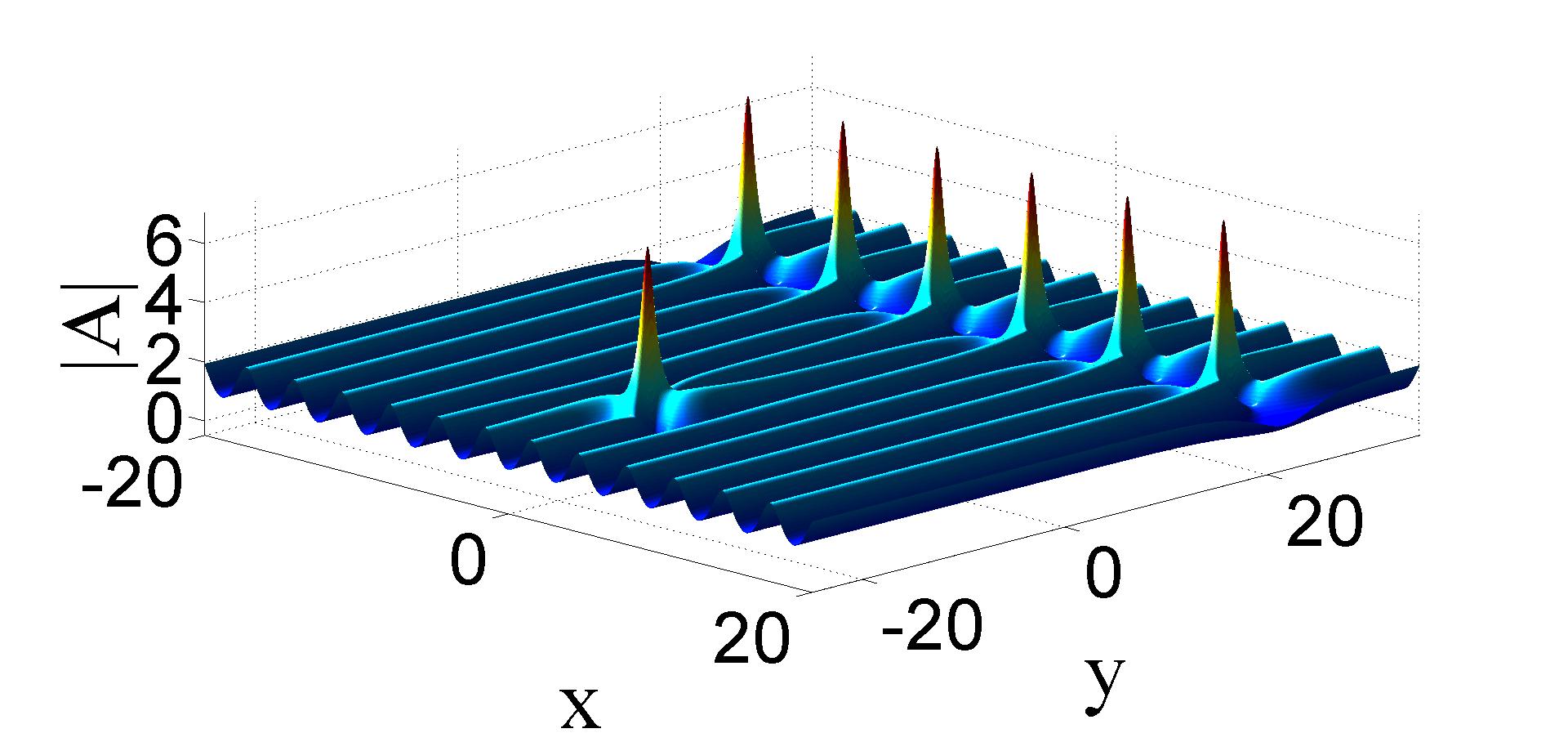}}
\subfigure[]{\includegraphics[height=3.5cm,width=5.5cm]{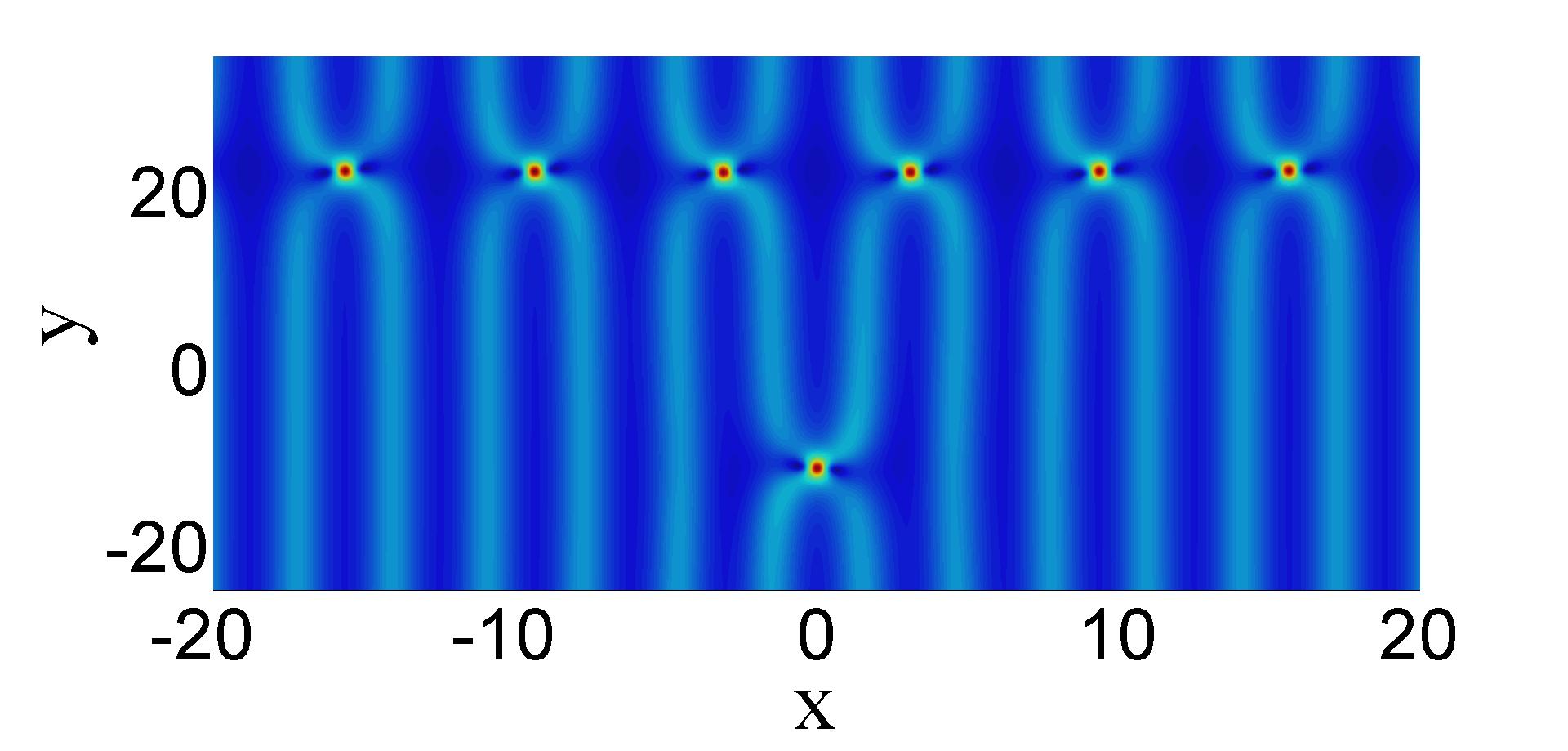}}
\caption{ Semi-rational solutions  of the $x$-nonlocal DSII equation
 constituting of a lump, a breather and periodic line waves with parameters $\gamma^2=-1,\epsilon=-1,\lambda_{0}=\frac{1}{2}\,,P_{3}=1\,,P_{4}=-1\,,P_{5}=2\,,Q_{3}=\frac{1}{4}\,,
 Q_{4}=\frac{1}{4}\,,Q_{5}=0\,,\eta_{3}^{0}=0\,,\eta_{4}^{0}=0\,,\eta_{5}^{0}=0$. The right panel is a density plot of the left.}\label{fig4}
\end{figure}

Lastly but interestingly, we derive a type of semi-rational solutions composed of a lump, a breather and periodic line waves for the $x$-nonlocal DS equations,  which could be generated from 5-soliton.  Setting
\begin{equation}\label{NLS-gao}
\begin{aligned}
N=5\,,Q_{1}=\lambda_{0}P_{1}\,,Q_{2}=-\lambda_{0}P_{2}\,,Q_{3}=Q_{4}\,,P_{4}=-P_{3}\,,\eta_{1}^{0}=i\pi\,,\eta_{2}^{0}=i\pi\,,
\end{aligned}
\end{equation}
and taking $P_{1}\,,P_{2}\rightarrow0$ in \eqref{rfg}, then functions $f$ and $g$ are translated into a combination of rational and exponential functions. The corresponding semi-rational solution is illustrated in Fig. \ref{fig4}.  As can be seen that this solution describes a fundamental lump coexistence with a breather on a background of periodic line waves. Both of the breather and periodic line waves are periodic in $x$ direction and localized in $y$ direction. As to the best of our knowledge, this type of hybrid solutions has not been shown even in the usual DS equations before.

\section{Solutions of the nonlocal DSII equation }\label{3}

Solutions of the fully $PT$ symmetric nonlocal DSI equation, namely $\gamma^2=1$ in (\ref{DSIIeq}) and (\ref{v2}) were discussed in our earlier work \cite{jigu}, including line breathers, two types of rogue waves, and semi-rational solutions. Thus we mainly focus on solutions of the fully $PT$ symmetric nonlocal DSII equation, i.e., $\gamma^2=-1$ in  (\ref{DSIIeq}) and (\ref{v2}) in this section.  The  nonlocal DSII equation defined by (\ref{DSIIeq}) and (\ref{v2}) can be written in the following bilinear form:
\begin{equation}\label{DSbi-f}
\begin{aligned}
&(D_{x}^{2}-D_{y}^{2}-iD_{t})\widetilde{g} \cdot \widetilde{f} =0,\\
&(D_{x}^{2}+D_{y}^{2})\widetilde{f} \cdot \widetilde{f} =2\epsilon \widetilde{f}^{2}-2\epsilon \widetilde{g}\widetilde{g}^*(-x,-y,t),\\
\end{aligned}
\end{equation}
where
\begin{equation} \label{DS-f}
\begin{aligned}
A=\sqrt{2}\frac{\widetilde{g}}{\widetilde{f}}\,,Q=\epsilon-( {2 \rm log} \widetilde{f})_{xx},
\end{aligned}
\end{equation}
and $\widetilde{f}\,,\widetilde{g}$ are functions of the three variables $x\,,y$ and $t$, and $f$ satisfies the condition
\begin{equation} \label{t-gh}
\begin{aligned}
&[\widetilde{f}(-x,-y,t)]^{*}=\widetilde{f}(x,y,t).
\end{aligned}
\end{equation}

To figure out the difference between solutions of the the partially and fully $PT$ symmetric nonlocal DS equations, we consider solutions of the fully $PT$ symmetric nonlocal DSII equation in the coming part of this paper.

\subsection{ The period and line breather solutions of the nonlocal DSII}$\\$

By the Hirota's bilinear method with the perturbation expansion, the periodic line wave solutions defined in \eqref{DS-f} with functions $\widetilde{f}$ and $\widetilde{g}$ are given as
 \begin{equation}\label{tfg}
\begin{aligned}
\widetilde{f}=&\sum_{\mu=0,1}\exp(\sum_{k<j}^{(N)}\mu_{k}\mu_{j}\widetilde{A}_{kj}+\sum_{k=1}^{N}\mu_{k}\widetilde{\eta_{k}}),\\
\widetilde{g}=&\sum_{\mu=0,1}\exp(\sum_{k<j}^{(N)}\mu_{k}\mu_{j}\widetilde{A_{kj}}+\sum_{k=1}^{N}\mu_{k}(\widetilde{\eta_{k}}+i\widetilde{\phi_{k}})),\\
\end{aligned}
\end{equation}
where
 \begin{equation}\label{tcs1}
\begin{aligned}
\widetilde{\eta_{k}}&=iP_{k}\,x+iQ_{k}\,y+\widetilde{\Omega_{k}}t+\eta_{k}^{0}\,,\\
\widetilde{\Omega_{k}}&=(P_{k}^{2}-Q_{k}^{2})\sqrt{-1-\frac{4\epsilon}{P_{k}^{2}+Q_{k}^{2}}}\,,\\
\exp(\widetilde{A_{kj}})&=-\frac{2\epsilon\cos(\widetilde{\phi_{k}}-\widetilde{\phi_{j}})-(P_{k}-P_{j})^{2}-(Q_{k}-Q_{j})^{2}-2\epsilon}{2\epsilon\cos(\widetilde{\phi_{k}}+\widetilde{\phi_{j}})-(P_{k}+P_{j})^{2}-(Q_{k}+Q_{j})^{2}-2\epsilon}\,,\\
\cos(\widetilde{\phi_{k}})&=1+\frac{P_{k}^{2}+Q_{k}^{2}}{2\epsilon}\,,\sin(\widetilde{\phi_{k}})=-\frac{(P_{k}^{2}+Q_{i}^{2})\sqrt{-1-\frac{4\epsilon}{P_{k}^{2}+Q_{k}^{2}}}}{2\epsilon}.
\end{aligned}
\end{equation}
Here, it is noted that the constraint $-1-\frac{4\epsilon}{P_{k}^{2}+Q_{k}^{2}}\geq0$ must be hold for $\widetilde{\Omega_{k}}$ to be real and $|\sin(\widetilde{\phi_{k}})|\leq1$.  Specifically, when choosing $|P_{k}|=|Q_{k}|$ in \eqref{tcs1}, namely $\widetilde{\Omega_{k}}=0$, the corresponding solutions are independent of $t$.

\begin{figure}[!htbp]
\centering
\subfigure[$t=-5$]{\includegraphics[height=2.5cm,width=2.9cm]{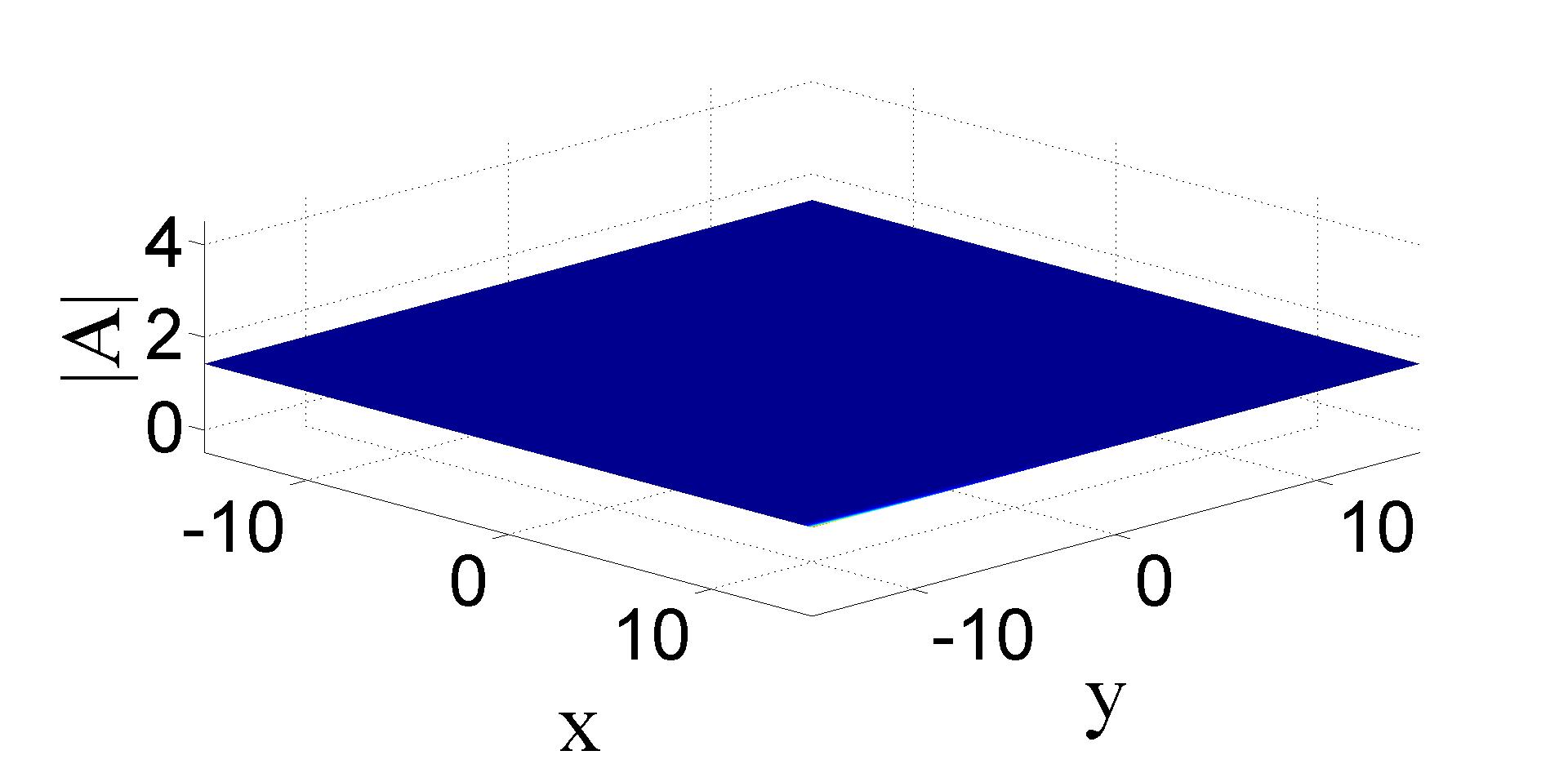}}
\subfigure[$t=-1$]{\includegraphics[height=2.5cm,width=2.9cm]{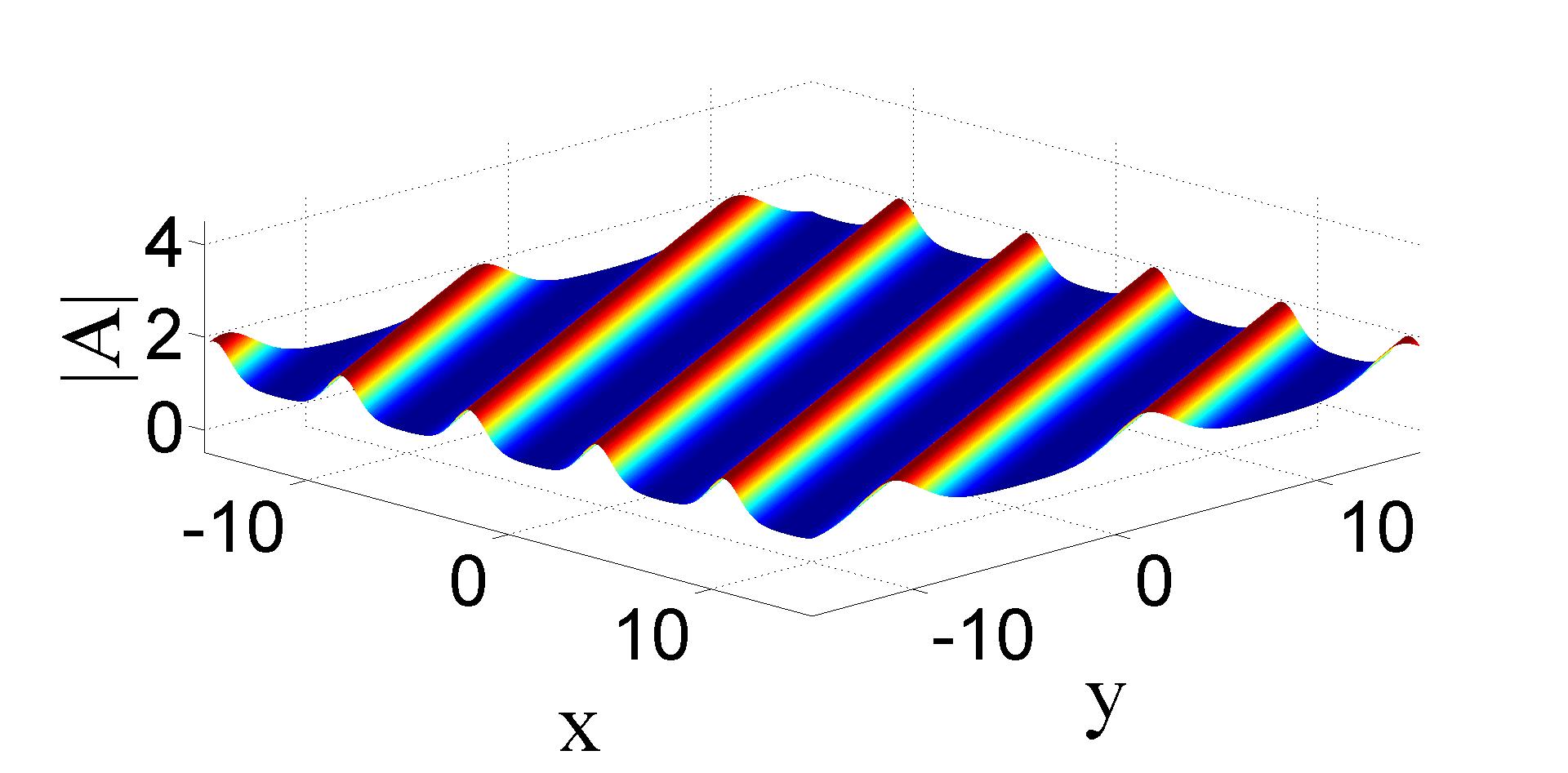}}
\subfigure[$t=0$]{\includegraphics[height=2.5cm,width=2.9cm]{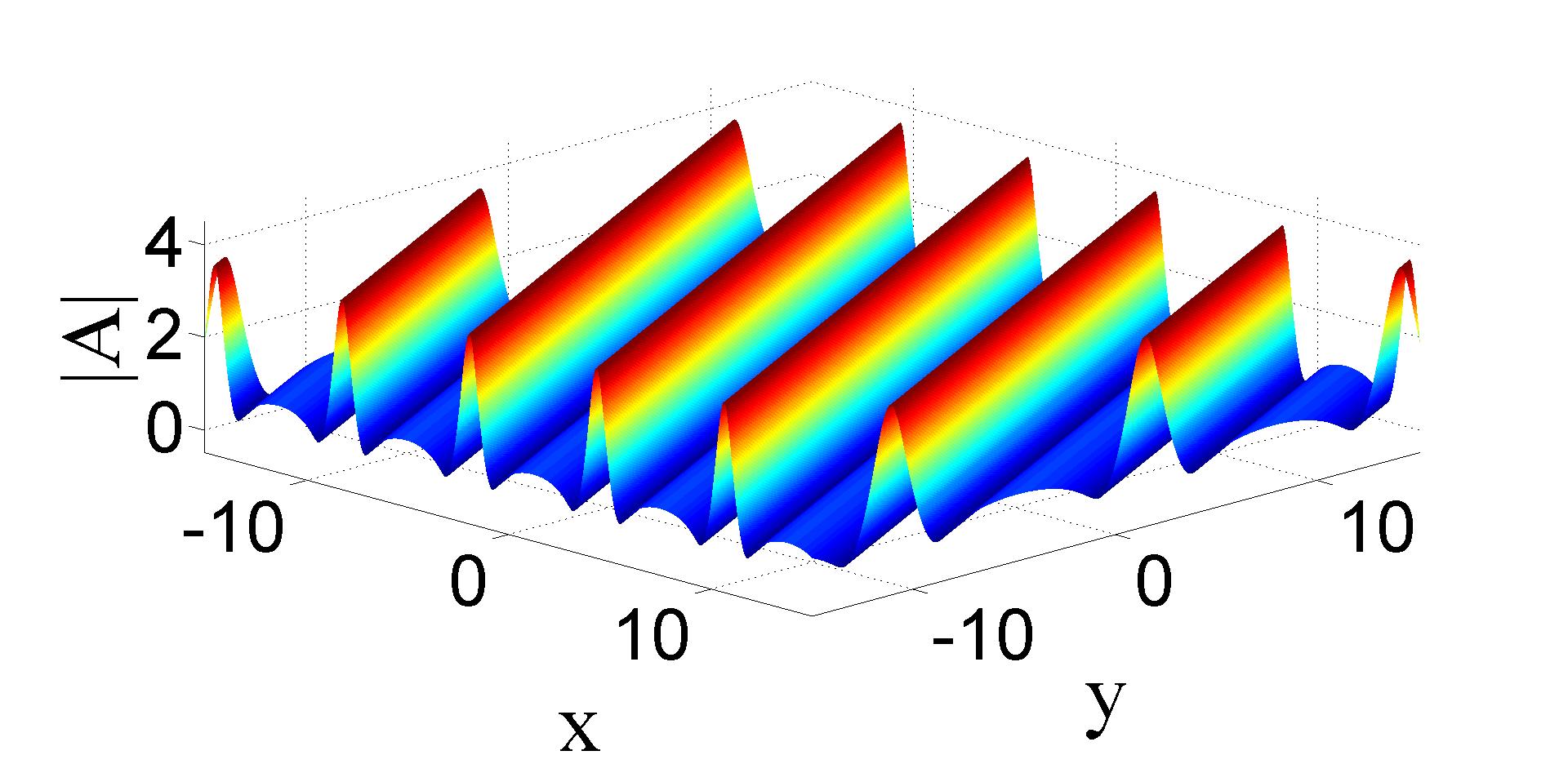}}
\subfigure[$t=1$]{\includegraphics[height=2.5cm,width=2.9cm]{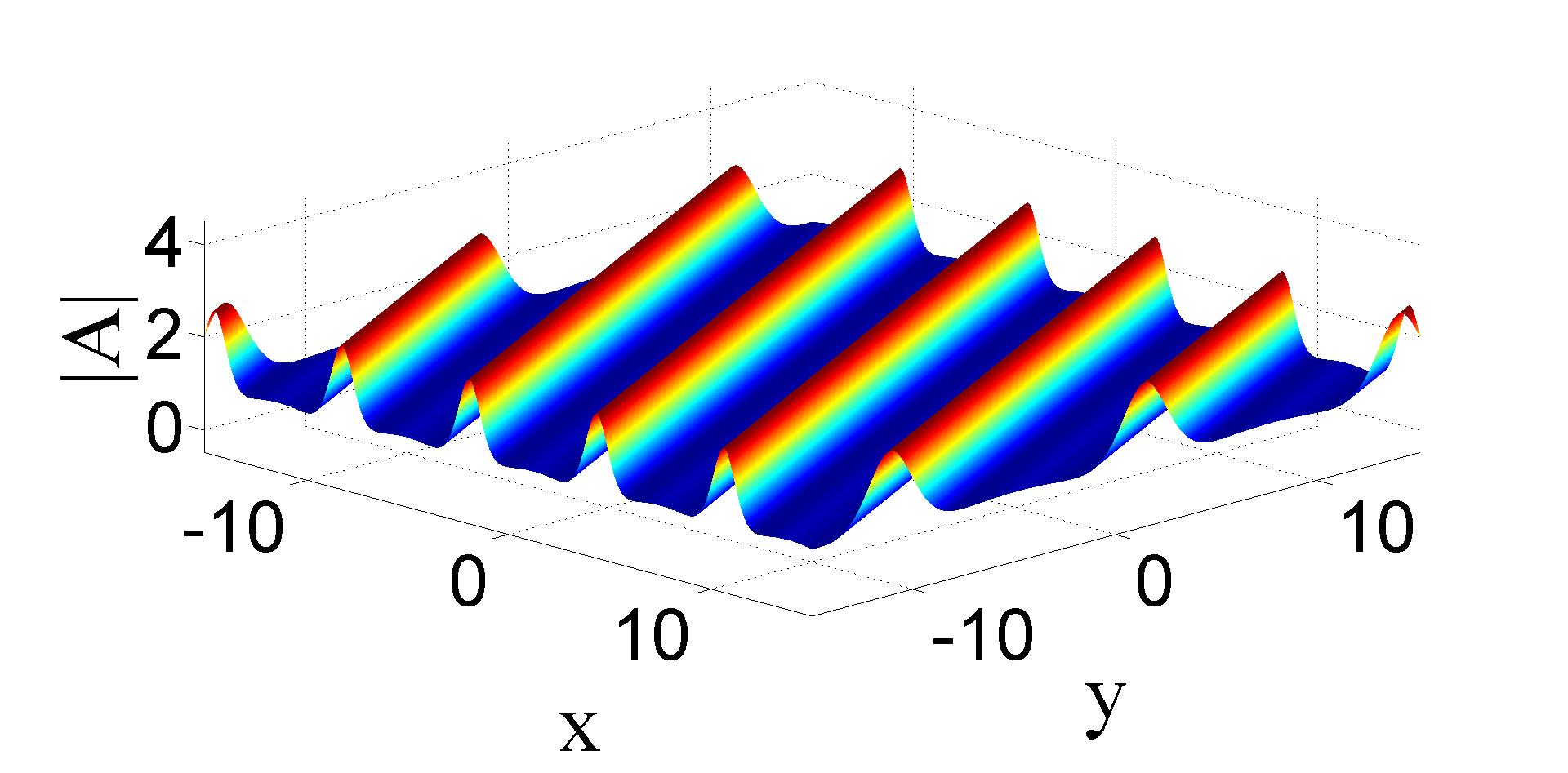}}
\subfigure[$t=5$]{\includegraphics[height=2.5cm,width=2.9cm]{lbr-1.jpg}}
\caption{The time evolution of line breathers of the nonlocal DSII equation in the $(x,y)$-plane with parameters $\epsilon=-1,N=2,P_{1}=1,P_{2}=-1,Q_{1}=\frac{1}{2},Q_{2}=-\frac{1}{2},\eta_{1}^{0}=\eta_{2}^{0}=0$.~}\label{ll-br}
\end{figure}

 Similar to our  works \cite{3DKP, jigu} on other equations, a subclass of analytical solutions termed $n$th-order line breathers (i.e., growing-and-decaying model \cite{PRE} ) can be generated by taking parameters in \eqref{tfg}
 \begin{equation}\label{pa-1}
\begin{aligned}
N=2n\,,P_{n+k}=-P_{k}\,,Q_{n+k}=-Q_{k}\,,\eta_{n+k}^{0*}=\eta_{k}^{0}\,, \widetilde{\Omega_{k}}\neq0,
\end{aligned}
\end{equation}
where $n$ is an arbitrary positive integer. For example, when one takes parameters in \eqref{tfg}
 \begin{equation}\label{line-br}
\begin{aligned}
\epsilon=-1,N=2,P_{1}=-P_{2},Q_{1}=-Q_{2},\eta_{1}^{0*}=\eta_{2}^{0}\,,
\end{aligned}
\end{equation}
then a first-order line breather solution is obtained
through (\ref{DS-f}), which is illustrated in Fig.\ref{ll-br}.  As can be seen that, the line breather solution has a parallel line profile with a varying height in the $(x,y)$-plane. It describes periodic line waves arise from a constant background and then disappear into the constant background again.  A crucial and visual difference between the first-order
breather solution of the $x$-nonlocal DS equations and the nonlocal DSII is that the latter is a usual ($2+1$)-dimensional breather (see Fig. \ref{fig1}), which is periodic in $x$ direction and localized in $y$ direction.  Furthermore, as fundamental line rogue waves \cite{3DKP,DS1,DS2} are treated as a limit case of line breathers, the existence of line breathers indicates that line rogue waves may exist in the nonlocal DSII equation given by \eqref{DSIIeq} and \eqref{v2}.  To make this point clear, we proceed to derive rational solutions of this equation in the next subsection.

\subsection{Rational solutions of the  nonlocal DSII equation} $\\$

Similar to the results  of the $x$-nonlocal DS equations expressed in section \ref{2}, rational solutions of the nonlocal DSII equation can also be derived by taking a long wave limit of $\widetilde{f}$ and $\widetilde{g}$, and which  can also be given in the following theorem.
$\\$

\noindent\textbf{Theorem 2.} {\sl  The nonlocal DSII equation possesses two $Nth-$order rational solutions
 $A$ and $Q$ given by (\ref{DS-f}) with the two following polynomials  $\widetilde{f}$ and $\widetilde{g}$:
\begin{equation}\label{tt-fg}
\begin{aligned}
\widetilde{f}=&\prod_{k=1}^{N}\widetilde{\theta_{k}}+\frac{1}{2}\sum_{k,j}^{(N)}\widetilde{\alpha_{kj}}\prod_{l\neq k,j}^{N}\widetilde{\theta_{l}}+\cdots\\&+\frac{1}{M!2^{M}}\sum_{k,j,...,m,n}^{(N)}\overbrace{\widetilde{\alpha_{kj}}\widetilde{\alpha_{sl}}\cdots\widetilde{\alpha_{mn}}}^{M}\prod_{p\neq k,j,...m,n}^{N}\widetilde{\theta_{p}}+\cdots,\\
\widetilde{g}=&\prod_{k=1}^{N}(\widetilde{\theta_{i}}+\widetilde{b_{i}})+\frac{1}{2}\sum_{k,j}^{(N)}\widetilde{\alpha_{kj}}\prod_{l\neq k,j}^{N}(\widetilde{\theta_{l}}+\widetilde{b_{l}})
+\cdots\\&+\frac{1}{M!2^{M}}\sum_{k,j,...,m,n}^{(N)}\overbrace{\widetilde{\alpha_{kj}}\widetilde{\alpha_{sl}}\cdots\widetilde{\alpha_{mn}}}^{M}\prod_{p\neq k,j,...m,n}^{N}(\widetilde{\theta_{p}}+\widetilde{b_{p}})+\cdots,\\
\end{aligned}
\end{equation}
with
\begin{equation}\label{tt-rt}
\begin{aligned}
&\widetilde{\theta_{j}}=i\,x+i\,\lambda_{j}y+2\gamma_{j}\sqrt{\frac{\epsilon}{\lambda_{j}^{2}+1}}(1-\lambda_{j}^{2})t\,,
\widetilde{b_{j}}=-\frac{i\gamma_{j}(\lambda_{j}^{2}+1)}{\epsilon}\sqrt{-\frac{\epsilon}{\lambda_{j}^{2}+1}}\,,\\
&\widetilde{a_{kj}}=\frac{(\lambda_{k}^{2}+1)(\lambda_{j}^{2}+1)}{2\gamma_{k}\gamma_{j}\sqrt{-\frac{\epsilon}
{\lambda_{k}^{2}+1}}\sqrt{-\frac{\epsilon}{\lambda_{j}^{2}+1}}(\lambda_{k}^{2}+1)(\lambda_{j}^{2}+1)
+2\epsilon(1+\lambda_{k}\lambda_{j})},
\end{aligned}
\end{equation}
and $\gamma_{j}=-1\,,1$. }\\

 \noindent\textbf{Remark 2.} The constraint $-\frac{\epsilon}{\lambda_{j}^{2}+1}>0$ must be hold for $\sqrt{-\frac{\epsilon}{\lambda_{j}^{2}+1}}$ to be real. Hereafter, we just discuss the $\epsilon=-1$ case.

\noindent\textbf{Remark 3.} When $N=2n-1\,,\lambda_{k}\neq\lambda_{j}$, the corresponding
rational solutions are kink-shaped line rogue waves.  However, these solutions are singular at a certain  finite time, thus we do not analyze this subclass of singular rational solutions of the nonlocal DSII.  Note that the nonlocal DSI equation also has this kind of singular solutions \cite{jigu}.

 \noindent\textbf{Remark 4.} Setting $N=2n\,,\lambda_{n+k}=\lambda_{k}\,,\gamma_{k}\gamma_{n+k}=-1$ in \eqref{tt-fg}, the corresponding rational solutions are nonsingular,  which are $W$-shaped line rogue waves.  The nonsingularity can be  proved by a similar way as in Ref. \cite{long2}, thus the proof is omitted.

\begin{figure}[!htbp]
\centering
\subfigure[$t=-20$]{\includegraphics[height=3cm,width=2.8cm]{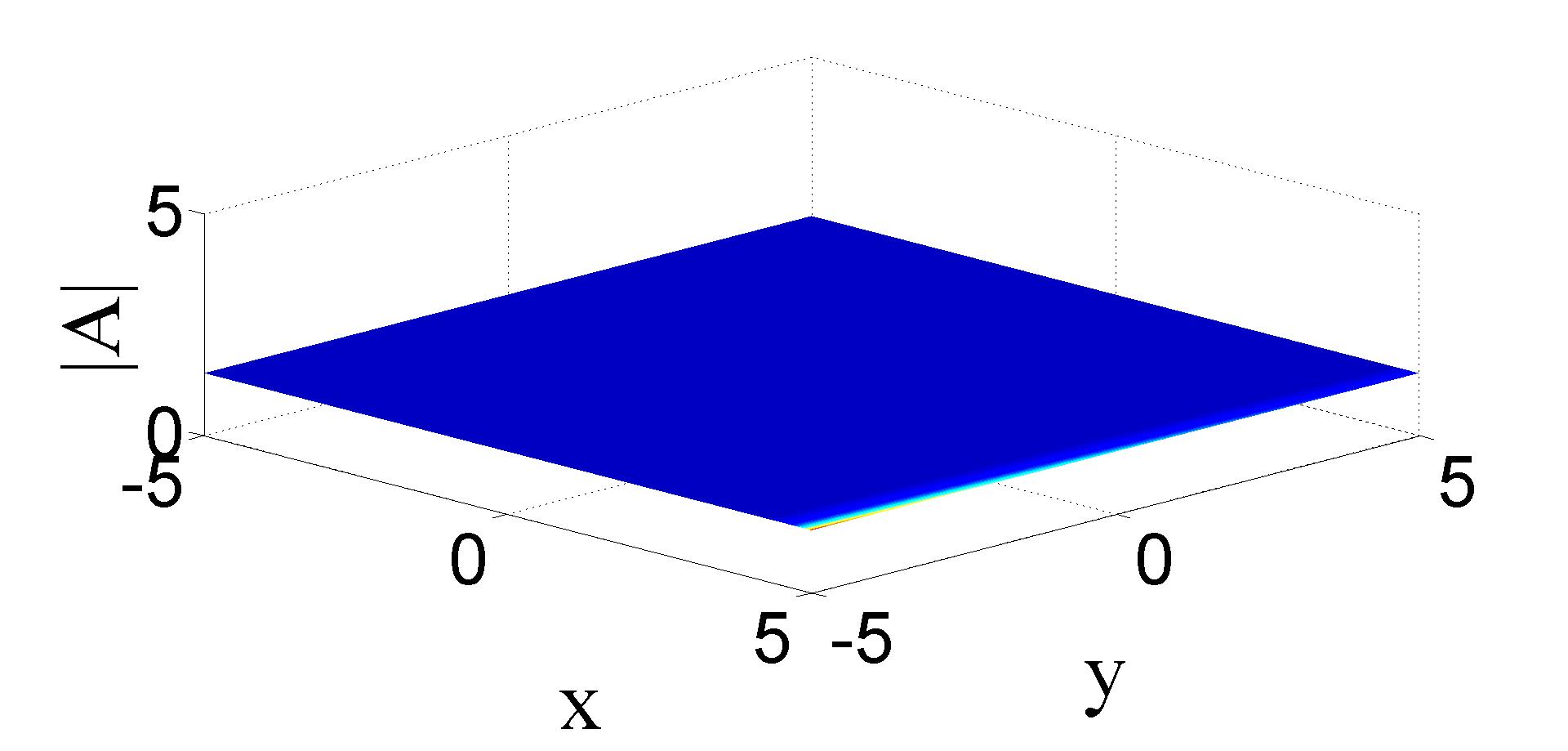}}\quad
\subfigure[$t=-1$]{\includegraphics[height=3cm,width=2.8cm]{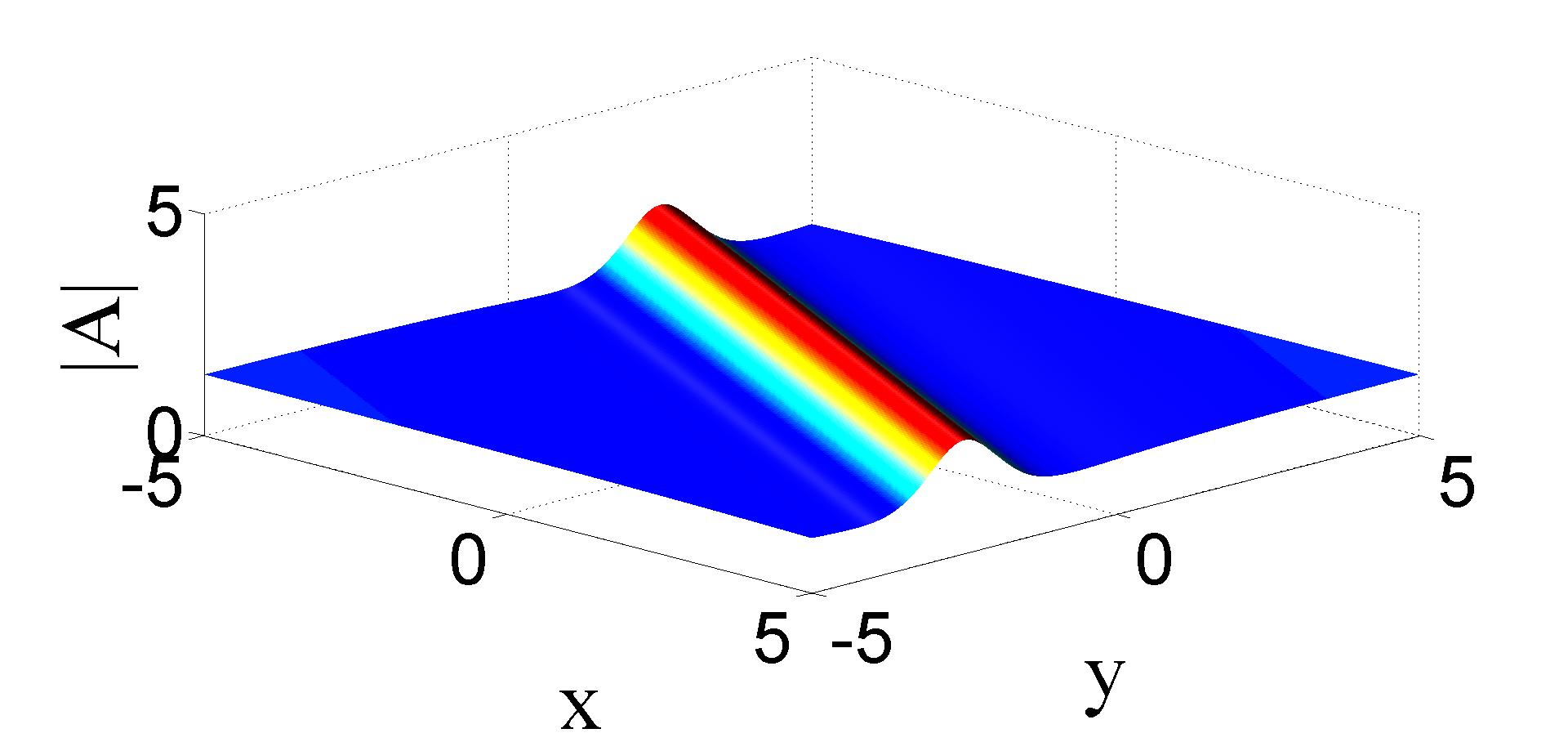}}\quad
\subfigure[$t=0$]{\includegraphics[height=3cm,width=2.8cm]{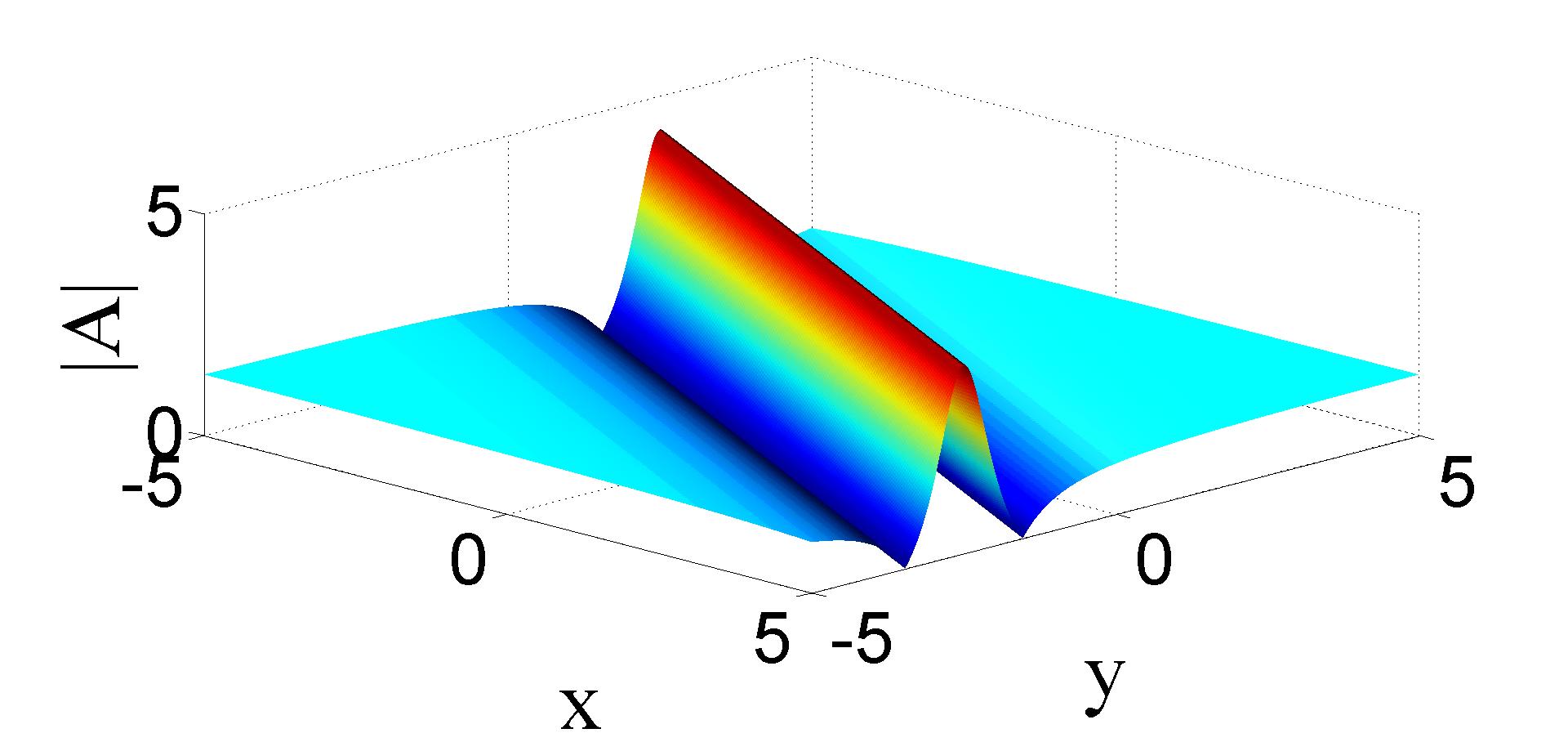}}\quad
\subfigure[$t=1$]{\includegraphics[height=3cm,width=2.8cm]{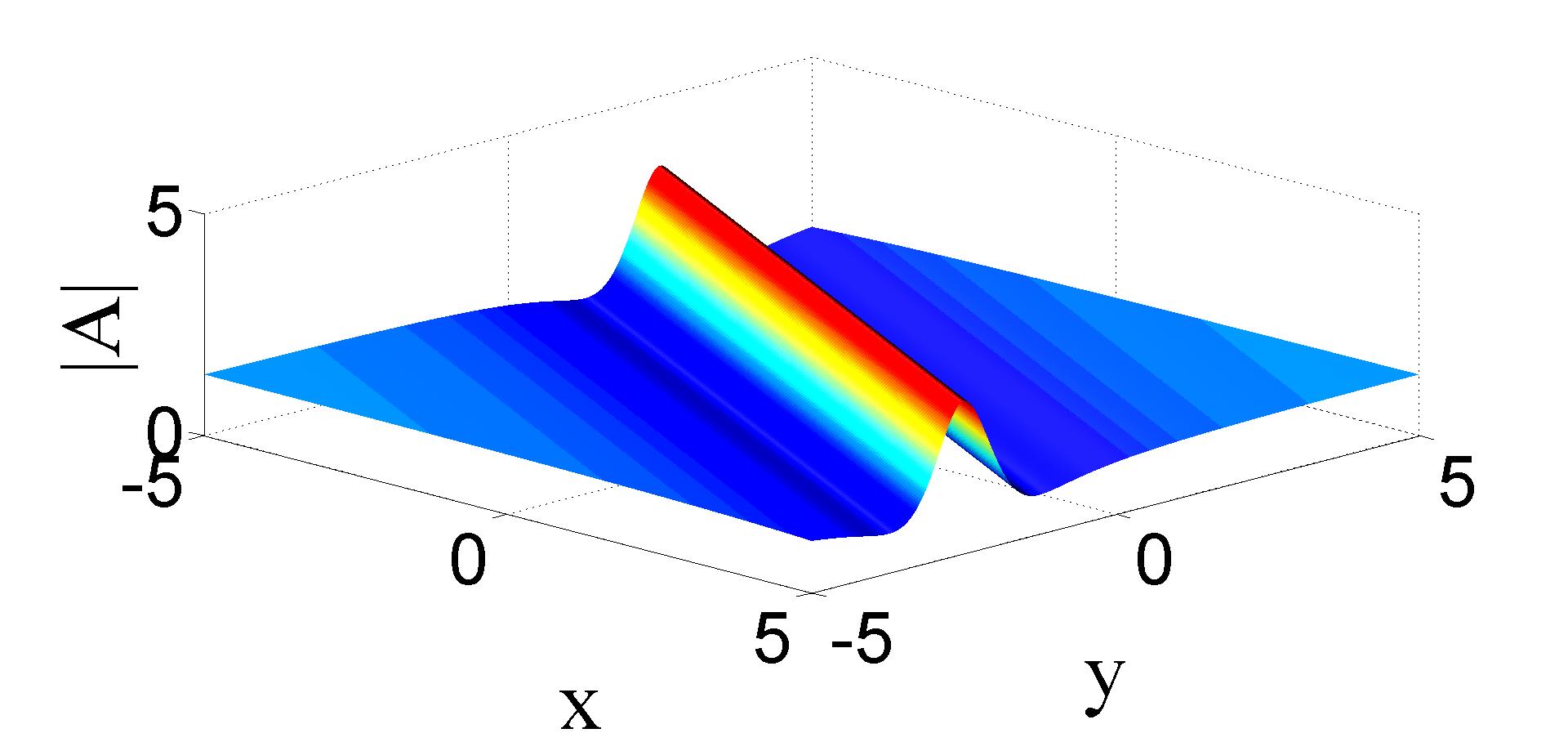}}\quad
\subfigure[$t=20$]{\includegraphics[height=3cm,width=2.8cm]{1rw-1.jpg}}
\caption{The time evolution of fundamental line rogue waves of the  nonlocal DSII given by \eqref{trw-1} in the  $(x,y)$-plane with parameter $a=2$.~}
\label{1-lrw}
\end{figure}

Next, we proceed to analyze typical dynamics of rogue waves constructed through the polynomials defined in (\ref{DS-f}) and  (\ref{tt-fg}).
The fundamental rogue waves in nonlocal DSII equation can be obtained by taking parameters in \eqref{tt-fg}
\begin{equation}\label{pa-4}
\begin{aligned}
N=2\,,\gamma_{1}\gamma_{2}=-1,,\lambda_{2}=\lambda_{1}=a\,,\epsilon=-1.
\end{aligned}
\end{equation}
Then, the final expression of fundamental rogue waves reads
\begin{equation}\label{trw-1}
\begin{aligned}
A(x,y,t)&=\sqrt{2}\frac{(\widetilde{\theta_{1}}+\widetilde{b_{1}})(\widetilde{\theta_{2}}+\widetilde{b_{2}})+\widetilde{a_{12}}}{\widetilde{\theta_{1}}\widetilde{\theta_{2}}+\widetilde{a_{12}}}\\
&=\frac{(\widetilde{\theta_{1}}+\widetilde{b_1})(\widetilde{\theta^{*}_{1}}-\widetilde{b^*_{1}})+\widetilde{a_{12}}}{\widetilde{\theta_{1}}\widetilde{\theta^*_{1}}+\widetilde{a_{12}}}\\
&=\sqrt{2}[1-\frac{4i(a^{2}+1)t+(a^{2}+1)}{(x+ay)^{2}+4\frac{(1-a^{2})^{2}}{1+a^{2}}t^{2}+\frac{1}{4}(1+a^{2})}]\,,\\
Q(x,y,t)&=\epsilon-2( {\rm log} (\widetilde{\theta_{1}}\widetilde{\theta_{2}}+\widetilde{a_{12}}))_{xx}\\
&=\epsilon-2( {\rm log} (\widetilde{\theta_{1}}\widetilde{\theta^*_{1}}+\widetilde{a_{12}}))_{xx}\\
&=-1+\frac{4(x+ay)^{2}-16\frac{(1-a^{2})^{2}}{1+a^{2}}t^{2}-(1+a^{2})}{[(x+ay)^{2}+4\frac{(1-a^{2})^{2}}{1+a^{2}}t^{2}+\frac{1}{4}(1+a^{2})]^{2}}\,,
\end{aligned}
\end{equation}
where $a$ is a freely real constant. This rational solution describes a line wave with the line oriented in the $(a,-1)$ direction of the $(x,y)$-plane. As the slop of the line wave is $-a$, which means the orientation angle of the line wave can rang from $0^{\circ}$ to $180^{\circ}$. Thus the orientation direction of line wave \eqref{trw-1} is almost arbitrary. As shown in Fig. \ref{1-lrw}, when $t\rightarrow \pm \infty$, this line wave $|A|$ uniformly goes to a constant background $\sqrt{2}$ in the $(x,y)$-plane. Interestingly, in this process, the height of the line waves is varying. In a intermediate times, $|A|$ rises to maximum values $3\sqrt{2}$ (i.e., three times the background amplitude) along the center ( $x+ay=0$ ) of the line wave at $t=0$. Thus these line waves possess a typical  characteristic of the rogue wave: appearing from no where and disappearing without trace, hence they are line rogue waves\cite{DS1,DS2}. Note that when $a=0$, the fundamental rogue waves defined in equation \eqref{trw-1} are independent of $y$, thus they reduce to the Peregrine rogue waves of the nonlocal NLS equation.

High-order rogue waves are constructed by selecting parameters in Theorem 2
\begin{equation}\label{t-high}
\begin{aligned}
N = 2n , \lambda_{n+k} = \lambda_{k} \,, \gamma_{n+k}\gamma_{k} = -1,\,(n\geq 2,1\leq k\leq n).
\end{aligned}
\end{equation}
 These solutions describe various senses of superimposition between $n$ individual fundamental line rogue waves. In the far field of the $(x,y)$-plane, $N$ separated line rogue waves arise from the constant
background. Then they interact with each other, and generate interesting curvy wave patterns. Finally, they disappear uniformly into the background again as $t\rightarrow +\infty$.

\begin{figure}[!htbp]
\centering
\subfigure[$t=-20$]{\includegraphics[height=3cm,width=2.8cm]{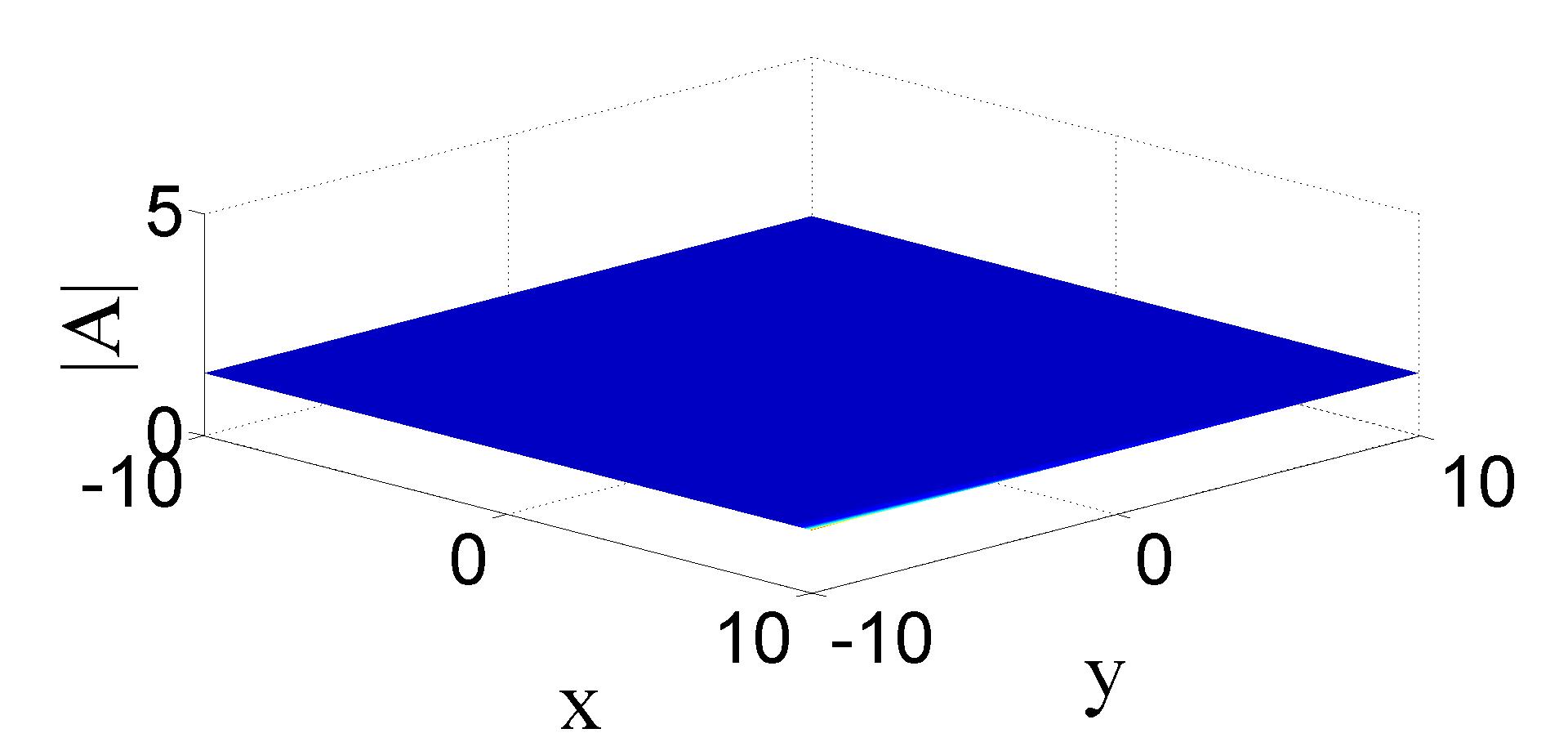}}\quad
\subfigure[$t=-1$]{\includegraphics[height=3cm,width=2.8cm]{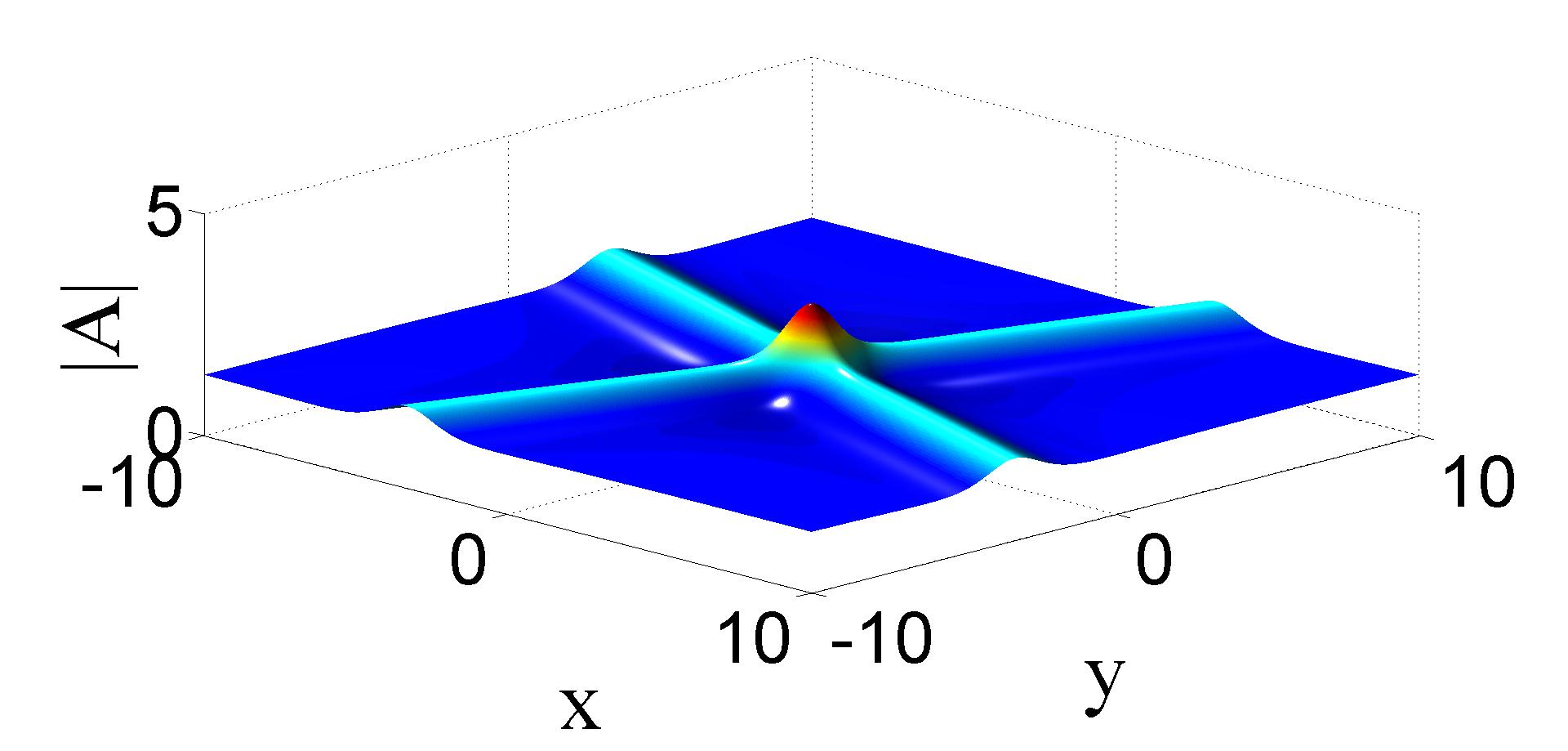}}\quad
\subfigure[$t=0$]{\includegraphics[height=3cm,width=2.8cm]{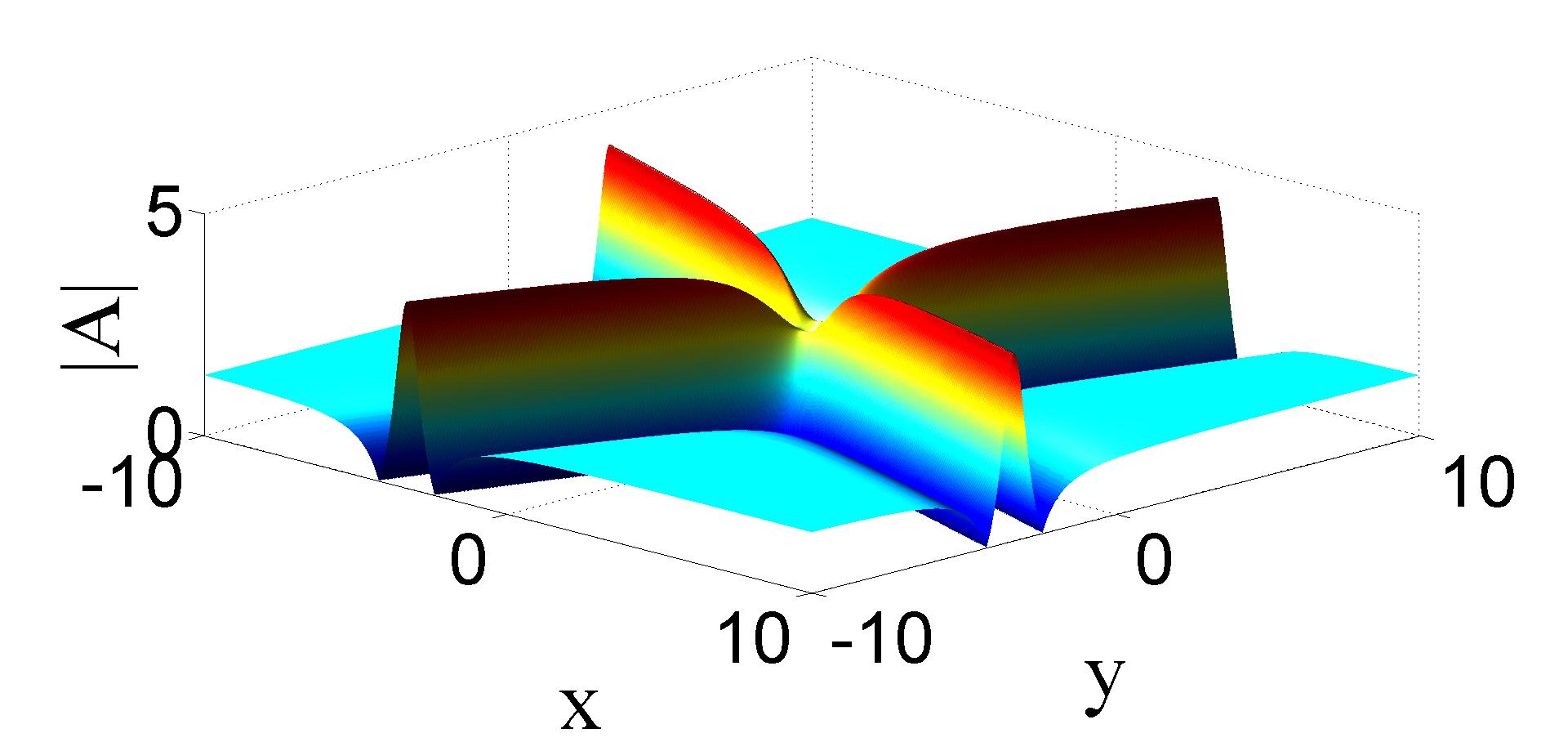}}\quad
\subfigure[$t=1$]{\includegraphics[height=3cm,width=2.8cm]{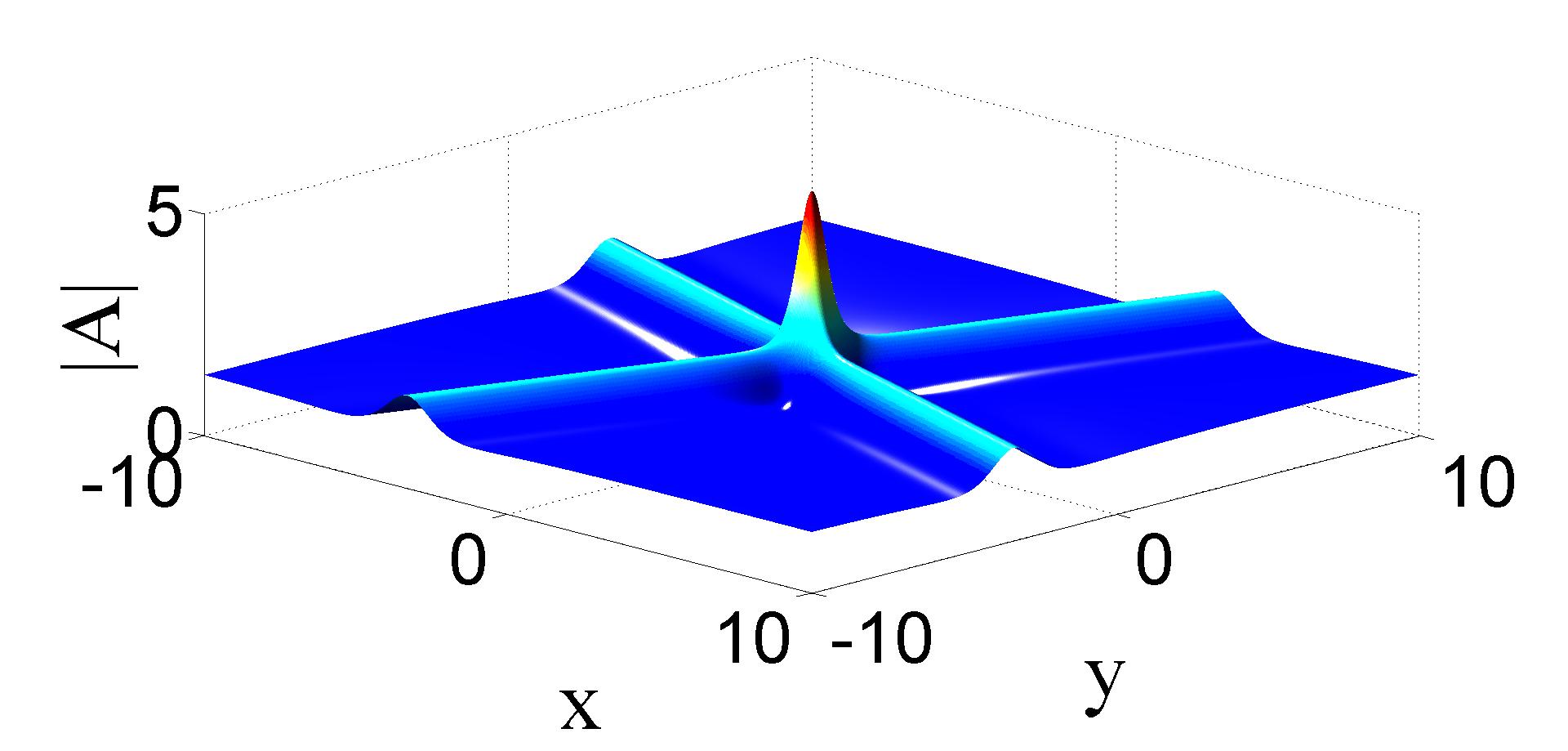}}\quad
\subfigure[$t=20$]{\includegraphics[height=3cm,width=2.8cm]{2rw-1.jpg}}
\caption{The time evolution of a second-order line rogue wave of the nonlocal DSII equation in the $(x,y)$-plane with parameters given by \eqref{pa-6}.~}\label{fig7}
\end{figure}

For instance, taking $N=4,\lambda_1=\lambda_3,\lambda_2=\lambda_4,\gamma_1\gamma_3=-1,\gamma_2\gamma_4=-1$ in Theorem 2, the corresponding rational solutions are second-order rogue waves, and the explicit forms of functions $\widetilde{f}$ and $\widetilde{g}$ used to generate the second-order rogue wave solutions are given by the following formulae:
\begin{equation}\label{2t-fg}
\begin{aligned}
\widetilde{f}=&\widetilde{\theta_{1}}\widetilde{\theta_{2}}\widetilde{\theta_{3}}\widetilde{\theta_{4}}+\widetilde{\alpha_{12}}\widetilde{\theta_{3}}\widetilde{\theta_{4}}+\widetilde{\alpha_{13}}
\widetilde{\theta_{2}}\widetilde{\theta_{4}}+\widetilde{\alpha_{14}}\widetilde{\theta_{2}}\widetilde{\theta_{3}}+\widetilde{\alpha_{23}}\widetilde{\theta_{1}}\widetilde{\theta_{4}}
+\widetilde{\alpha_{24}}\widetilde{\theta_{1}}\widetilde{\theta_{3}}+\widetilde{\alpha_{34}}\widetilde{\theta_{1}}\widetilde{\theta_{2}}\\
&+\widetilde{\alpha_{12}}\widetilde{\alpha_{34}}+\widetilde{\alpha_{13}}\widetilde{\alpha_{24}}+\widetilde{\alpha_{14}}\widetilde{\alpha_{23}},\\
\widetilde{g}=&(\widetilde{\theta_{1}}+\widetilde{b_{1}})(\widetilde{\theta_{2}}+\widetilde{b_{2}})(\widetilde{\theta_{3}}+\widetilde{b_{3}})(\widetilde{\theta_{4}}+\widetilde{b_{4}})
+\widetilde{\alpha_{12}}(\widetilde{\theta_{3}}+\widetilde{b_{3}})(\widetilde{\theta_{4}}+\widetilde{b_{4}})+\widetilde{\alpha_{13}}(\widetilde{\theta_{2}}+\widetilde{b_{2}})(\widetilde{\theta_{4}}+\widetilde{b_{4}})\\
&+\widetilde{\alpha_{14}}(\widetilde{\theta_{2}}+\widetilde{b_{2}})(\widetilde{\theta_{3}}+\widetilde{b_{3}})+\widetilde{\alpha_{23}}(\widetilde{\theta_{1}}+\widetilde{b_{1}})(\widetilde{\theta_{4}}+\widetilde{b_{4}})+\widetilde{\alpha_{24}}(\widetilde{\theta_{1}}+\widetilde{b_{1}})(\widetilde{\theta_{3}}+\widetilde{b_{3}})\\
&+\widetilde{\alpha_{34}}(\widetilde{\theta_{1}}+\widetilde{b_{1}})(\widetilde{\theta_{2}}+\widetilde{b_{2}})+\widetilde{\alpha_{12}}\widetilde{\alpha_{34}}+\widetilde{\alpha_{13}}\widetilde{\alpha_{24}}+\widetilde{\alpha_{14}}\widetilde{\alpha_{23}},\\
\end{aligned}
\end{equation}
where $\widetilde{\theta_{i}}\,,\widetilde{\alpha_{ij}}$ and $\widetilde{b_{i}}$ are given by \eqref{tt-rt}.  For particular of
\begin{equation}\label{pa-6}
\begin{aligned}
\epsilon=-1\,,\lambda_{1}=\lambda_{3}=2\,,\lambda_{2}=\lambda_{4}=-\frac{1}{2}\,,\gamma_{1}\gamma_{3}=-1\,,
\gamma_{2}\gamma_{4}=-1\,,
\end{aligned}
\end{equation}
the corresponding solution $|A|$ is shown in Fig.\ref{fig7}.
It is shown that two crossed line rogue wave
arise from the constant background gradually in the $(x,y)$-plane. Visually, the region of the intersection reaches higher amplitude (see $t=-1$ panel).  However, these higher amplitudes in the area where these
two line rogue waves cross over each other fade quickly, and in the far field these amplitudes become higher and higher (see $t=0$ panel). At larger time, these two line rogue waves start to immerse into the constant
background uniformly, until they disappear into the constant background finally (see $t=20$ panel). It should be noted that these two line rogue waves have never separated during the short period that rogue waves appear, which is
different from the second-order rogue waves in the usual DSI equation, as the curvy waves of the later one are well separated at some moment (see t=0 panel of Fig.2 in Ref. \cite{DS1}).

Besides, when one takes parameters in \eqref{2t-fg}
\begin{equation}\label{pa-7}
\begin{aligned}
\epsilon=-1\,,\lambda_{1}=\lambda_{3}=1\,,\lambda_{2}=\lambda_{4}=-1\,,\gamma_{1}\gamma_{3}=-1\,,\gamma_{2}\gamma_{4}=-1\,,
\end{aligned}
\end{equation}
the corresponding second-order nonsingular rational solution is independent of $t$.

\subsection{Semi-rational solutions of the nonlocal DSII equation}$\\$

In this subsection, we take a long wave limit of the periodic solutions in \eqref{tfg} partially, and keep the other part of them still  in exponential functions, thus semi-rational solutions expressed in a combination of rational and exponential functions are generated through \eqref{DS-f}.  These semi-rational solutions demonstrate typical dynamics of rogue waves on a  background  of periodic line waves.
Indeed, setting $0<2j<N$ and $1\leq k\leq 2j$,
\begin{equation} \label{pa-8}
\begin{aligned}
Q_{k}=\lambda_{k}P_{k}\,\,,\eta_{k}^{0}=i\pi,
\end{aligned}
\end{equation}
and taking limit $P_{k}\rightarrow 0$ for all $k$,  then  two functions $\widetilde{f}$ and $\widetilde{g}$  in \eqref{tfg} becomes
 a combination of rational and exponential functions, which generate semi-rational solutions $A$ and $Q$ of
  the  nonlocal DSII equation through \eqref{DS-f}.  Specifically, when one takes $(P^2_k-Q_{k}^2)\sqrt{-1-\frac{4\epsilon}{P_{k}^{2}+Q_{k}^{2}}}=0$, i.e., $\widetilde{\Omega_{k}}=0$ in \eqref{tfg}, the exponential functions in semi-rational solutions are independent of $t$, thus the periodic line waves
maintain a perfect profile without any decay during their propagation in the $(x,y)$-plane, which provides a
significant periodic line waves background.  So the corresponding semi-rational solutions describe rogue waves on a background of periodic line waves.
\begin{figure}[!htbp]
\centering
\subfigure[$t=-20$]{\includegraphics[height=3cm,width=3cm]{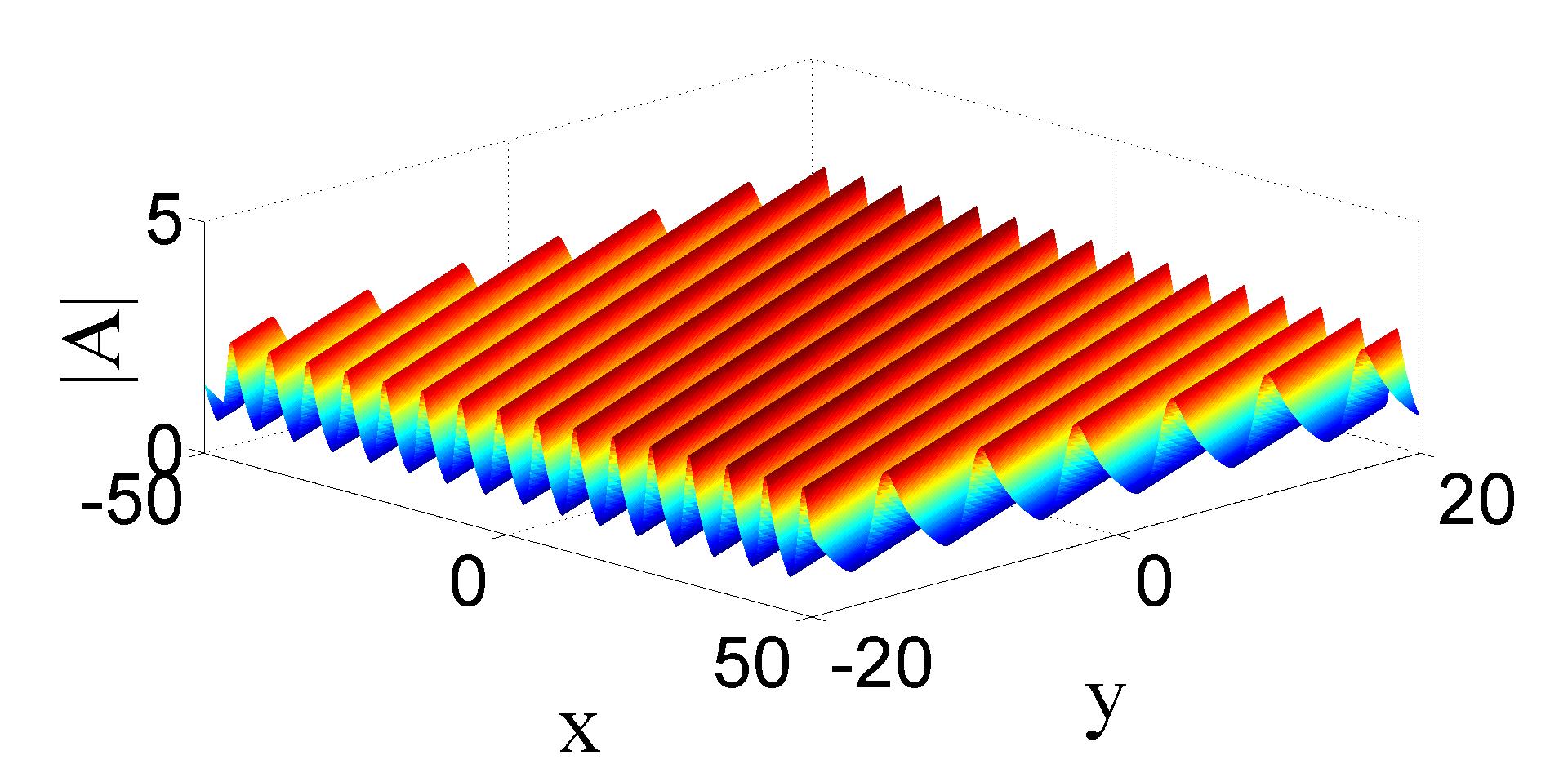}}
\subfigure[$t=-1$]{\includegraphics[height=3cm,width=3cm]{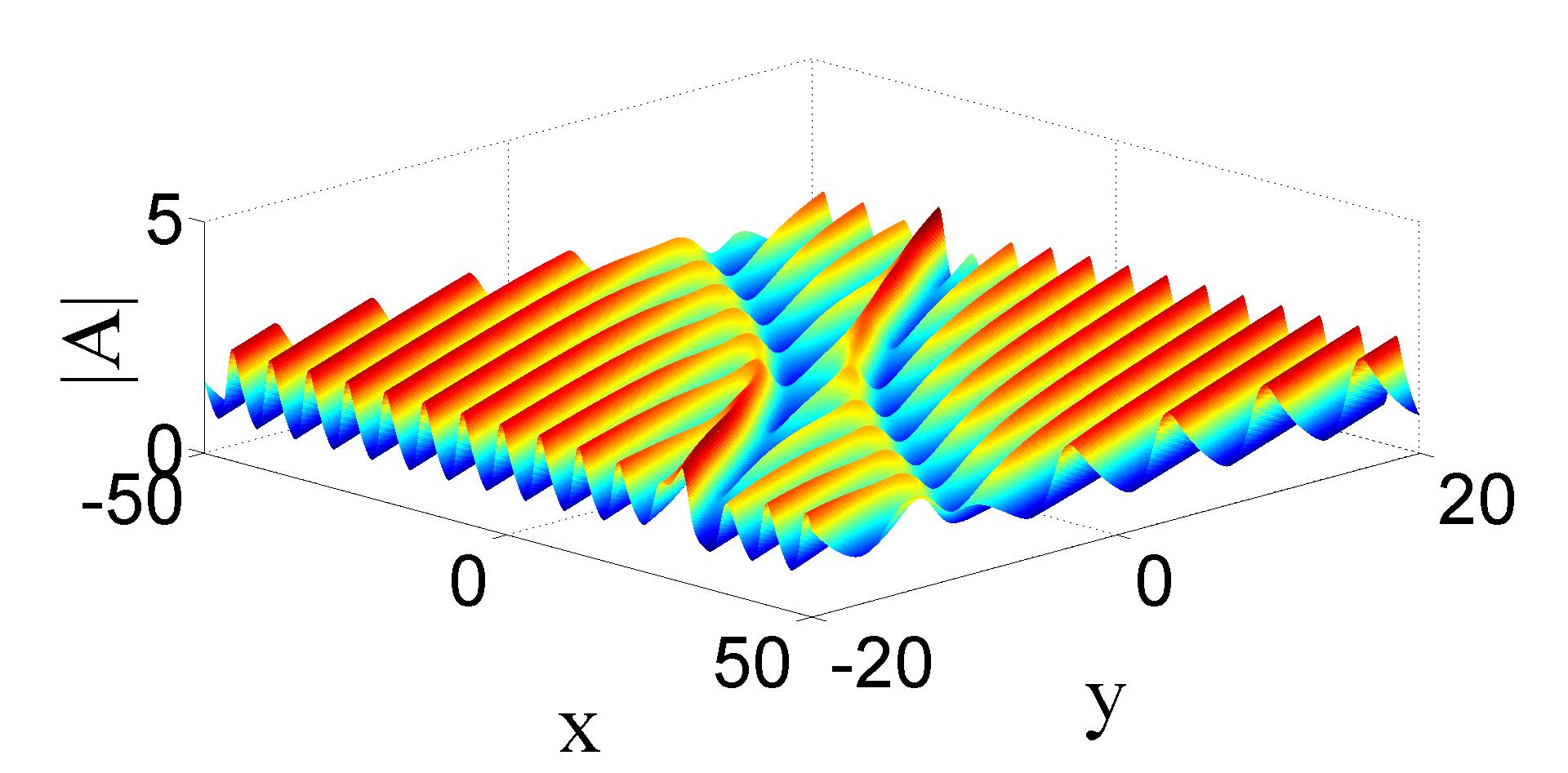}}
\subfigure[$t=0$]{\includegraphics[height=3cm,width=3cm]{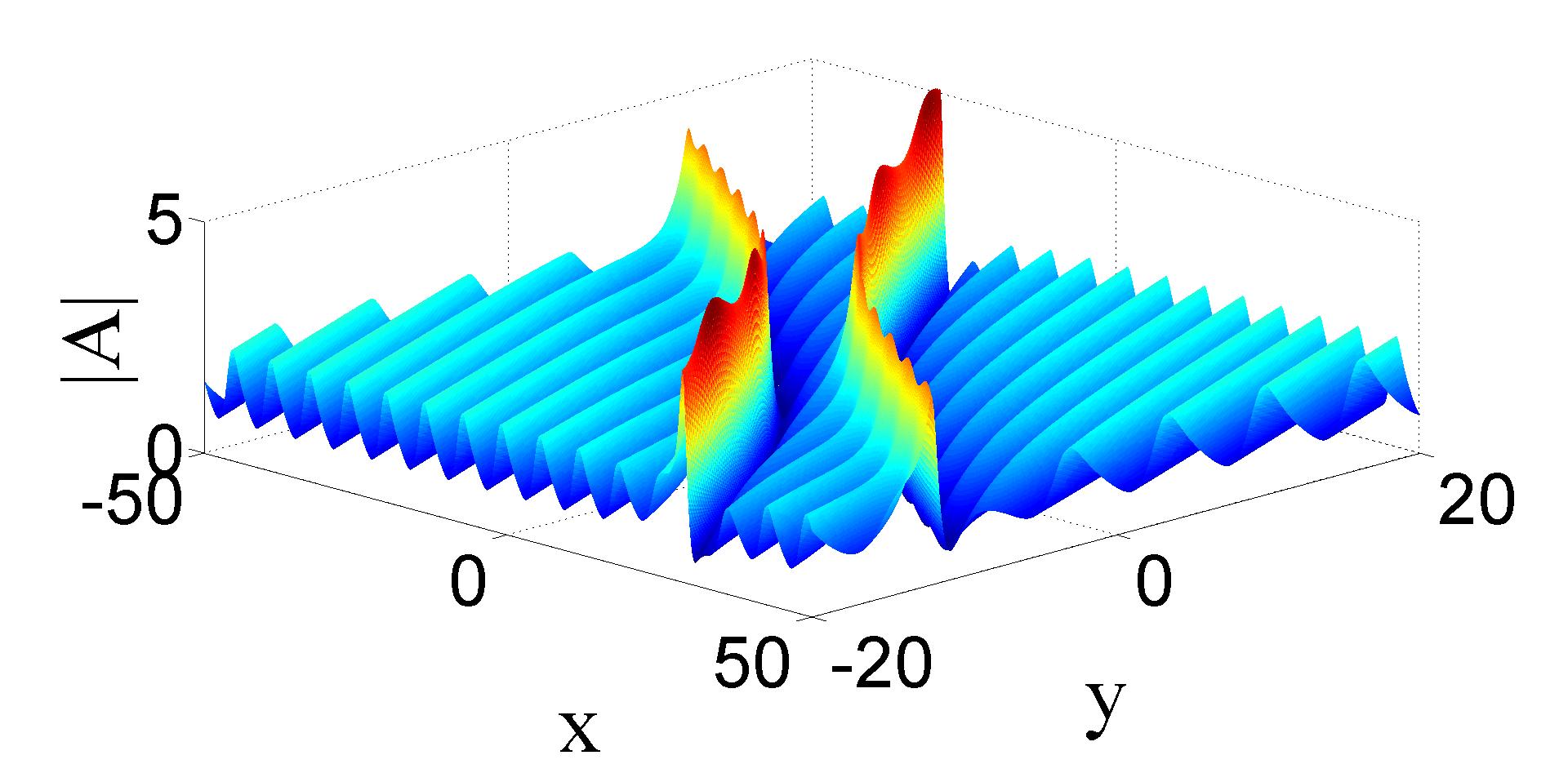}}
\subfigure[$t=1$]{\includegraphics[height=3cm,width=3cm]{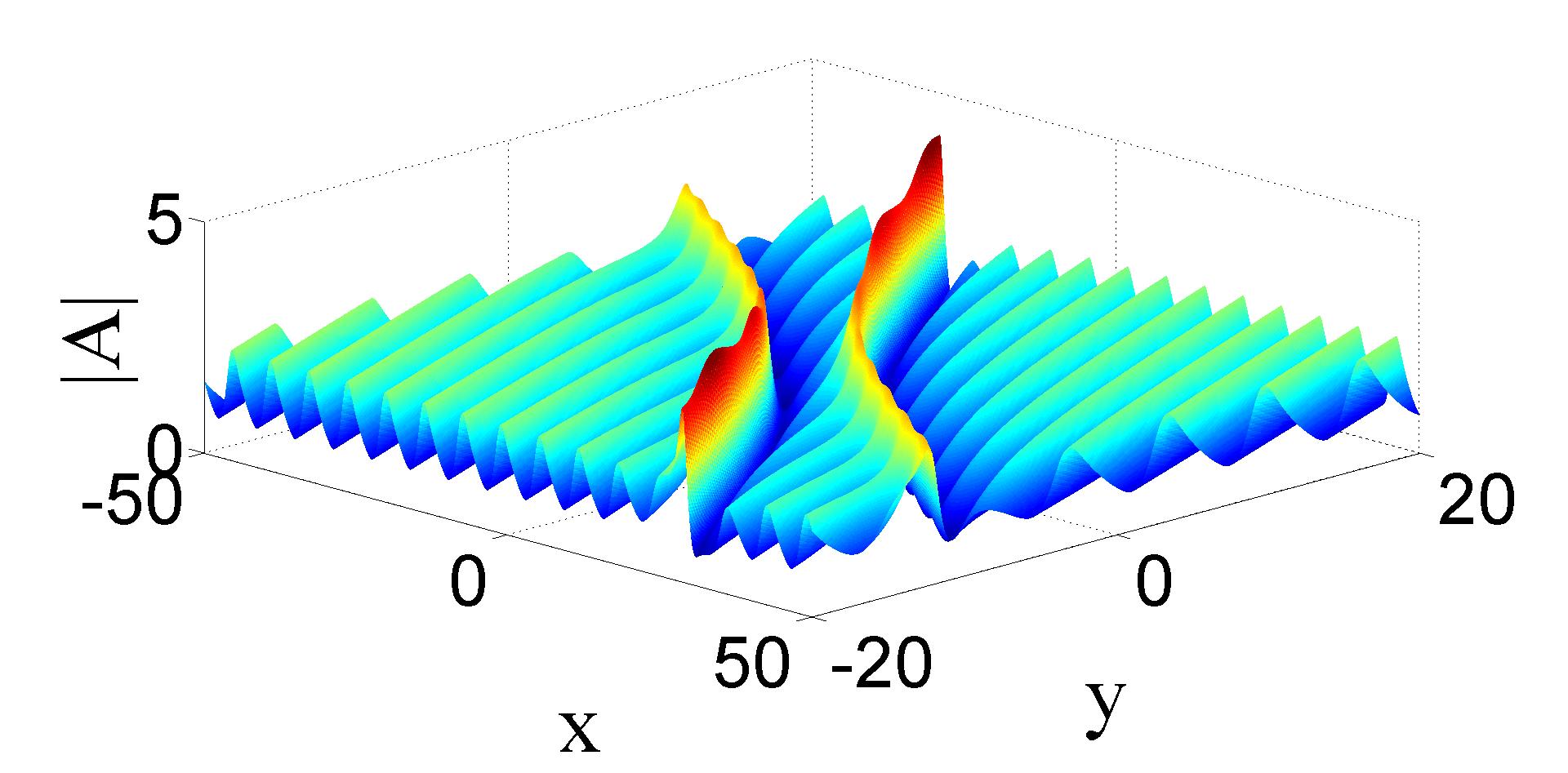}}
\subfigure[$t=20$]{\includegraphics[height=3cm,width=3cm]{f2hy-10.jpg}}
\caption{The time evolution of a second-order line rogue wave on a background of periodic line waves of  the  nonlocal DSII equation  in the  $(x,y)$-plane with parameters given by \eqref{pa-11}.~}\label{fig8}
\end{figure}
To illustrate this unique dynamic of rogue wave phenomenon, we consider the second-order rogue wave on a background of periodic line waves. This requirement can be done by
taking
\begin{equation} \label{pa-10}
\begin{aligned}
N=5\,,Q_{k}=\lambda_{k}P_{k}\,,\eta_{k}^{0}=i\pi\,(\,k=1,2,3,4\,)\,,\widetilde{\Omega_5}=0,
\end{aligned}
\end{equation}
and letting $P_{i}\rightarrow 0$ in \eqref{tfg}, then functions $\widetilde{f}$ and $\widetilde{g}$ become
\begin{equation} \label{ff3}
\begin{aligned}
\widetilde{f}=&(\widetilde{\theta_{1}}\widetilde{\theta_{2}}\widetilde{\theta_{3}}\widetilde{\theta_{4}}+\widetilde{a_{12}}\widetilde{\theta_{3}}\widetilde{\theta_{4}}+\widetilde{a_{13}}\widetilde{\theta_{2}}\widetilde{\theta_{4}}+\widetilde{a_{14}}\widetilde{\theta_{2}}\widetilde{\theta_{3}}+\widetilde{a_{23}}\widetilde{\theta_{1}}\widetilde{\theta_{4}}
+\widetilde{a_{24}}\widetilde{\theta_{1}}\widetilde{\theta_{3}}+\widetilde{a_{34}}\widetilde{\theta_{1}}\widetilde{\theta_{2}}+\widetilde{a_{12}}\widetilde{a_{34}}+\\
&\widetilde{a_{13}}\widetilde{a_{24}}+\widetilde{a_{14}}\widetilde{a_{23}})+e^{\widetilde{\eta_{5}}}[\widetilde{\theta_{1}}\,\widetilde{\theta_{2}}\,\widetilde{\theta_{3}}\,\widetilde{\theta_{4}}+\widetilde{a_{45}}\,\widetilde{\theta_{1}}\,\widetilde{\theta_{2}}\,\widetilde{\theta_{3}}+\widetilde{a_{35}}\,\widetilde{\theta_{1}}\,\widetilde{\theta_{2}}\,\widetilde{\theta_{4}}
+\widetilde{a_{25}}\,\widetilde{\theta_{1}}\,\widetilde{\theta_{3}}\,\widetilde{\theta_{4}}+\widetilde{a_{15}}\widetilde{\theta_{2}}\,\widetilde{\theta_{3}}\,\widetilde{\theta_{4}}+\\
&(\widetilde{a_{35}}\widetilde{a_{45}}+\widetilde{a_{34}})\widetilde{\theta_{1}}\,\widetilde{\theta_{2}}+(\widetilde{a_{25}} \widetilde{a_{45}}+\widetilde{a_{24}})\,\widetilde{\theta_{1}}\,\widetilde{\theta_{3}}+(\widetilde{a_{25}}\widetilde{a_{35}}+\widetilde{a_{23}})\,\widetilde{\theta_{1}}\,\widetilde{\theta_{4}}
+(\widetilde{a_{15}}\widetilde{a_{45}}+\widetilde{a_{14}})\,\widetilde{\theta_{2}}\,\widetilde{\theta_{3}}\\
&+(\widetilde{a_{15}}\widetilde{a_{35}}+\widetilde{a_{13}})\,\widetilde{\theta_{2}}\,\widetilde{\theta_{4}}+
(\widetilde{a_{15}}\widetilde{a_{25}}+\widetilde{a_{12}})\widetilde{\theta_{3}}\,\widetilde{\theta_{4}}+(\widetilde{a_{25}}\widetilde{a_{35}}\,\widetilde{a_{45}}
+\widetilde{a_{23}}\widetilde{a_{45}}+\widetilde{a_{25}}\widetilde{a_{34}}+\widetilde{a_{24}}\widetilde{a_{35}})\,\widetilde{\theta_{1}}\\&
+(\widetilde{a_{15}}\widetilde{a_{35}}\widetilde{a_{45}}+\widetilde{a_{14}}\widetilde{a_{35}}+\widetilde{a_{13}}\widetilde{a_{45}}+\widetilde{a_{15}}\widetilde{a_{34}})\,\widetilde{\theta_{2}}+(\widetilde{a_{15} } \widetilde{a_{25}}\widetilde{a_{45}}+\widetilde{a_{14}}\widetilde{a_{25}}+\widetilde{a_{15}}\widetilde{a_{24}}+\widetilde{a_{12}}\widetilde{a_{45}})\,\widetilde{\theta_{3}}\\
&+(\widetilde{a_{15}}\widetilde{a_{25}}\widetilde{a_{35}}+\widetilde{a_{15}}\widetilde{a_{23}}+\widetilde{a_{13}}\widetilde{a_{25}}+\widetilde{a_{12}}\widetilde{a_{35}})\,
\widetilde{\theta_{4}}+\widetilde{a_{12}}\widetilde{a_{34}}+\widetilde{a_{13}}\,\widetilde{a_{24}}+\widetilde{a_{14}}\widetilde{a_{23}}+\widetilde{a_{12}}\,\widetilde{a_{35}}\,\widetilde{a_{45}}+\\
&\widetilde{a_{13}}\widetilde{a_{25}}\widetilde{a_{45}}+\widetilde{a_{14}}\widetilde{a_{25}}\widetilde{a_{35}}+\widetilde{a_{15}}\widetilde{a_{24}}\widetilde{a_{35}}+\widetilde{a_{15}}\widetilde{a_{25}}\widetilde{a_{34}}\,+\widetilde{a_{15}}\widetilde{a_{23}}\widetilde{a_{45}}+\widetilde{a_{15}}\widetilde{a_{25} } \widetilde{a_{35}} \widetilde{a_{45}}]\,,\\
g=&[(\widetilde{\theta_{1}}+\widetilde{b_{1}})(\widetilde{\theta_{2}}+\widetilde{b_{2}})(\widetilde{\theta_{3}}+\widetilde{b_{3}})(\widetilde{\theta_{4}}+\widetilde{b_{4}})
+\widetilde{a_{12}}(\widetilde{\theta_{3}}+\widetilde{b_{3}})(\widetilde{\theta_{4}}+\widetilde{b_{4}})+\widetilde{a_{13}}(\widetilde{\theta_{2}}+\widetilde{b_{2}})(\widetilde{\theta_{4}}+\widetilde{b_{4}})+\\
&\widetilde{a_{14}}(\widetilde{\theta_{2}}+\widetilde{b_{2}})(\widetilde{\theta_{3}}+\widetilde{b_{3}})+\widetilde{a_{23}}(\widetilde{\theta_{1}}+\widetilde{b_{1}})(\widetilde{\theta_{4}}+\widetilde{b_{4}})
+\widetilde{a_{24}}(\widetilde{\theta_{1}}+\widetilde{b_{1}})(\widetilde{\theta_{3}}+\widetilde{b_{3}})+\widetilde{a_{34}}(\widetilde{\theta_{1}}+\widetilde{b_{1}})(\widetilde{\theta_{2}}\\
&+\widetilde{b_{2}})+\widetilde{a_{12}}\widetilde{a_{34}}+\widetilde{a_{13}}\widetilde{a_{24}}+\widetilde{a_{14}}\widetilde{a_{23}}]+e^{\widetilde{\eta_{5}}+
i\widetilde{\phi_{5}}}[(\widetilde{\theta_{1}}+\widetilde{b_{1}})(\widetilde{\theta_{2}}+\widetilde{b_{2}})(\widetilde{\theta_{3}}+\widetilde{b_{3}})(\widetilde{\theta_{4}}+\widetilde{b_{4}})+\\
&\widetilde{a_{45}}\,(\widetilde{\theta_{1}}+\widetilde{b_{1}})(\widetilde{\theta_{2}}+\widetilde{b_{2}})(\widetilde{\theta_{3}}+\widetilde{b_{3}})+\widetilde{a_{35}}\,
(\widetilde{\theta_{1}}+\widetilde{b_{1}})(\widetilde{\theta_{2}}+\widetilde{b_{2}})(\widetilde{\theta_{4}}+\widetilde{b_{4}})
+\widetilde{a_{25}}\,(\widetilde{\theta_{1}}+\widetilde{b_{1}})(\widetilde{\theta_{3}}+\\
&\widetilde{b_{3}})(\widetilde{\theta_{4}}+\widetilde{b_{4}})+\widetilde{a_{15}}(\widetilde{\theta_{2}}+\widetilde{b_{2}})(\widetilde{\theta_{3}}+\widetilde{b_{3}})(\widetilde{\theta_{4}}+\widetilde{b_{4}})+
(\widetilde{a_{35}}\widetilde{a_{45}}+\widetilde{a_{34}})(\widetilde{\theta_{1}}+\widetilde{b_{1}})(\widetilde{\theta_{2}}+\widetilde{b_{2}})+\\
&(a_{25} a_{45}+a_{24})\,(\theta_{1}+b_{1})(\theta_{3}+b_{3})+(a_{25}a_{35}+a_{23})(\theta_{2}+b_{2})(\theta_{4}+b_{4})+(a_{15}a_{45}+a_{14})\\
&\,(\widetilde{\theta_{2}}+\widetilde{b_{2}})(\widetilde{\theta_{3}}+\widetilde{b_{3}})+(\widetilde{a_{15}}\widetilde{a_{35}}+\widetilde{a_{13}})\,
(\widetilde{\theta_{2}}+\widetilde{b_{2}})(\widetilde{\theta_{4}}+\widetilde{b_{4}})+(\widetilde{a_{15}}\widetilde{a_{25}}+\widetilde{a_{12}})(\widetilde{\theta_{3}}+\widetilde{b_{3}})
(\widetilde{\theta_{4}}+\widetilde{b_{4}})\\
&+(\widetilde{a_{25}}\widetilde{a_{35}}\,\widetilde{a_{45}}+\widetilde{a_{23}}\widetilde{a_{45}}+\widetilde{a_{25}}\widetilde{a_{34}}+\widetilde{a_{24}}
\widetilde{a_{35}})\,(\widetilde{\theta_{1}}+\widetilde{b_{1}})+(\widetilde{a_{15}}\widetilde{a_{35}}\widetilde{a_{45}}+\widetilde{a_{14}}\widetilde{a_{35}}+\widetilde{a_{13}}\widetilde{a_{45}}+
\widetilde{a_{15}}\widetilde{a_{34}})\\
&(\widetilde{\theta_{2}}+\widetilde{b_{2}})+(\widetilde{a_{15}} \widetilde{a_{25}}\widetilde{a_{45}}+\widetilde{a_{14}}\widetilde{a_{25}}+\widetilde{a_{15}}\widetilde{a_{24}}+\widetilde{a_{12}}\widetilde{a_{45}})\,(\widetilde{\theta_{3}}\widetilde{+b_{3}})
+(\widetilde{a_{15}}\widetilde{a_{25}}\widetilde{a_{35}}+\widetilde{a_{15}}\widetilde{a_{23}}+\widetilde{a_{13}}\widetilde{a_{25}}\\
&+\widetilde{a_{12}}\widetilde{a_{35}})(\widetilde{\theta_{4}}+\widetilde{b_{4}})+\widetilde{a_{12}}(\widetilde{a_{34}}+\widetilde{a_{35}}
\,\widetilde{a_{45}})+\widetilde{a_{13}}(\,\widetilde{a_{24}}+\widetilde{a_{25}}\widetilde{a_{45}})+\widetilde{a_{14}}(\widetilde{a_{23}}+\widetilde{a_{25}}\widetilde{a_{35}})+\\
&\widetilde{a_{15}}(\widetilde{a_{24}}\widetilde{a_{35}}+\widetilde{a_{25}}\widetilde{a_{34}}\,+\widetilde{a_{23}}\widetilde{a_{45}}+\widetilde{a_{25} }\widetilde{a_{35}} \widetilde{a_{45}})].
\end{aligned}
\end{equation}
Here $\widetilde{\theta_{j}}\,,\widetilde{b_{j}}\,,\widetilde{\eta_{j}}\,,\widetilde{a_{kj}}\,(\,1\leq k<j\leq4)$ are given by \eqref{tt-fg}, $\widetilde{\eta_{5}}\,,\widetilde{\phi_{5}}$ are given by \eqref{tcs1},  and $$\widetilde{a_{k5}}=\frac{(\lambda_{k}^{2}+1)(P_{5}^{2}+Q_{5}^{2})}{\sqrt{-\frac{P_{5}^{2}+Q_{5}^{2}-4}{(\lambda_{k}^{2}+1)(P_{5}^{2}+Q_{5}^{2})}}(\lambda_{k}^{2}+1)(P_{5}^{2}+Q_{5}^{2})-2(\lambda_{k}Q_{5}+P_{5})}.$$

This solution for parameter choices
\begin{equation}\label{pa-11}
\begin{aligned}
\lambda_{1}=\lambda_{3}=4\,,\lambda_{2}=\lambda_{4}=\frac{3}{2}\,,\gamma_{1}\gamma_{3}=-1\,,\gamma_{2}\gamma_{4}=-1\,,
P_{5}=1\,,Q_{5}=1\,,\eta_{5}^{0}=\frac{\pi}{3}
\end{aligned}
\end{equation}
is shown in Fig.\ref{fig8}. As can be seen, the corresponding solution describes a second-order line rogue wave on a background of periodic line waves.  When $t\rightarrow \pm \infty$, line waves approach to the background of periodic line waves (see $t= \pm 20$ panel). In the intermediate time,  two fundamental line rogue waves arise from the periodic line waves, and interact with these periodic line waves, the region of their
intersection acquires higher amplitude (see $t=0$ panel).   Interestingly, these two fundamental line rogue waves in the second-order rogue wave are separated remarkably. That is different from the second-order rogue waves shown in Fig. \ref{fig7}, which are never separated in whole evolution even at intermediate time (see the panel at $t=0$ of Fig.\ref{fig7}).  Specifically, the maximum amplitudes of this second-order rogue wave does not exceed $5\sqrt{2}$ for all time, thus rogue waves on a background of periodic line waves do not generate high peaks. As to best of the authors' knowledge, these unique dynamics of rogue wave on a  background with periodic line waves have not been shown in the nonlocal DSII equation before, even in the usual DSII equation.
Note that, for particular case of
\begin{equation}\label{pa-12}
\begin{aligned}
\lambda_{1}=\lambda_{3}=1\,,\lambda_{2}=\lambda_{4}=-1\,,\gamma_{1}\gamma_{3}=-1\,,\gamma_{2}\gamma_{4}=-1\,,
P_{5}=Q_{5}
\end{aligned}
\end{equation}
in \eqref{ff3}, the corresponding solutions $A$ and $Q$ are independent of $t$.

\section{Summary and discussion}\label{4}
In summary, we have derived $N$-soliton solutions and periodic line wave solutions for the $x$-nonlocal DS equations and the nonlocal DSII equation respectively by employing the Hirota's bilinear method.  Under suitable selections of parameters, usual breathers and line breathers for these two types of nonlocal DS equations are generated respectively.
Taking a long wave limit of soliton solutions and periodic line wave solutions under special parameter constraints, nonsingular rational and semi-rational solutions for these two equation are given explicitly.  For the $x$-nonlocal DS equations, the nonsingular rational solutions are lumps, and the semi-rational solutions are a hybrid of lumps, usual breathers and periodic line waves. However, for the  nonlocal DSII equation, the nonsingular rational solutions are line rogue waves, and the semi-rational solutions are a mixture of line rogue waves and periodic line waves.

By comparing solutions of the $x$-nonlocal DS equations and the nonlocal DSII equation in detail, the main differences can be summarized in following items:
\begin{itemize}
\item Breather solution. The $x$-nonlocal DS equations have usual breathers which are periodic in $x$ direction and localized in $y$ direction, see Fig.\ref{fig1}. The nonlocal DSII equation has  line breathers which are periodic line waves and periodic in both $x$ and $y$ directions. Another great difference is that the line breathers just exist for a short period (see Fig.\ref{ll-br}), while the usual breathers are not disappeared.
\item Nonsingular rational solution. The simplest (fundamental) nonsingular rational solutions are lumps in the $x$-nonlocal DS equations (see Fig.\ref{1-lump}), and the high-order nonsingular rational solutions are composed of several fundamental lumps, see Fig.\ref{fig3}.  However, for the nonlocal DSII equation, the nonsingular rational solutions are line rogue waves, see Fig. \ref{1-lrw} and Fig.\ref{fig7}.
\item Semi-rational solutions. The semi-rational solutions of the $x$-nonlocal DS equations describe the interaction of lumps, breathers and periodic line waves, see Figs.\ref{fig5},\ref{fig6},\ref{lump+br},\ref{fig4}.  On the other hand, the semi-rational solutions of the nonlocal DSII equation are composed of line rogue waves and periodic line wave, see Fig.\ref{fig8}.
\end{itemize}
These featured properties of the $x$-nonlocal and nonlocal DS equations show that the partially $PT$ and fully $PT$ systems deserve  to be more explored  and to be better understood in the future.

As discussed in\cite{yang,par1,par2,par3,par4,par5,par6}, both of partially $PT$ symmetric potentials and $PT$ symmetric potentials in multi-dimensions have an important role in optics, our results may have interesting applications in optics.
Moreover, it is interesting to reduce the above solutions to other (1+1)-dimensional nonlocal equations, besides the nonlocal NLS equation.  Finally, it is worthy to emphasize that the technique presented in
this paper may also succeed to other nonlinear systems with different kinds of fully or partially PT symmetric potentials, or reverse space-time nonlocal equations\cite{fokas-pt,ab-pt}. The related results will be reported elsewhere later.

\section*{Acknowledgments}
This work is supported by the NSF of China under Grant No. 11671219,  and the K.C. Wong Magna Fund in Ningbo University. We thank Mr. Chao Qian, Mr. Yaobin Liu for help on the plotting, and other members in our group
at Ningbo University for many useful discussions and suggestions on the paper.


\begin{thebibliography}{10}
\providecommand{\url}[1]{\texttt{#1}}
\providecommand{\urlprefix}{URL }
\expandafter\ifx\csname urlstyle\endcsname\relax
  \providecommand{\doi}[1]{doi:\discretionary{}{}{}#1}\else
  \providecommand{\doi}{doi:\discretionary{}{}{}\begingroup
  \urlstyle{rm}\Url}\fi
\providecommand{\eprint}[2][]{\url{#2}}

\bibitem{prl-1}
\textsc{C.~M. Bender} and \textsc{S.~Boettcher}, Real spectra in non--Hermitian
  Hamiltonians having PT symmetry, \emph{Phys. Rev. Lett.} 80:5243 (1998).

\bibitem{prl-2}
\textsc{C.~M. Bender}, \textsc{S.~Boettcher}, and \textsc{P.~N. Meisinger},
  PT--symmetric quantum mechanics, \emph{J. Math. Phys.} 40:2201 (1999).

\bibitem{prl-3}
\textsc{C.~M. Bender}, \textsc{D.~C. Brody}, and \textsc{H.~F. Jones}, Complex
  extension of quantum mechanics, \emph{Phys. Rev. Lett.} 89:270401 (2003).

\bibitem{prl-4}
\textsc{A.~Mostafazadeh}, Exact PT--symmetry is equivalent to Hermiticity,
  \emph{J.Phys.A:Math Gen.} 36:7081 (2003).

\bibitem{prl-5}
\textsc{C.~M. Bender}, \textsc{D.~C. Brody}, \textsc{H.~F. Jones}, and
  \textsc{B.~K. Meister}, Faster than Hermitian quantum mechanics, \emph{Phys.
  Rev. Lett.} 98:040403 (2007).

\bibitem{2}
\textsc{A.~Ruschhaupt}, \textsc{F.~Delgado}, and \textsc{J.~Muga}, Physical
  realization of PT--symmetric potential scattering in a planar slab waveguide,
  \emph{J.Phys.A:Math Gen.} 38: 171 (2005).

\bibitem{3}
\textsc{R.~El-Ganainy}, \textsc{K.~Makris}, \textsc{D.~Christodoulides}, and
  \textsc{Z.~H. Musslimani}, Theory of coupled optical PT--symmetric structures,
  \emph{Opt. Lett.} 32:2632 (2007).

\bibitem{4}
\textsc{H.~Cartarius} and \textsc{G.~Wunner}, Model of a PT--symmetric
  Bose--Einstein condensate in a $\delta$--function double--well potential,
  \emph{Phys.Rev.A} 86:013612 (2012).

\bibitem{5}
\textsc{J.~Schindler}, \textsc{A.~Li}, \textsc{M.~C. Zheng}, \textsc{F.~M.
  Ellis}, and \textsc{T.~Kottos}, Experimental study of active LRC circuits
  with PT symmetries, \emph{Phys.Rev.A} 84:040101 (2011).

\bibitem{6}
\textsc{C.~M. Bender}, \textsc{B.~K. Berntson}, \textsc{D.~Parker}, and
  \textsc{E.~Samuel}, Observation of PT phase transition in a simple mechanical
  system, \emph{Am. J. Phys} 81:173 (2013).

\bibitem{magnetics}
\textsc{T.~Gadzhimuradov} and \textsc{A.~Agalarov}, Towards a gauge--equivalent
  magnetic structure of the nonlocal nonlinear Schr{\"o}dinger equation,
  \emph{Phys.Rev.A} 93:062124 (2016).

\bibitem{yang}
\textsc{J.~Yang}, Partially PT symmetric optical potentials with all--real
  spectra and soliton families in multidimensions, \emph{Opt. Lett.}
  39:1133--1136 (2014).

\bibitem{par1}
\textsc{Y.~V. Kartashov}, \textsc{V.~V. Konotop}, and \textsc{L.~Torner},
  Topological states in partially--PT--symmetric azimuthal potentials,
  \emph{Phys. Rev. Lett.} 115:193902 (2015).

\bibitem{par2}
\textsc{J.~Yang}, Symmetry breaking of solitons in two--dimensional complex
  potentials, \emph{Phys.Rev.E} 91:023201 (2015).

\bibitem{par3}
\textsc{A.~Beygi}, \textsc{S.~Klevansky}, and \textsc{C.~M. Bender}, Coupled
  oscillator systems having partial PT symmetry, \emph{Phys.Rev.A} 91:062101
 (2015).

\bibitem{par4}
\textsc{C.~Huang} and \textsc{L.~Dong}, Stable vortex solitons in a ring--shaped
  partially--PT--symmetric potential, \emph{Opt. Lett.} 41:5194 (2016).

\bibitem{par5}
\textsc{S.~V. Suchkov}, \textsc{A.~A. Sukhorukov}, \textsc{J.~Huang},
  \textsc{S.~V. Dmitriev}, \textsc{C.~Lee}, and \textsc{Y.~S. Kivshar},
  Nonlinear switching and solitons in PT--symmetric photonic systems,
  \emph{Laser Photonics Rev.} 10:177 (2016).

\bibitem{par6}
\textsc{V.~V. Konotop}, \textsc{J.~Yang}, and \textsc{D.~A. Zezyulin},
  Nonlinear waves in PT--symmetric systems, \emph{Rev. Mod. Phys.} 88:035002
  (2016).

\bibitem{fokas-pt}
\textsc{A.~Fokas}, Integrable multidimensional versions of the nonlocal
  nonlinear Schr{\"o}dinger equation, \emph{Nonlinearity} 29:319 (2016).

\bibitem{ab-pt}
\textsc{M.~J. Ablowitz} and \textsc{Z.~H. Musslimani}, Integrable nonlocal
  nonlinear equations, \emph{Stud. Appl. Math.} (2016), n/a--n/a.

\bibitem{ablowitz}
\textsc{M.~J. Ablowitz} and \textsc{Z.~H. Musslimani}, Integrable nonlocal
  nonlinear Schr{\"o}dinger equation, \emph{Phys. Rev. Lett.} 110:064105
  (2013).

\bibitem{jigu}
\textsc{J.~Rao}, \textsc{Y.~Zhang}, \textsc{A.~Fokas}, and \textsc{J.~He},
  Rogue waves of the nonlocal Davey--Stewartson I equation, \emph{Summitted}
  (2016).

\bibitem{zzx2}
\textsc{Z. Zhou}, Darboux transformations and global explicit solutions for
  nonlocal Davey--Stewartson I equation, \emph{arXiv:1612.05689} (2016).

\bibitem{zhang}
\textsc{Y.~Zhang}, \textsc{D.~Qiu}, \textsc{Y.~Cheng}, and \textsc{H.~J.S},
  Rational solution of the nonlocal nonlinear Schr\"odinger equation and its
  application, \emph{Rom. J. Phys.} 62:108 (2017).

\bibitem{chow}
\textsc{Z.~Xu} and \textsc{K.~Chow}, Breathers and rogue waves for a third
  order nonlocal partial differential equation by a bilinear transformation,
  \emph{Appl. Math. Lett.} 56:72 (2016).

\bibitem{xutao}
\textsc{M.~Li} and \textsc{T.~Xu}, Dark and antidark soliton interactions in
  the nonlocal nonlinear Schr{\"o}dinger equation with the self-induced
  parity--time--symmetric potential, \emph{Phys.Rev.E} 91:033202 (2015).

\bibitem{zuo2}
\textsc{X. Wen}, \textsc{Z.~Yan}, and \textsc{Y.~Yang}, Dynamics of
  higher--order rational solitons for the nonlocal nonlinear Schr{\"o}dinger
  equation with the self--induced parity--time--symmetric potential, \emph{Chaos}
  26:063123 (2016).

\bibitem{zuo1}
\textsc{X.~Huang} and \textsc{L.~Ling}, Soliton solutions for the nonlocal
  nonlinear Schr{\"o}dinger equation, \emph{Eur.Phys.J.Plus}
  131:148 (2016).



\bibitem{ab-pt1}
\textsc{M.~J. Ablowitz}, \textsc{X.-D. Luo}, and \textsc{Z.~H. Musslimani},
  Inverse scattering transform for the nonlocal nonlinear
  Schr\"odinger equation with nonzero boundary conditions,
  \emph{arXiv:1612.02726} (2016).


\bibitem{ab-pt2}
\textsc{M.~J. Ablowitz} and \textsc{Z.~H. Musslimani}, Inverse scattering
  transform for the integrable nonlocal nonlinear Schr{\"o}dinger equation,
  \emph{Nonlinearity} 29:915 (2016).


\bibitem{zzx1}
\textsc{Z.~X. Zhou}, Darboux transformations and global solutions for a
  nonlocal derivative nonlinear Schr\"odinger equation, \emph{arXiv:1612.04892}
  (2016).

\bibitem{lou}
\textsc{S.~Lou}, Alice-bob systems, $ Ps $--$ Td $-$ C$ principles and
  multi--soliton solutions, \emph{arXiv:1603.03975} (2016).

\bibitem{zhiwei}
\textsc{Z.~Wu} and \textsc{J.~He}, New hierarchies of derivative nonlinear
Schr\"odinger--type equation, \emph{Rom. Rep. Phys.} 68:79 (2016).

\bibitem{wei}
\textsc{W.~Liu}, \textsc{D. Qiu}, \textsc{Z. Wu}, and \textsc{J. He},
  Dynamical behavior of solution in integrable nonlocal
  Lakshmanan--Porsezian--Daniel equation, \emph{Commun. Theor. Phys.} 65:671
  (2016).

\bibitem{taojpsj}
\textsc{M.~Li}, \textsc{T.~Xu}, and \textsc{D.~Meng}, Rational solitons in the
  parity--time--symmetric nonlocal nonlinear Schr{\"o}dinger model, \emph{J.
  Phys. Soc. Jpn.} 85:124001 (2016).

\bibitem{hirota}
\textsc{R.~Hirota}, \emph{The Direct Method in Soliton Theory}, Cambridge
  University Press, 2004.

\bibitem{long1}
\textsc{M.~J. Ablowitz} and \textsc{J.~Satsuma}, Solitons and rational
  solutions of nonlinear evolution equations, \emph{J. Math. Phys.}
  19:2180(1978).

\bibitem{long2}
\textsc{J.~Satsuma} and \textsc{M.~J. Ablowitz}, Two--dimensional lumps in
  nonlinear dispersive systems, \emph{J. Math. Phys.} 20:1496 (1979).

\bibitem{3-NLS}
\textsc{L.~C. Zhao} and \textsc{J.~Liu}, Rogue--wave solutions of a
  three--component coupled nonlinear Sch\"odinger equation, \emph{Phys.Rev.E}
  87:013201 (2013).

\bibitem{KP1}
\textsc{P.~Dubard} and \textsc{V.~B. Matveev}, Multi--rogue waves solutions:
  from the NLS to the KP--I equation, \emph{Nonlinearity} 26:93 (2013).

\bibitem{KP2}
\textsc{J.~Rao}, \textsc{L.~Guo}, \textsc{C.~Qian}, and \textsc{J.~He}, Dynamics
  of high--order rogue waves on a multi--solitons background in the KP--I
  equation, \emph{Submmitted} (2016).

\bibitem{3D}
\textsc{S.~Chen}, \textsc{J.~M. Soto-Crespo}, \textsc{F.~Baronio},
  \textsc{P.~Grelu}, and \textsc{D.~Mihalache}, Rogue wave bullets in a
  composite (2+1)D nonlinear medium, \emph{Opt. Express} 24:15251
  (2016).

\bibitem{kkp}
\textsc{P.~Gaillard}, Fredholm and Wronskian representations of solutions to
  the KPI equation and multi--rogue waves, \emph{J. Math. Phys.} 57:063505
  (2016).

\bibitem{eur}
\textsc{D.~J. Kedziora}, \textsc{A.~Ankiewicz}, and \textsc{N.~Akhmediev},
  Rogue waves and solitons on a cnoidal background, \emph{Eur. Phys. J. Special
  Topics} 223:43 (2014).

\bibitem{2-nls}
\textsc{L. Zhao}, \textsc{B.~Guo}, and \textsc{L.~Ling}, High--order rogue
  wave solutions for the coupled nonlinear Schr\"odinger equations--II, \emph{J.
  Math. Phys.} 57:043508 (2015).

\bibitem{3DKP}
\textsc{C.~Qian}, \textsc{J.~Rao}, \textsc{R.~Liu}, and \textsc{J.~He}, Rogue
  waves in the Three-Dimensional Kadomtsev--Petviashvili equation, \emph{Chin.Phys.Lett} 33:110201 (2016).

\bibitem{PRE}
\textsc{M.~Tajiri} and \textsc{T.~Arai}, Growing--and--decaying mode solution to
  the Davey--Stewartson equation, \emph{Phys.Rev.E} 60:2297 (1999).

\bibitem{DS1}
\textsc{Y.~Ohta} and \textsc{J.~Yang}, Rogue waves in the Davey--Stewartson I
  equation, \emph{Phys.Rev.E} 86:036604 (2012).

\bibitem{DS2}
\textsc{Y.~Ohta} and \textsc{J.~Yang}, Dynamics of rogue waves in the
  Davey--Stewartson II equation, \emph{J.Phys.A: Math. Phys.} 46:105202 (2013).

\end{thebibliography}
 \end{document}